%% file: aanda.tex
\documentclass{aa}  

\usepackage{graphicx}
\usepackage{txfonts}
\usepackage{makecell}
\usepackage{gensymb}
\usepackage{booktabs}
\usepackage{float}
\usepackage{hyperref}
\hypersetup{colorlinks, linkcolor=blue, citecolor=blue, urlcolor=blue}

\makeatletter

\begin{document}

   \title{Photometric properties of nuclear star clusters and their host galaxies in the Fornax cluster\thanks{Tables \ref{tab:galaxy_sample} and \ref{tab:nucleus_sample} are only available in electronic form at the CDS via anonymous ftp to cdsarc.u-strasbg.fr (130.79.128.5) or via \url{http://cdsweb.u-strasbg.fr/cgi-bin/qcat?J/A+A/}}}

  \subtitle{}

   \author{Alan H. Su\inst{1}
          \and
          Heikki Salo\inst{1}
          \and
          Joachim Janz\inst{2,1,3}
          \and
          Aku Venhola\inst{1}
          \and
          Reynier F. Peletier\inst{4}
}

   \institute{Space Physics and Astronomy Research Unit, University of Oulu, Pentti Kaiteran katu 1, FI-90014 Oulu, Finland\\
              \email{hung-shuo.su@oulu.fi}
              \and
              Finnish Centre of Astronomy with ESO (FINCA), Vesilinnantie 5, University of Turku, FI-20014 Turku, Finland
              \and
              Specim, Spectral Imaging Ltd., Elektroniikkatie 13, FI-90590 Oulu, Finland
              \and
              Kapteyn Institute, University of Groningen, Landleven 12, 9747 AD Groningen, The Netherlands
             }

   \date{Received - / Accepted - }

 
  \abstract
   {}
   {We investigate the relations between nuclear star clusters (NSCs) and their host galaxies, as well as conduct a comparison between the structural properties of nucleated and non-nucleated galaxies. We also address the environmental influences on the nucleation of galaxies in the Fornax main cluster and the Fornax A group. }
   {We select 557 galaxies ($10^{5.5} M_{\odot} < M_{\rm *,galaxy} < 10^{11.5} M_{\odot} $) for which structural decomposition models and non-parametric morphological measurements are available from our previous work. We determine the nucleation of galaxies based on a combination of visual inspection of galaxy images and residuals from multi-component decomposition models, as well as use a model selection statistic, the Bayesian information criterion (BIC), to avoid missing any faint nuclei. We also test the BIC as an unsupervised method to determine the nucleation of galaxies. We characterise the NSCs using the nucleus components from the multi-component models conducted in the $g'$, $r'$, and $i'$ bands. }
   {Overall, we find a dichotomy in the properties of nuclei which reside in galaxies more or less massive than $M_{\rm *,galaxy} \approx 10^{8.5} M_{\odot}$. In particular, we find that the nuclei tend to be bluer than their host galaxies and follow a scaling relation of $M_{\rm *,nuc} \propto {M_{\rm *,galaxy}}^{0.5}$ for $M_{\rm *,galaxy} < 10^{8.5} M_{\odot}$. In galaxies with $M_{\rm *,galaxy} > 10^{8.5} M_{\odot}$ we find redder nuclei compared to the host galaxy which follow $M_{\rm *,nuc} \propto M_{\rm *,galaxy}$. Comparing the properties of nucleated and non-nucleated early-type galaxies, we find that nucleated galaxies tend to be redder in global ($g'-r'$) colour, have redder outskirts relative to their own inner regions ($\Delta (g'-r')$), be less asymmetric ($A$) and exhibit less scatter in the brightest second order moment of light ($M_{20}$) than their non-nucleated counterparts at a given stellar mass. However, with the exception of $\Delta (g'-r')$ and the Gini coefficient ($G$), we do not find any significant correlations with cluster-centric distance. Yet, we find the nucleation fractions to be typically higher in the Fornax main cluster than in the Fornax A group, and that the nucleation fraction is highest towards the centre of their respective environments. Additionally, we find that the observed ultra-compact dwarf (UCD) fraction (i.e. the number of UCDs over the number of UCDs and nucleated galaxies) in Fornax and Virgo peak at the cluster centre, and is consistent with the predictions from simulations. Lastly, we find that the BIC can recover our labels of nucleation up to an accuracy of 97\% without interventions. }
   {The different trends in NSC properties suggest that different processes are at play at different host stellar masses. A plausible explanation is that the combination of globular cluster in-spiral and in-situ star formation play a key role in the build-up of NSCs. In addition, the environment is clearly another important factor in the nucleation of galaxies, particularly at the centre of the cluster where the nucleation and UCD fractions peak. Nevertheless, the lack of significant correlations with the structures of the host galaxies is intriguing. Finally, our exploration of the BIC as a potential method of determining nucleation have applications for large scale future surveys, such as Euclid. }

    
   \maketitle
%

\section{Introduction}
The nucleation of galaxies typically refers to a bright compact object located at the central region of the galaxy, composed of a massive stellar cluster also known as a nuclear star cluster (NSC). These objects typically appear as an excess of light in the innermost part of the surface brightness profiles of galaxies. Although nuclei typically only take up a few percent of the total light of a given galaxy \citep{cote2006, turner2012, georgiev2016, sanchezjanssen2019_nsc}, their prevalence across galaxies in our Local Universe cannot be underestimated. Galaxies which host nuclei can exhibit a range of properties, from low to high stellar masses and between early- and late-type galaxies \citep[see e.g.][]{georgiev2016, neumayer2020}. Indeed, even our own Milky Way has been observed to host a NSC \citep[see][and references therein]{fritz2016}. Furthermore, given the proximity of nuclei to the centre of the galaxies, it is possible that there can be some interplay with the central black holes in larger mass galaxies, which have comparable masses \citep{cote2006, antonini2013, arcasedda2016, greene2020}. 

How NSCs form and grow is an active topic of study. Recent studies attribute two main mechanisms to the growth of NSCs. The first is the in-spiral of globular clusters (GCs) towards the centre of a galaxy due to dynamical friction \citep{tremaine1975, capuzzodolcetta1993, oh2000, lotz2001}. Observationally, there are several works which found links between the GCs and NSCs of early-type galaxies, such as the lack of GCs in the central regions where NSCs reside \citep{lotz2001} and the similarity in NSC and GC occupation fraction \citep{sanchezjanssen2019_nsc}. This mechanism appears to dominate for lower mass galaxies ($M_* \lesssim 10^9 M_{\odot}$), given the similarity between the predicted and observed NSC-to-host stellar mass scaling relations \citep{antonini2013, gnedin2014, neumayer2020}. Another mechanism is the in-situ star formation from gas in the central region of a galaxy \citep{seth2006}. For example, the compression of gas in the central region of galaxies via tidal compression \citep{emsellem2008}, or fuelled by gas infall \citep{maciejewski2004, hunt2008, emsellem2015}, coalescence of star forming clumps \citep{bekki2006, bekki2007}, or due to magnetorotational instability \citep{milosavljevi2004}. Whilst the timescales vary between the aforementioned processes, in-situ star formation is considered to be the dominant channel of NSC growth for massive galaxies \citep[$M_* \gtrsim 10^9 M_{\odot}$; e.g.][]{paudel2011, johnston2020, fahrion2021, pinna2021}. Combining both cluster in-spiral and in-situ star formation, as well as effects from black holes, \citet{antonini2015} found that NSCs tend to dominate over the central black holes for dwarfs, and vice versa for massive galaxies \citep{graham2016}. 

In terms of the stellar population of NSCs, one NSC which has been observed in great detail is that of the Milky Way. The Milky Way NSC has a half-light radius of $7\pm2\,$pc and a mass of $4.2\times 10^7 M_{\odot}$ \citep[][]{fritz2016}. Analyses from infrared surveys suggest that stars within the central parsec tend to be old ($9\pm 2$\,Gyr; \citealt{genzel2010}, their Sect.~6.1), similar to those from the Galactic Bulge. Stars belonging to the NSC (and its periphery) tend to be metal-rich \citep{schultheis2019, thorsbro2020}, although there is evidence that young \citep{feldmeierkrause2015}, and metal-poor \citep{ryde2016, feldmeierkrause2017} stars are also present. In principle, the metal-poor stars could have come from the in-spiral of GCs. However, a young stellar population is unlikely to originate from GCs and points to in-situ star formation \citep{nishiyama2016}. Another hypothetical scenario is the infall of young stellar clusters into the NSC, which is unlikely in the case of the Milky Way due to the predominantly old stellar population in the inner Galactic Bulge \citep{nogueraslara2018} and nuclear stellar disk \citep[NSD,][]{nogueraslara2019}\footnote{In actuality, the Milky Way's NSC is embedded within a NSD, which is flatter and more extended than the NSC itself \citep[see][]{launhardt2002}. For reference, the NSD region ($\lesssim 120$\,pc) contains a mass of $(8\pm2)\times10^8 M_{\odot}$ \citep{nogueraslara2019}.}. 

For extragalactic NSCs, stars can no longer be resolved individually, which hampers efforts to study the age and metallicity distributions. Nonetheless, space-based observations with high spatial resolution can still resolve NSCs as a whole, which led to several studies of NSC luminosity functions, nucleation fractions, scaling relations, and colours for galaxies in the Virgo \citep{cote2006, ferrarese2006}, Fornax \citep{turner2012}, and Coma \citep{denbrok2014} clusters. More recently, spectroscopic studies using integral field units (IFUs) have uncovered a mix of ages and metallicities for NSCs in galaxies in the Fornax cluster \citep[e.g.][]{johnston2020, fahrion2021}. The varying star formation histories suggest that both cluster in-spiral and in-situ star formation play a role in the formation and growth of most NSCs, depending on the host galaxy mass. 

A wealth of photometric data on nucleated galaxies have come from recent ground based, deep, and wide-field observations in cluster and group environments (e.g. FDS, \citealt{iodice2016}; NGVS, \citealt{ferrarese2012}; NGFS, \citealt{munoz2015}; ELVES, \citealt{carlsten2021_dwarf}; MATLAS, \citealt{habas2020}). In particular, \citet{venhola2019} found that nucleated and non-nucleated dwarfs in the Fornax cluster can show significantly different luminosity functions and structural scaling relations. For dwarfs in the Virgo cluster, \citet{sanchezjanssen2019_nsc} found that the nucleus occupation fraction peaks at $M_* \approx 10^9 M_{\odot}$. Recently, \citet{su2021} conducted multi-component decompositions on both the dwarfs \citep[from][]{venhola2018} and massive galaxies \citep[from][]{iodice2019, raj2019, raj2020} in the Fornax main cluster and the nearby Fornax A group. The multi-component decompositions, which included the nucleus components, provide structural parameters in multiple bands which we make use of in this work. 

This work is structured as follows. We describe the data and the sample we use in Sect.~\ref{sect:sample}, including our label of nucleation for each galaxy in our sample. In Sect.~\ref{sect:auto_mod_select} we test methods of automatically selecting the most appropriate decomposition models and apply them to determine the nucleation of galaxies. In Sect.~\ref{sect:nuc_frac} we show the nucleation fraction as a function of the environment (i.e. between Fornax main cluster and Fornax A group). In Sect.~\ref{sect:photo_prop} we present the properties of nucleated and non-nucleated galaxies in our sample and compare between them. In Sect.~\ref{sect:discussion} we compare the nucleated galaxies to mechanisms of NSC growth from the literature, discuss the role of environment on the structural properties of our sample galaxies, and compare the nuclei to ultra-compact dwarfs (UCDs). Finally, we summarise our results in Sect.~\ref{sect:conclusion}. Throughout this study we use a distance modulus of 31.5\,mag, which is equivalent to a distance of 20\,Mpc \citep{blakeslee2009}. At this distance 1\,arcsec corresponds to $\sim 0.097$\,kpc.


\section{Sample}\label{sect:sample}
\subsection{Data}\label{sect:data}
We utilise data from the Fornax Deep Survey (FDS), a combination of two guaranteed time observations surveys with the OmegaCAM instrument \citep{kuijken2002} at the VLT Survey Telescope (VST): FOCUS (P.I. R.~F.~Peletier) and VEGAS \citep[P.I. E.~Iodice,][]{capaccioli2015}. The FDS covers 26\,deg$^2$ around the Fornax main cluster and Fornax A group in $u'$, $g'$, $r'$, and $i'$ bands, with average seeing full width at half maximum (FWHM) of 1.2\,arcsec, 1.1\,arcsec, 1.0\,arcsec, and 1.0\,arcsec, respectively. The images have a pixel scale of 0.2\,arcsec\,pix$^{-1}$. In terms of the data depth, the 1$\sigma$ signal-to-noise per pixel can be converted to surface brightness of 26.6, 26.7, 26.1, 25.5\,mag\,arcsec$^{-2}$ for $u'$, $g'$, $r'$, and $i'$ bands, respectively. Alternatively, when averaged over an area of 1\,arcsec$^{2}$, the surface brightness correspond to 28.3, 28.4, 27.8, 27.2\,mag\,arcsec$^{-2}$ for $u'$, $g'$, $r'$, and $i'$ bands, respectively. The images are calibrated in the AB magnitude system. The full observation strategy can be found in \citet{iodice2016} and the reduction steps can be found in \citet{venhola2018}. The FDS data can be obtained from the ESO Science Archive \citep{peletier2020}. 

\begin{figure}
\centering
\includegraphics[width=\hsize]{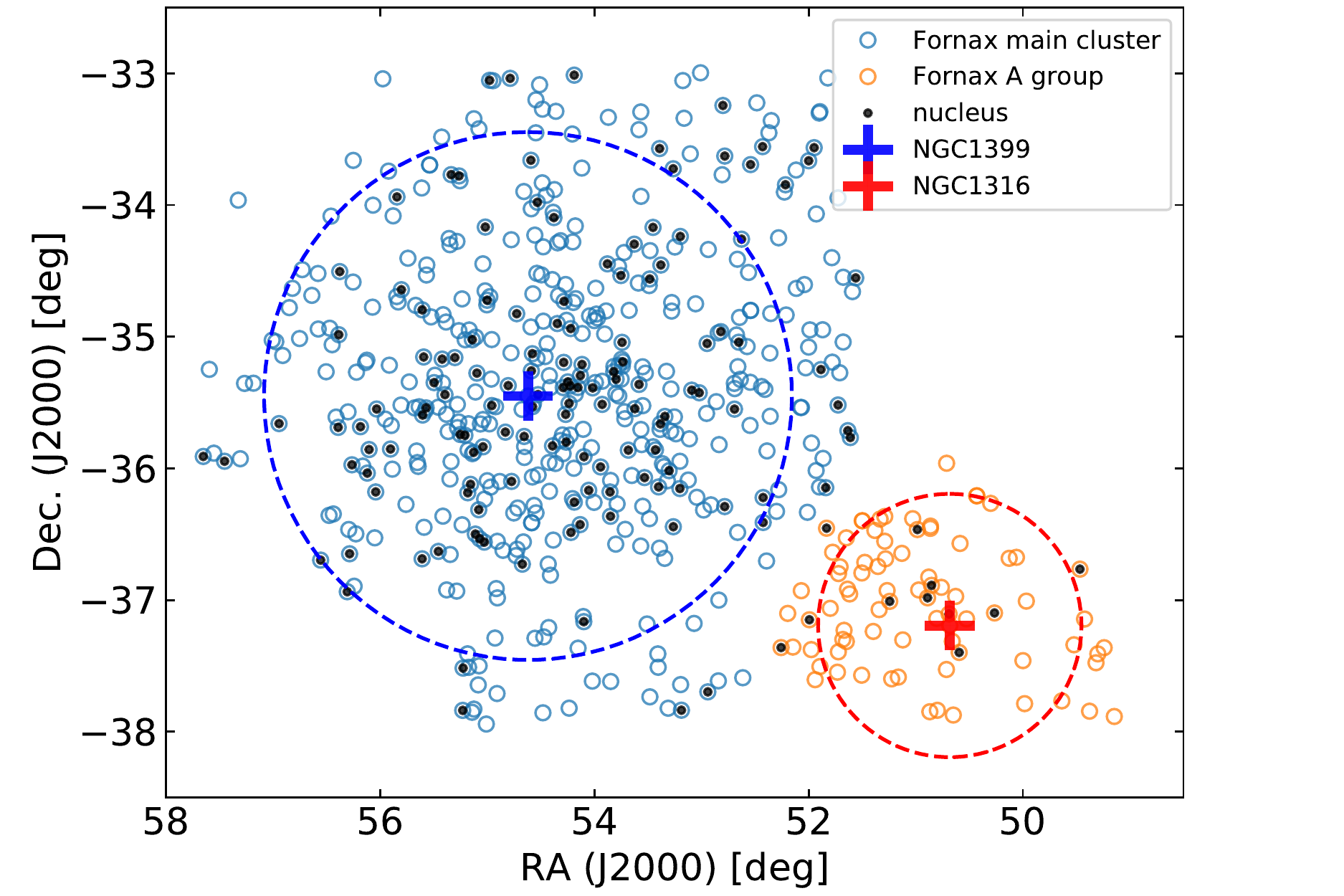}
\caption{Overview of Fornax galaxies, split between Fornax main cluster (blue) and Fornax A group (orange). The dashed circles denote the virial radius for the Fornax main cluster ($\sim$2\,deg=0.7\,Mpc) and the Fornax A group ($\sim$1\,deg=0.35\,Mpc) \citep{drinkwater2001}. The black dots denote galaxies with a nucleus.}
\label{fig:radec}
\end{figure}

To tackle the topic of galaxy nucleation, we make use of the analysis of FDS galaxies from \citet{su2021}, which consists of 582 galaxies (dwarfs from FDSDC; \citealt{venhola2018}, and massive galaxies from \citealt{iodice2019} and \citealt{raj2019, raj2020}) in the Fornax main cluster and Fornax A group (see Fig.~\ref{fig:radec}). These galaxies were deemed likely cluster and group members based on the selection cuts in \citet{venhola2018}. To briefly describe the data preprocessing steps from \citet{su2021}: postage stamp images for each member galaxy were cut in each band; the images were sky-subtracted using a constant value, except select cases where a strong gradient requires a plane subtraction; a separate PSF was created for each FDS field by averaging the radial flux profiles of point sources in the field and interpolating, which produced axisymmetric PSFs \citep[for more details see][their Sect.~3]{su2021}. The compilation in \citet{su2021} contains quantities derived from structural decompositions as well as non-parametric morphological indices. The structural decompositions were conducted using GALFIT \citep{peng2010_galfit}, with two types of models for each galaxy: Sérsic+PSF and multi-component. The Sérsic+PSF model fit the unresolved nucleus (if present) with the PSF and the galaxy with a Sérsic function. Constraints were applied such that the integrated magnitude of the nucleus component cannot be fainter than 35\,mag. This allowed the Sérsic parameters to remain accurate even for non-nucleated galaxies and prevented GALFIT from failing. 

From the multi-component models we obtain the morphological structures present in the galaxies, such as bulges, bars, disks, and nuclei using the Sérsic, Ferrers, exponential, and the PSF functions to model them, respectively in \citet{su2021}. The multi-component decompositions were first conducted using the $r'$ band images, with the parameters from each component allowed to vary to determine their best-fit values. To obtain the magnitudes of individual components in the $g'$ and $i'$ bands, only the magnitude parameter was allowed to vary for each component, whilst the other parameters were fixed to the best-fit values obtained from the $r'$ band model. This ensured that an equivalent aperture was used for each galaxy across different bands. In conjunction, the non-parametric morphological indices provide a model independent perspective of the galaxies, including deviations from smooth, elliptical light distributions. The combination of quantities allows us to probe the properties of the nuclei themselves, along with the properties of the host galaxies. The analyses and relevant images can be found in \citet{su2021} as well as on our web page\footnote{\url{https://www.oulu.fi/astronomy/FDS_DECOMP/main/index.html}}.

The compilation also includes the stellar mass of the galaxies. The stellar mass was estimated using a combination of $g'$, $r'$, and $i'$ band magnitudes and the empirical relation based on \citet{taylor2011}
\begin{equation}
    \log_{10}\left( \frac{M_*}{M_{\odot}} \right)=1.15+0.70(g'-i')-0.4M_{r'}+0.4(r'-i'), \label{eqn:mstar}
\end{equation}
where $M_{r'}$ is the absolute $r'$ band total magnitude, and the $g'-i'$ and $r'-i'$ are the total colours based on multi-component decompositions \citep[see also][their Sect.~5]{su2021}. Moreover, we use the galaxy coordinates to calculate the projected separation of the galaxies to NGC1399 and NGC1316, as proxies for the centre of the Fornax main cluster and Fornax A group environments, respectively. We assign each galaxy as a member of either the cluster or group environment following \citet{su2021}, as well as adopt an early- and late-type classification from \citet{venhola2018}. 

\subsection{Nucleus detection}\label{sect:nuc_detect}
The distinction between nucleated and non-nucleated galaxies in \citet{su2021} was based on the visual identification of components which were then modelled in multi-component decompositions. Whilst this generally worked well for the bright nuclei in our galaxies, it may fail to capture the faintest nuclei. To check that we did not miss any nuclei, we now add a nucleus component to the multi-component decomposition models of all non-nucleated galaxies (i.e. those without a nucleus component in their multi-component models in \citealt{su2021}) in our sample and rerun the decomposition models via GALFIT. Similar to the Sérsic+PSF decompositions, we apply a lower limit on the nucleus component so that their integrated magnitude cannot be fainter than 35\,mag, otherwise GALFIT would crash for (truly) non-nucleated galaxies. 

To evaluate whether the added nucleus component provides an improvement for each galaxy, we use the Bayesian Information Criterion \citep[BIC, see][]{schwarz1978} as a quantitative indicator. The $BIC$ can be formulated \citep[according to][]{kass1995} as 
\begin{equation}
    BIC=\chi ^2+k\ln{n_{\textup{pix}}},\label{eqn:bic}
\end{equation}
where $k$ is the number of free parameters (e.g. our Sérsic+PSF model has six free parameters: five from the Sérsic function and one from the PSF function), $n_{pix}$ is the number of pixels used in model fitting (excluding masked pixels), and
\begin{equation}
    \chi^2=\sum_{x}\sum_{y}\frac{\left[O(x,y)-M(x,y)\right]^2}{\sigma(x,y)^2},
\end{equation}
where $O$ is the galaxy image, $M$ is the PSF-convolved decomposition model, $\sigma$ is the uncertainty in the pixels (i.e. sigma image), and $x$ and $y$ denote the pixel index in the x- and y-axes of the images, respectively. In essence, the first term in Eq.~\ref{eqn:bic} describes how well the model fits the data whilst the second term provides a penalty for model complexity, essentially averting models with more components than necessary.

\citet{head2014} \citep[see also][]{simard2011} argue that the pixels in images are not independent due to effects of the PSF. Hence, an estimate of the resolution element is given as the circular area with radius as the half-width-at-half-maximum (HWHM) in pixels (i.e. $A_{\textup{res}}=\pi\,$HWHM$^2$)\footnote{The mean PSF HWHM over all FDS fields in the $r'$ band is $\sim 2.4$\,pix}. This modified BIC can be calculated as
\begin{equation}
    BIC_{\textup{res}}=\frac{\chi ^2}{A_{\textup{res}}} +k\ln{\frac{n_{\textup{pix}}}{A_{\textup{res}}}}.\label{eqn:bic_res}
\end{equation}
The additional $A_{\textup{res}}$ term means that in general, $BIC_{\textup{res}}$ penalises complex models more than the $BIC$. 

To use the BIC for model selection, each model requires its BIC value to be calculated. The model with the lowest BIC is favoured. This can be formulated as $\Delta$BIC$=$BIC$_{\textup{no nuc}}-$BIC$_{\textup{nuc}}$ for both $BIC$ and $BIC_{\textup{res}}$, such that a positive $\Delta$BIC means the nucleated model is preferred. We note that $\Delta$BIC is the key quantity in determining whether a nucleated model is preferred, rather than the BIC itself. In fact, by definition, the $\chi^2$ (and hence the BIC) increases with larger image size. Therefore, for a given galaxy model, the BIC for a postage stamp image which contains a large portion of the background sky would be higher than for a postage stamp image which includes less of the sky. However, as the sky region does not affect the decomposition parameters, the increase in the BIC is mainly driven by the change in the image size, which cancels out when the $\Delta$BIC is calculated. We test and confirm that the $\Delta$BIC is insensitive to the image size used, provided that the image includes the entire galaxy.

\subsection{Final sample}\label{sect:final_sample}

\begin{figure}
\centering
\includegraphics[width=\hsize]{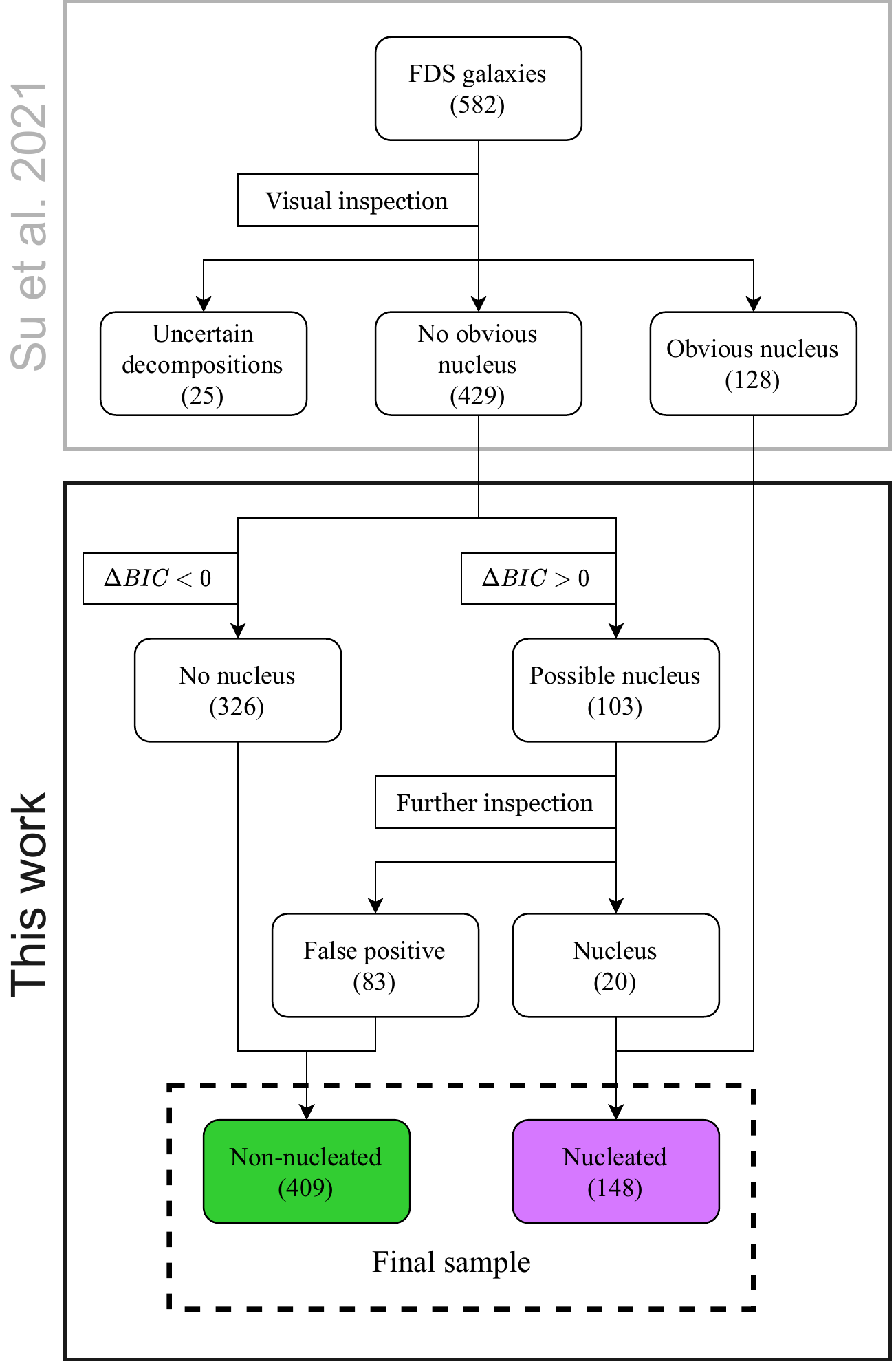}
\caption{Flow chart of our final galaxy sample. See Sect.~\ref{sect:final_sample} for details.}
\label{fig:sample_flowchart}
\end{figure}

To summarise the process of deriving our final working sample, we show the steps in Fig.~\ref{fig:sample_flowchart}. From the compilation sample of 582 galaxies, 25 were removed due to unsatisfactory decomposition models (for more details see Appendix~B and footnote~10 of \citealt{su2021}), leaving 557 galaxies, of which 128 were determined as nucleated in \cite{su2021}. Given the greater sensitivity of $BIC$ in detecting faint nuclei over $BIC_{\textup{res}}$, for each non-nucleated galaxy we calculate the $BIC$ for the original multi-component model and the corresponding model with the added nucleus component. Hence, galaxies which have a positive $\Delta BIC$ are those which are potentially nucleated, even if they were not treated as nucleated in \citet{su2021}. Of these 103 galaxies, we visually inspect the central 10\,arcsec radius of the original galaxy images and residuals (from both types of decomposition models) and found 20 galaxies which, in fact, host a nucleus\footnote{In Appendix~\ref{app:new_nucleated} we show the galaxy and residual images.}. For the purpose of comparing between nucleated and non-nucleated galaxies in this study, we also label these 20 galaxies as nucleated. In total, our sample includes 148 nucleated galaxies and 409 non-nucleated galaxies which we use for further analysis. 

\begin{figure}
\centering
\includegraphics[width=\hsize]{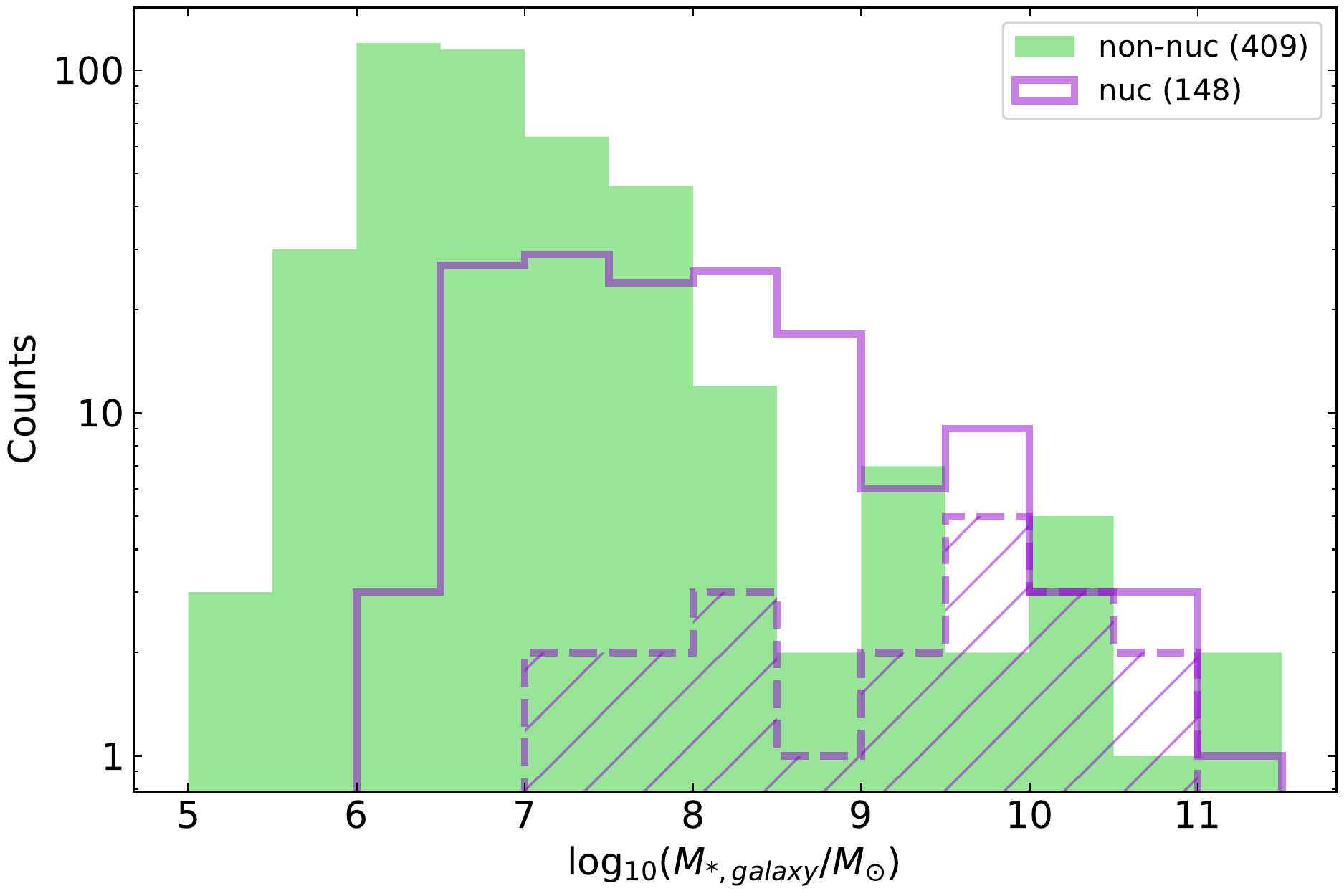}
\caption{Histograms of host galaxy stellar masses for nucleated (violet) and non-nucleated (green) galaxies in the Fornax main cluster and Fornax A group. The dashed histogram denotes the distribution of newly identified nucleated galaxies (see Sect.~\ref{sect:final_sample}). Bins with widths of 0.5\,dex were used.}
\label{fig:hist_nucleation}
\end{figure}

In Fig.~\ref{fig:hist_nucleation} we show the distribution of stellar mass for nucleated and non-nucleated galaxies in our sample. Of the newly identified nuclei, many belong to massive ($M_* > 10^9 M_{\odot}$) galaxies with additional structures (e.g. bulges and bars). In total, 140 out of 466 early-type galaxies and 8 out of 91 late-type galaxies are nucleated. Based on our completeness tests, our nucleus sample should be complete to around $M_{\rm *,nuc} \sim 10^{4.5}\,M_{\odot}$ (see Appendix~\ref{sect:completeness} for details on the completeness). In Table~\ref{tab:galaxy_sample} we present the photometric properties of a few of our galaxies as examples; the full table can be found online. 

\begin{table*}[!ht]
    \centering
    \caption{Photometric properties of our galaxies}
    \input{table_galaxy_short.txt}
    \tablefoot{We identify each galaxy by their FDS identification number (1) and their Fornax cluster catalogue \citep[FCC,][]{ferguson1989} designation (2), where possible, as well as their right ascension (3) and declination (4). Using the multi-component decomposition models, we determine the stellar mass (5), absolute $r'$ band magnitude (6) and $g'-r'$ (7) and $g'-i'$ (8) colours for the galaxies. Here only an excerpt is shown; the full table can be found online.}
    \label{tab:galaxy_sample}
\end{table*}

\begin{figure}
\centering
\includegraphics[width=\hsize]{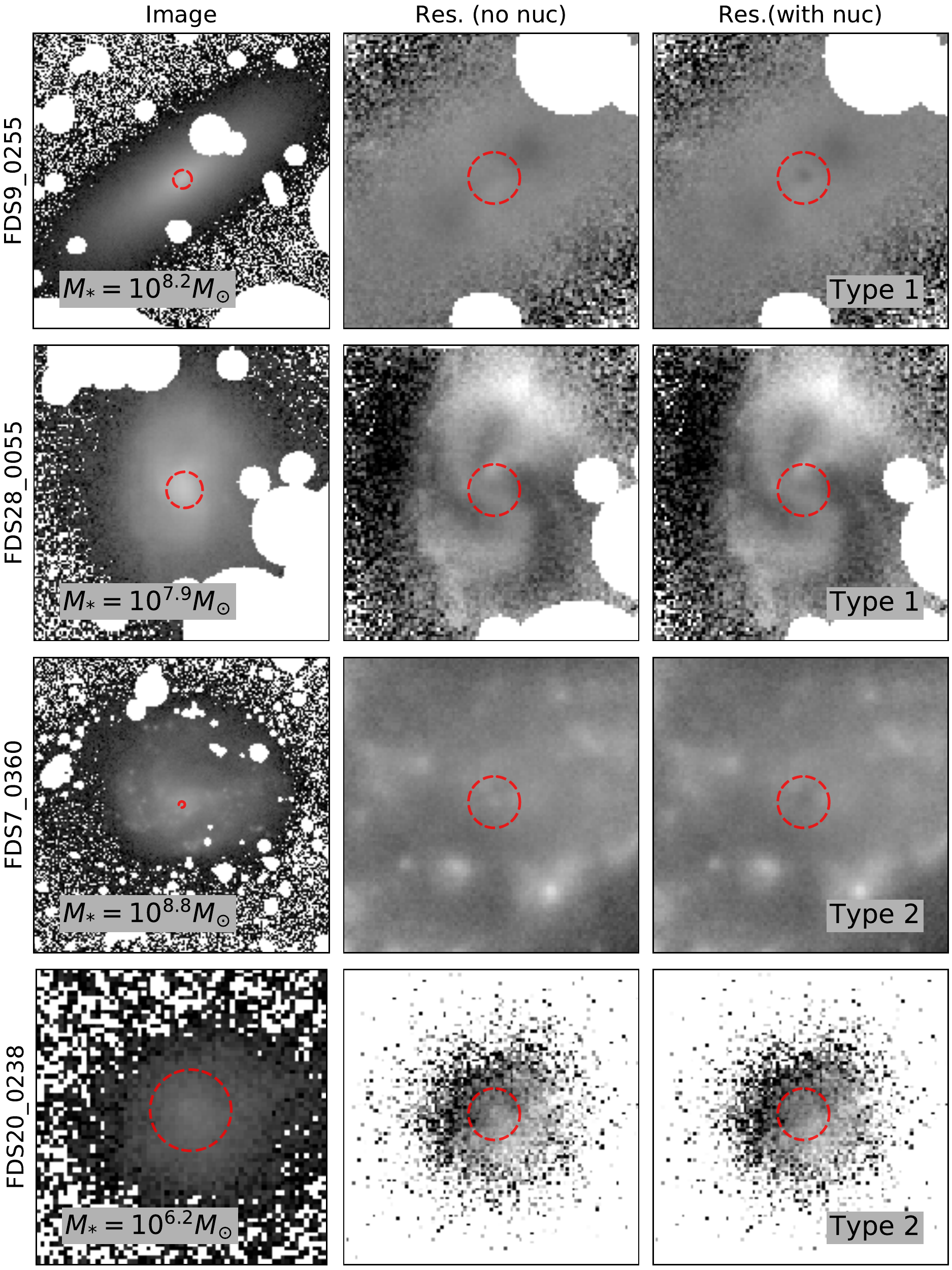}
\caption{Examples of $BIC$ false positives (i.e. $\Delta BIC > 0$ but no nucleus). We show the images (left) and residuals based on the multi-component model without (centre) and with (right) a nucleus component. The white regions denote masked regions. The images in the first column have widths of  $4R_e$ and surface brightness limits between 18--27\,mag\,arcsec$^2$, whereas the residual images have widths of 20\,arcsec and are limited from -1 to 1\,mag\,arcsec$^2$ in surface brightness. The red circles have a radius of the average PSF FWHM$\times 2$. The galaxy stellar masses are annotated in the first column. }
\label{fig:bic_falsepositive}
\end{figure}

From detailed inspections, we find that the 83 false positive galaxies (i.e. $\Delta BIC > 0$ but not nucleated, see Fig.~\ref{fig:sample_flowchart}) broadly fall under one of two types. The first type (type 1) encompasses galaxies which do not host a nucleus but have additional sub-structures in the central regions of the galaxy, such that the inclusion of a nucleus component can reduce the residuals (and hence the corresponding $BIC$). These sub-structures are generally small deviations from a single Sérsic model (70), with some (12) showing signs of more complex and asymmetric structures (e.g. star-forming clumps, spirals). A few of the galaxies (13) fall into the second type of false positives (type 2). They generally have an unresolved compact object at the centre of their image, but the Sérsic component of the models are clearly off-centred. This can be due to the asymmetric shape of the galaxy on the whole, or the proximity of GCs in the central region of the galaxy. Due to the implied offset of the potential nucleus from the centre of the galaxy, it is somewhat ambiguous whether they are truly NSCs. As such, in these cases we do not label them as nucleated, resulting in their false positive label. In Fig.~\ref{fig:bic_falsepositive} we show examples of both types of false positives in terms of their images and residuals from multi-component models with and without a nucleus component. 

The ramification of our false positive classifications is that our nucleation sample is likely biased against clearly off-centred NSCs and, by implication, late-types. Of the 83 false positives, we find that 36 are late-types, and that 10 out of 13 type 2 false positives are late-types. If we assume that between 10 to 36 late-type false positives actually host an offset nucleus, the late-type nucleation fraction would increase from 8/91$\approx$0.09 to between 0.20 and 0.48, with the latter as an upper limit. This would broadly be comparable to the early-type nucleation fraction ($\approx$0.3). 

\section{Nucleation fraction}\label{sect:nuc_frac}
\begin{figure}
\centering
\includegraphics[width=\hsize]{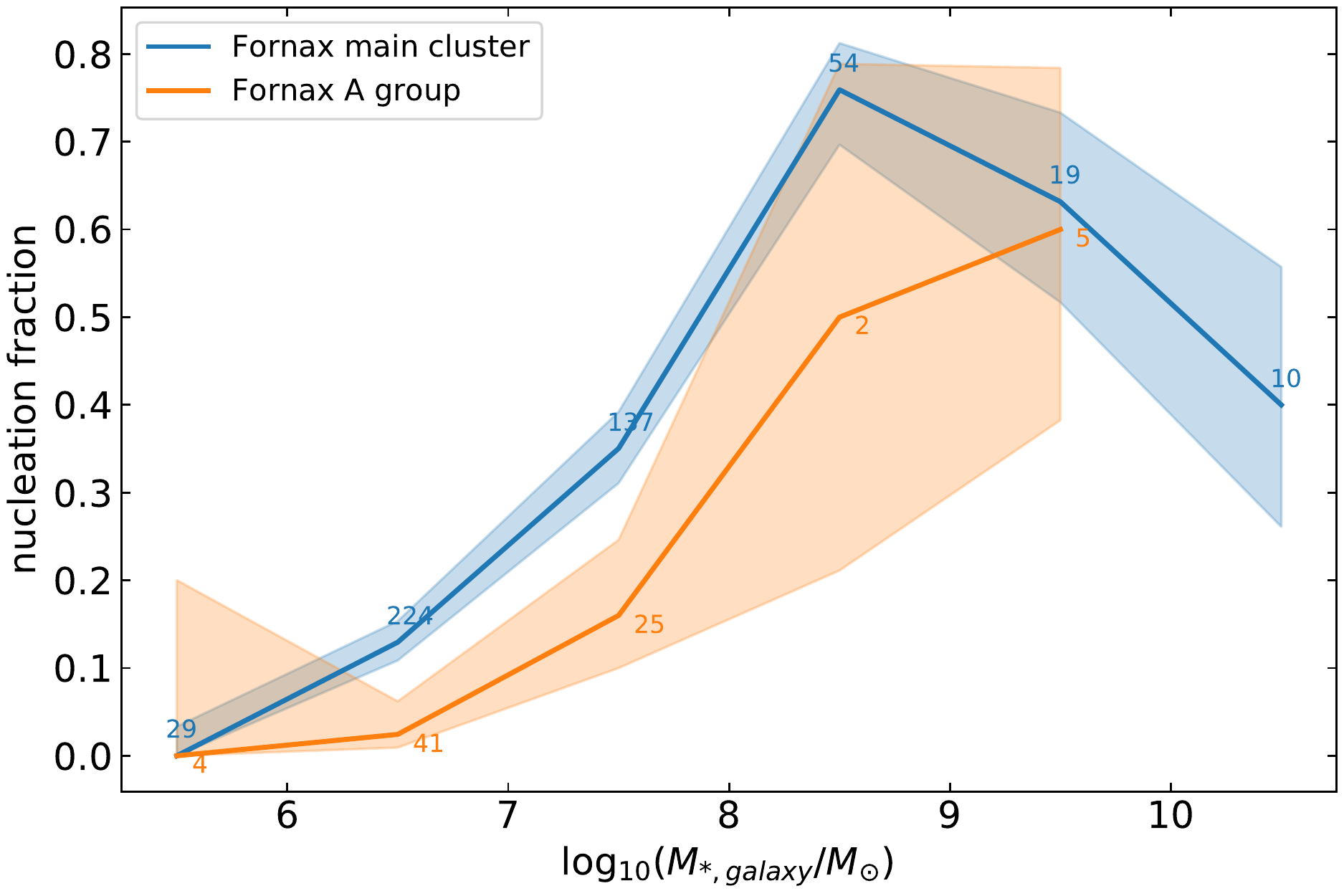}
\caption{Nucleation fraction as a function of host galaxy stellar mass for the Fornax main cluster (blue) and the Fornax A group (orange). The shaded regions denote the 68\% binomial confidence interval based on the Wilson score interval \citep{wilson1927}. For both sub-samples a bin width of 1\,dex was used and the number of galaxies per bin is labelled in the plot. Due to the low number of galaxies, we omit bins with $M_* > 10^{10} M_{\odot}$ from the Fornax A group and $M_* > 10^{11} M_{\odot}$ from the Fornax main cluster. }
\label{fig:nuc_frac_maingroup}
\end{figure}

The nucleation of galaxies have been observed to depend on both the stellar mass and the environment that the host galaxy resides in \citep[e.g.][]{binggeli1987, cote2006, lisker2007, sanchezjanssen2019_nsc}; nucleated galaxies tend to be more massive and reside in denser environments than their non-nucleated counterparts. Here we investigate the nucleation fraction as a function of the galaxy stellar mass and the environment, specifically between the Fornax main cluster and Fornax A group. From Fig.~\ref{fig:nuc_frac_maingroup} we see that generally the nucleation fractions are lower in the Fornax A group than in the Fornax main cluster for all stellar masses. This is particularly clear around $M_* \sim 10^7 M_{\odot}$, where the nucleation fraction is significantly lower in the Fornax A group, potentially hinting towards a dependence of the environment on nucleation. Recently, \citet{carlsten2021_nuc} found that the nucleation fraction of dwarfs in the Local Volume appear to be lower than in the Fornax and Virgo clusters at $M_* \sim 10^7 M_{\odot}$. This is in agreement with our finding that the nucleation fraction is higher in the cluster environment. Additionally, we find that the nucleation fraction for the Fornax main cluster peaks at $M_{*,\textup{galaxy}} \sim 10^{8.5} M_{\odot}$, which is very similar to the peak found by \citet{sanchezjanssen2019_nsc} at $M_{*,\textup{galaxy}} \sim 10^{9} M_{\odot}$ for the Virgo cluster. 

Interestingly, the peak of nucleation fraction for the Fornax A group appears to be at higher galaxy masses than in the Fornax main cluster, although the modest sample size in the bins leads to the large uncertainties. Recently, \citet{zanatta2021} found a negative relation between the halo mass of the environment and the dwarf galaxy luminosity at a given nucleation fraction. In other words, the peak of the nucleation fraction occurs at a higher galaxy stellar mass for galaxies residing in a low halo mass environment, which corroborates our results. 

At the low-mass end, as the galaxy stellar mass decreases, the nucleation fraction also drops. This feature could be due to our detection limit (visual limit at $M_{\rm *,nuc} \approx 10^{4.5} M_{\odot}$), where we cannot always detect the lowest mass NSCs. On the other hand, it is possible that there is a limit in the mass of GCs which can survive the in-spiral process due to dynamical friction to form NSCs (see e.g. \citet{leaman2021}). As mentioned, upcoming surveys and data with a low enough seeing and resolution should be able to shed light on this 

\begin{figure}
\centering
\includegraphics[width=\hsize]{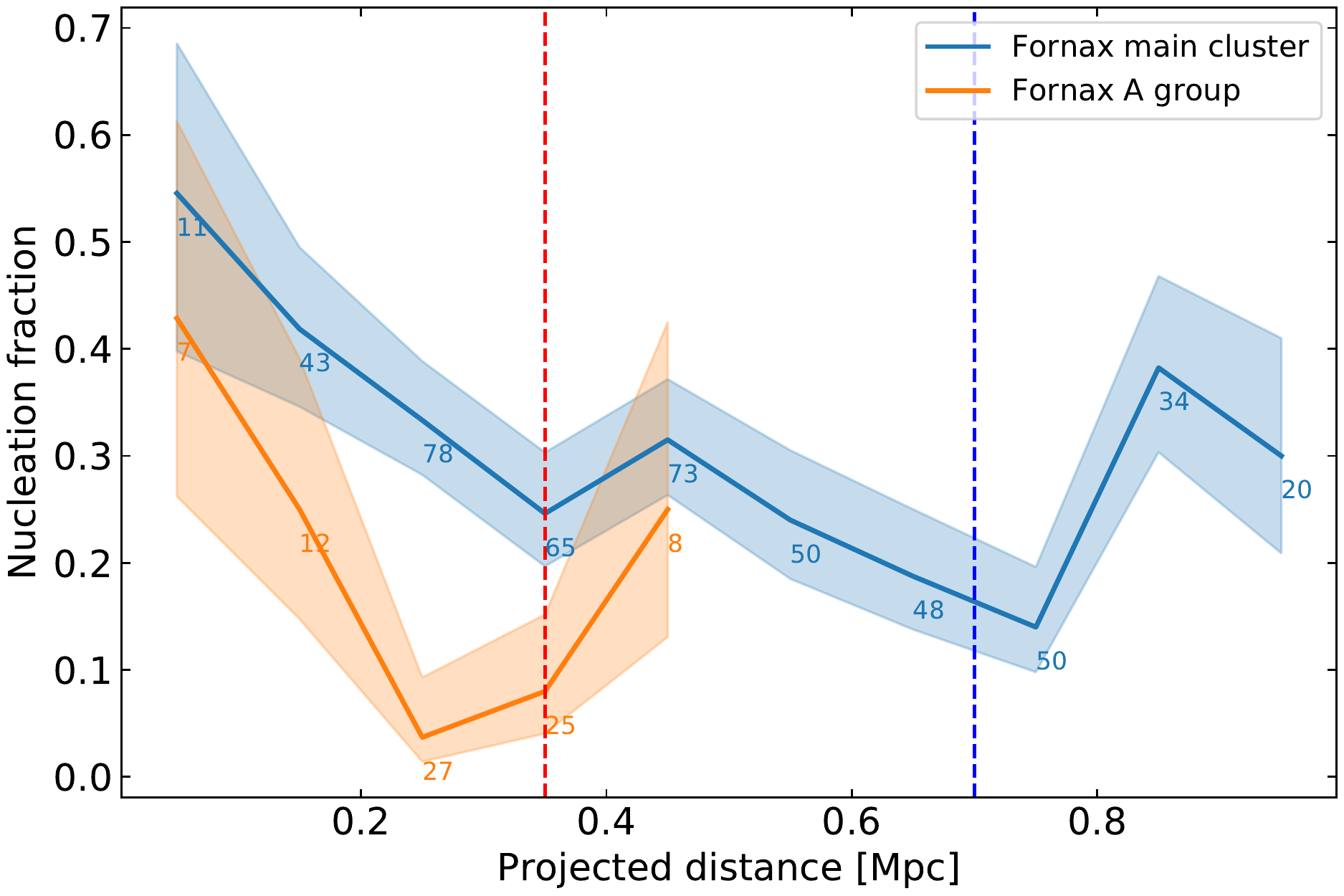}
\caption{Nucleation fraction between the Fornax main cluster (blue) and the Fornax A group (orange) as a function of projected distance (using NGC1399 and NGC1316 as the centre for the Fornax main cluster and Fornax A group, respectively). The shaded regions denote the 68\% binomial confidence interval. The number of galaxies in each bin are annotated in the plot. The dashed vertical lines denote the same limits as in Fig.~\ref{fig:radec}.}
\label{fig:nuc_frac_distance}
\end{figure}

Clearly, the environment plays a role in the nucleation of galaxies. As such, we also consider the nucleation fraction as a function of projected cluster- and group-centric distance between the Fornax main cluster and the Fornax A group. In Fig.~\ref{fig:nuc_frac_distance} we find the nucleation fractions peak towards the cluster centre and decrease with increasing distance, and that the nucleation fractions are generally lower in the Fornax A group than in the Fornax main cluster across bins of projected distance. This is in line with the earlier Fornax study of \citet{venhola2019}, who also estimated the 3D deprojected radial density distribution of dwarfs and found that those in the core of the Fornax main cluster are almost exclusively nucleated early-type dwarfs, with the non-nucleated dwarfs residing at further distances. 

Surprisingly, we find that the nucleation fractions appear to increase beyond the virial radius for both environments. In the case of the Fornax A group, the increase in the nucleation fraction towards the outskirts appears to be due to two (of three) nucleated galaxies with the furthest group-centric distances, located between the Fornax main cluster and the Fornax A group. In comparison, nucleated galaxies appear in all directions beyond the virial radius around the Fornax main cluster. If the two nucleated galaxies were instead considered as members of the Fornax main cluster, the nucleation fraction of the outermost bin for Fornax main cluster would increase from $6/20=0.3$ to $8/22=0.36$. At the same time, the outermost bin for the Fornax A group would decrease from $2/8=0.25$ to $0/6=0$. This suggests that the nucleation fraction beyond the virial radius for the Fornax A group is tenuous at best, but appears to be real for the Fornax main cluster. Additionally, to check if the distribution is dependent on the mass of the galaxies (i.e. more massive dwarfs tend to be located towards the centre of the cluster, which could result in the peak in nucleation fraction), we limit the galaxies in Fig.~\ref{fig:nuc_frac_distance} by their stellar masses. We find that the distributions generally retain their shape, but the nucleation fractions generally decrease with lower galaxy stellar mass limits. 

Given that the increase in the nucleation fraction occurs beyond the virial radii of both environments, it is prudent to compare these galaxies to those in the field environment. \citet{baldassare2014} found a global nucleation fraction of $\sim 0.26$ for $10^9 M_{\odot} \lesssim M_{*,\textup{galaxy}} \lesssim 10^{11} M_{\odot}$ early-type galaxies. Recently, \citet{poulain2021} studied the nucleation of nearby dwarfs \citep[10\,Mpc$<d<$45\,Mpc, and $10^{5.5} M_{\odot} <M_{*,\textup{galaxy}}< 10^9 M_{\odot}$; see][]{habas2020} in low to moderate density environments and found a global nucleation fraction of $\sim 0.23$ (including both early- and late-types). The Local Volume ($d\lesssim 12$\,Mpc) sample of galaxies ($10^{2.5} M_{\odot} < M_{*,\textup{galaxy}} < 10^{11.5} M_{\odot}$) studied in \citet{hoyer2021} has a global nucleation fraction of $\sim 0.24$. Similarly, the analysis of \citet{carlsten2021_nuc} of Local Volume dwarfs ($10^{5.5} M_{\odot} < M_{*,\textup{galaxy}} < 10^{8.5} M_{\odot}$) determined a global nucleation fraction of 0.23. Overall, studies in the literature find the nucleation fraction in the field to be around 0.23--0.26, which is comparable to the nucleation fractions we find beyond the virial radius, within their uncertainties.


\section{Photometric properties}\label{sect:photo_prop}
Here we present properties of the nuclei and their host galaxies. We also compare the properties of nucleated galaxies with non-nucleated galaxies in our sample. Given that our sample of nucleated galaxies are overwhelmingly early-types, this can lead to any differences between the host properties to be due to the early- and late-type, rather than due to nucleation. Therefore, here we compare the quantities of nucleated and non-nucleated galaxies for early-types only. We present some of the derived properties of the nuclei in Table~\ref{tab:nucleus_sample} for a few galaxies as examples; the full table with our whole sample can be found online. 

\begin{table*}[!ht]
    \centering
    \caption{Photometric properties of our nuclei from early-type hosts}
    \input{table_nucleus_short.txt}
    \tablefoot{For each nucleus we denote their host galaxy FDS identification number (1) and FCC designation (2). From the multi-component decomposition models, we determine the galaxy and nucleus stellar masses (3 and 4), as well as the nucleus $g'-r'$ (5) and $g'-i'$ (6) colours. We also include the $r'$ band nucleus flux fraction (i.e. the fraction of light from the nucleus component compared to the total galaxy light; 7) Here only an excerpt is shown; the full table can be found online.}
    \label{tab:nucleus_sample}
\end{table*}

\subsection{Colours}\label{sect:nuc_colours}
\begin{figure}
\centering
\includegraphics[width=\hsize]{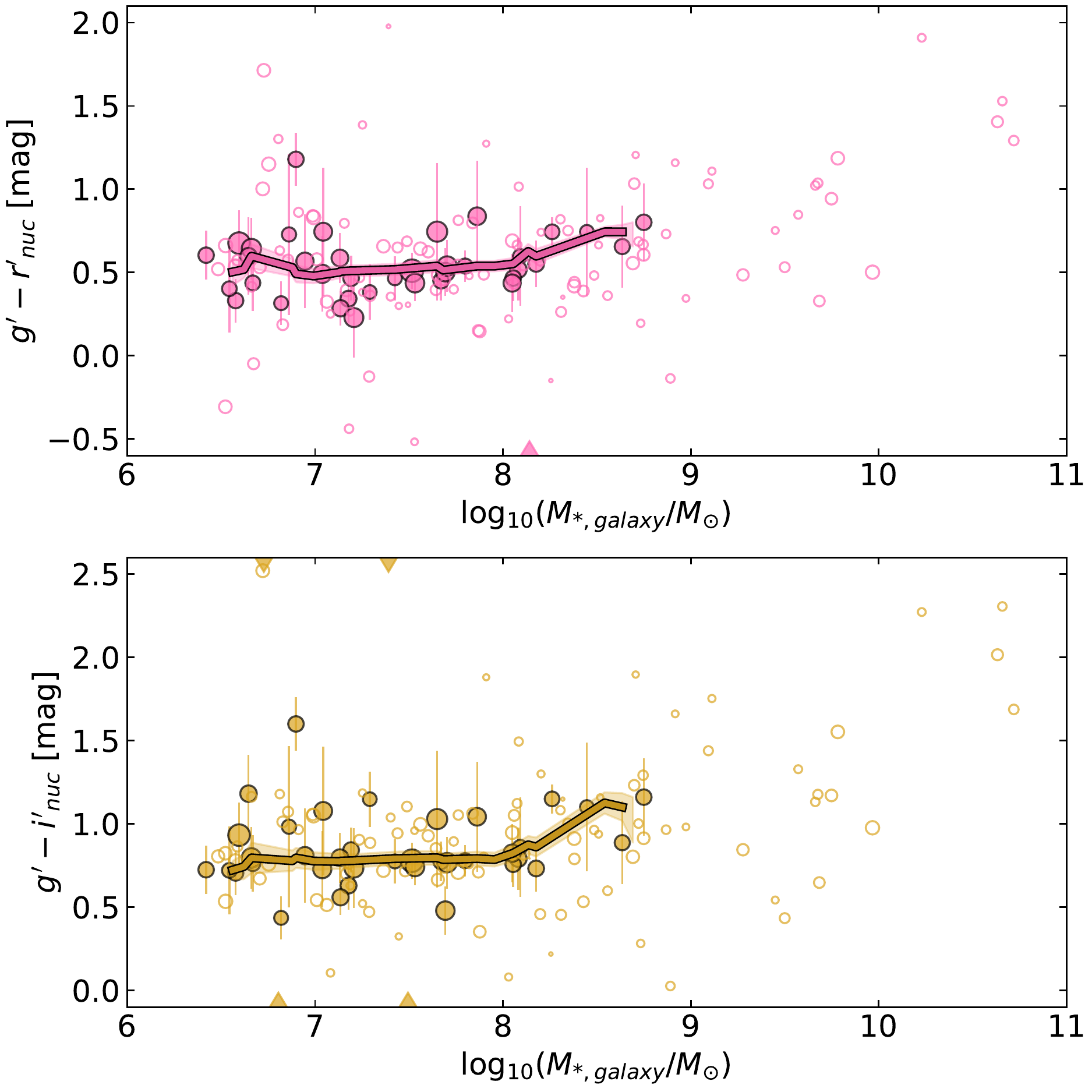}
\caption{$g'-r'$ (upper) and $g'-i'$ (lower) as a function of host galaxy stellar mass. Filled circles are galaxies with nucleus contrast $>1$, whereas open circles denote those with nucleus contrast $<1$. The marker size denotes the nucleus contrast (see Eq.~\ref{eqn:nuc_contrast}), with bigger points denoting higher nucleus contrast. The solid lines denote the moving (median) averages of the solid points, which were calculated using a bin width of 2 dex, and moved in steps of 0.2 dex. The shaded regions denote the corresponding standard error of the mean (SEM) in the moving averages. Nuclei with values beyond the axis limits are denoted as triangles along the x-axes. The individual error bars were estimated in Appendix~\ref{app:nuc_uncertainties}.}
\label{fig:nuc_colour}
\end{figure}

\begin{figure}
\centering
\includegraphics[width=\hsize]{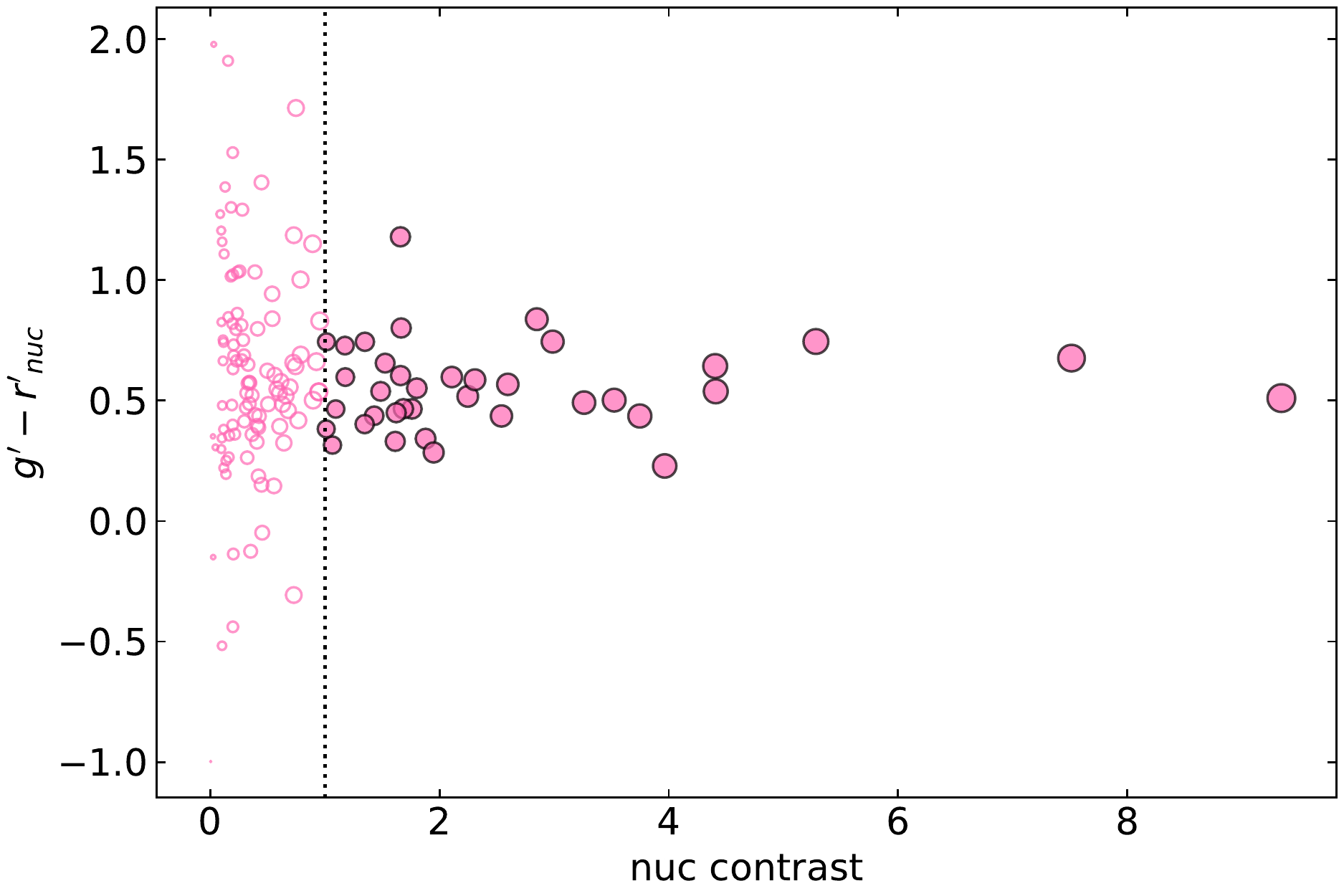}
\caption{Nucleus $g'-r'$ colour as a function of nucleus contrast (see Eq.~\ref{eqn:nuc_contrast}) for nuclei of early-type galaxies. The marker size denotes the nucleus contrast. The dotted vertical line denotes the limit of nucleus contrast $=1$, which loosely separates nuclei with high and low scatter in colour. }
\label{fig:nuc_colour_contrast}
\end{figure}

Based on multi-component decompositions, we use the magnitudes from the nucleus component to derive their integrated colours (see Sect.~\ref{sect:data}). In Fig.~\ref{fig:nuc_colour} we show the $g'-r'$ and $g'-i'$ colours of the nuclei in the early-type galaxies in our sample. For two of the galaxies which we identified as hosting faint nuclei in Sect.~\ref{sect:final_sample}, FDS12\_0367 and FDS15\_0232, their Sérsic+PSF decompositions fail to fit the nuclei (i.e. the PSF magnitudes reach the limiting 35\,mag) in the $g'$ and $i'$ bands, respectively. As such, we exclude the corresponding galaxies from figures which show the nucleus colours. From Fig.~\ref{fig:nuc_colour} it is clear that the scatter in nucleus colours is large across a range of host galaxy stellar masses. In Fig.~\ref{fig:nuc_colour_contrast} we show that the scatter in nucleus $g'-r'$ colour is a function of the nucleus contrast (see Eq.~\ref{eqn:nuc_contrast}), where low contrast nuclei tend to have higher scatter\footnote{We find the same behaviour for the $g'-i'$ colour.}. To reduce the scatter in colour from low contrast nuclei, we limit the bulk of the colour analysis to nuclei with nucleus contrast $>1$ (solid points in Fig.~\ref{fig:nuc_colour}). In Appendix~\ref{app:nuc_uncertainties} we estimate the uncertainties in the nucleus colours due to the PSF model used in the decompositions. Overall, the nuclei colours are rather similar across for host galaxies with $M_* < 10^9\,M_{\odot}$, with a weak or marginal increase in average colour at higher masses. 

\begin{figure}
\centering
\includegraphics[width=\hsize]{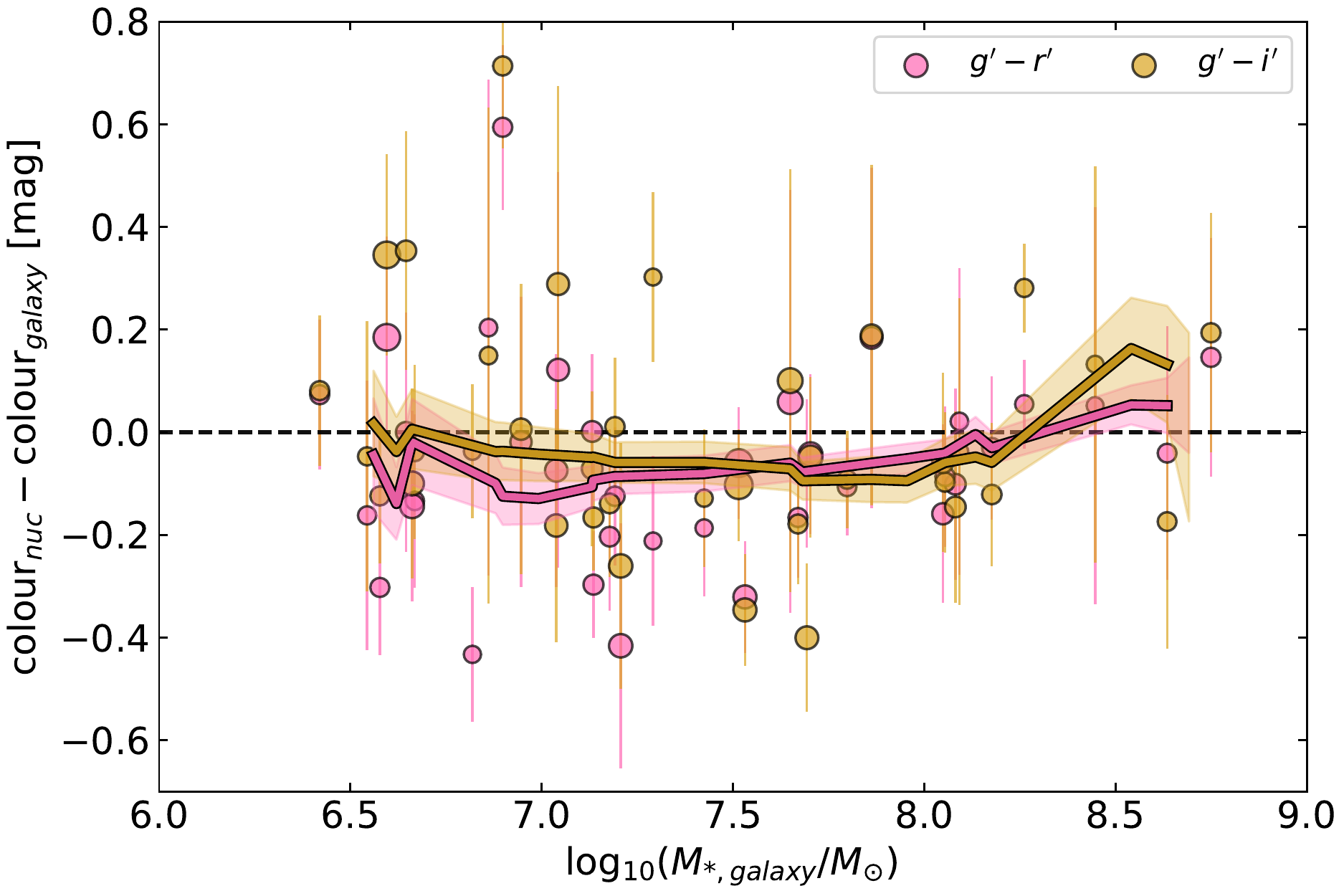}
\caption{Difference in colour between the nucleus and its host galaxy as a function of the host galaxy stellar mass. We show both $g'-r'$ (pink) and $g'-i'$ (gold) colours based on the multi-component decomposition models. The moving averages were calculated using a bin width of 2 dex, and moved in steps of 0.2 dex. The shaded regions denote the SEM. The marker size denotes the nucleus contrast. The uncertainties were estimated by propagating the nuclei colour uncertainties (see Appendix~\ref{app:nuc_uncertainties}) and the galaxy colour uncertainties calculated in \citet{su2021} (see their Table~1). }
\label{fig:nuc_host_colour}
\end{figure}

In Fig.~\ref{fig:nuc_host_colour} we show the difference in $g'-r'$ and $g'-i'$ colours between the nucleus component and the host galaxy for the nucleated early-type galaxies with nucleus contrast $>1$ in our sample. On the whole, there is scatter about zero, suggesting more or less similar colours between the nucleus and the host galaxies. The moving averages in $g'-r'$ and $g'-i'$ appear to support somewhat bluer ($\lesssim 0.1$\,mag) nuclei for $10^7 M_{\odot} \lesssim M_* \lesssim 10^8 M_{\odot}$ host galaxies. 

\begin{figure}
\centering
\includegraphics[width=\hsize]{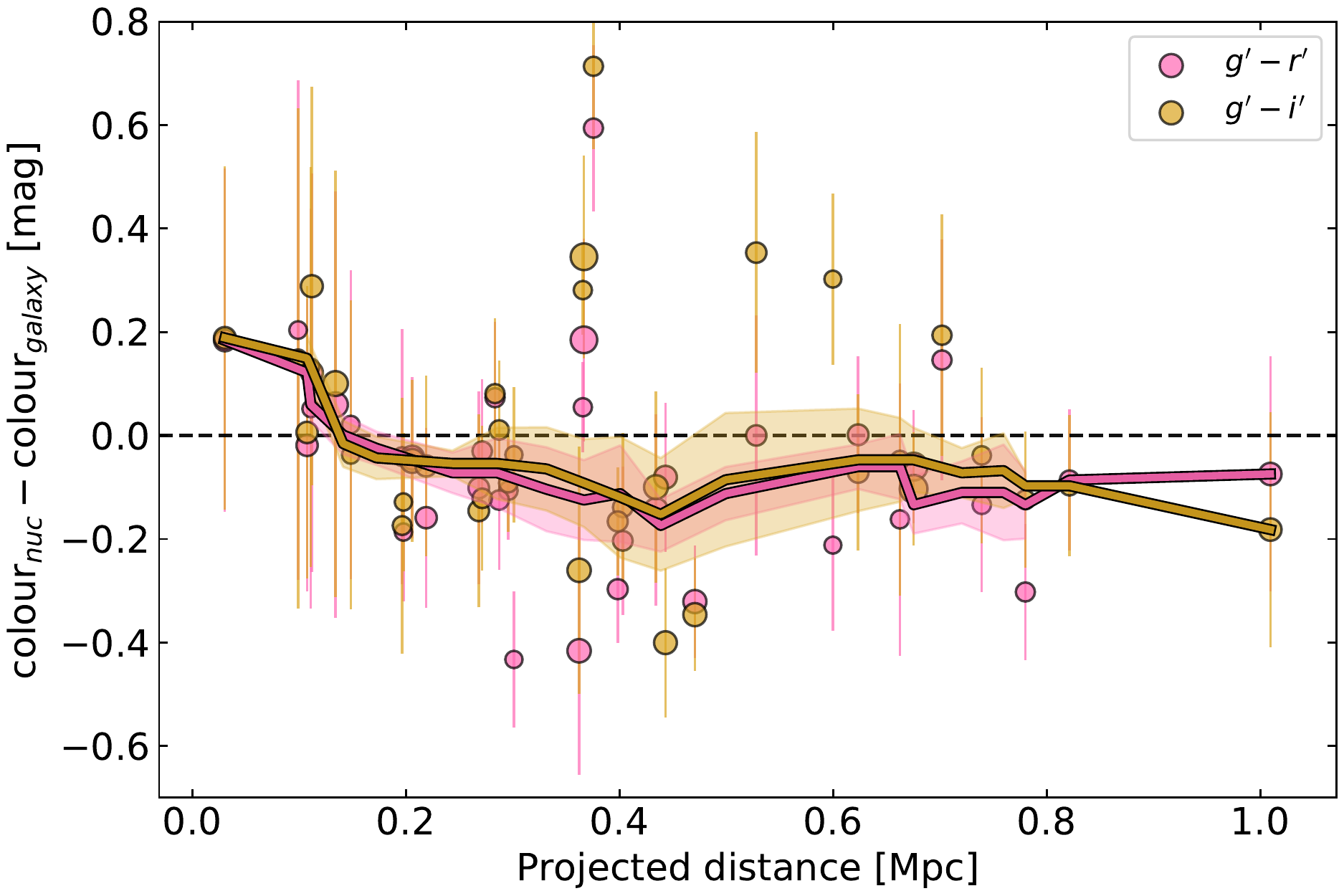}
\caption{Similar to Fig.~\ref{fig:nuc_colour}, but showing the difference in colour between the nucleus and its host galaxy as a function of the projected distance. The marker size denotes the nucleus contrast. Unlike Fig.~\ref{fig:nuc_colour}, here we only show galaxies in the Fornax main cluster. }
\label{fig:nuc_colour_dists}
\end{figure}

Aside from the host stellar mass, we also investigate the difference in colour with projected distance. Given that the Fornax A group has a lack of nucleated galaxies, here we only focus on the Fornax main cluster. From the moving averages of Fig.~\ref{fig:nuc_colour_dists} we find that the nuclei of galaxies residing within the inner ($\sim 0.1$\,Mpc) region of the Fornax main cluster appear to be marginally redder than their host galaxy, whereas the median values appear to suggest bluer nuclei at higher projected cluster-centric distances. We check the stellar masses of the galaxies in the inner $0.1$\,Mpc region and found a mix of values, which suggests that the redder colour in the inner region is not due to the most massive galaxies which have redder colours. 

\subsection{Stellar mass}\label{sect:nuc_stellar_mass}
\begin{figure}
\centering
\includegraphics[width=\hsize]{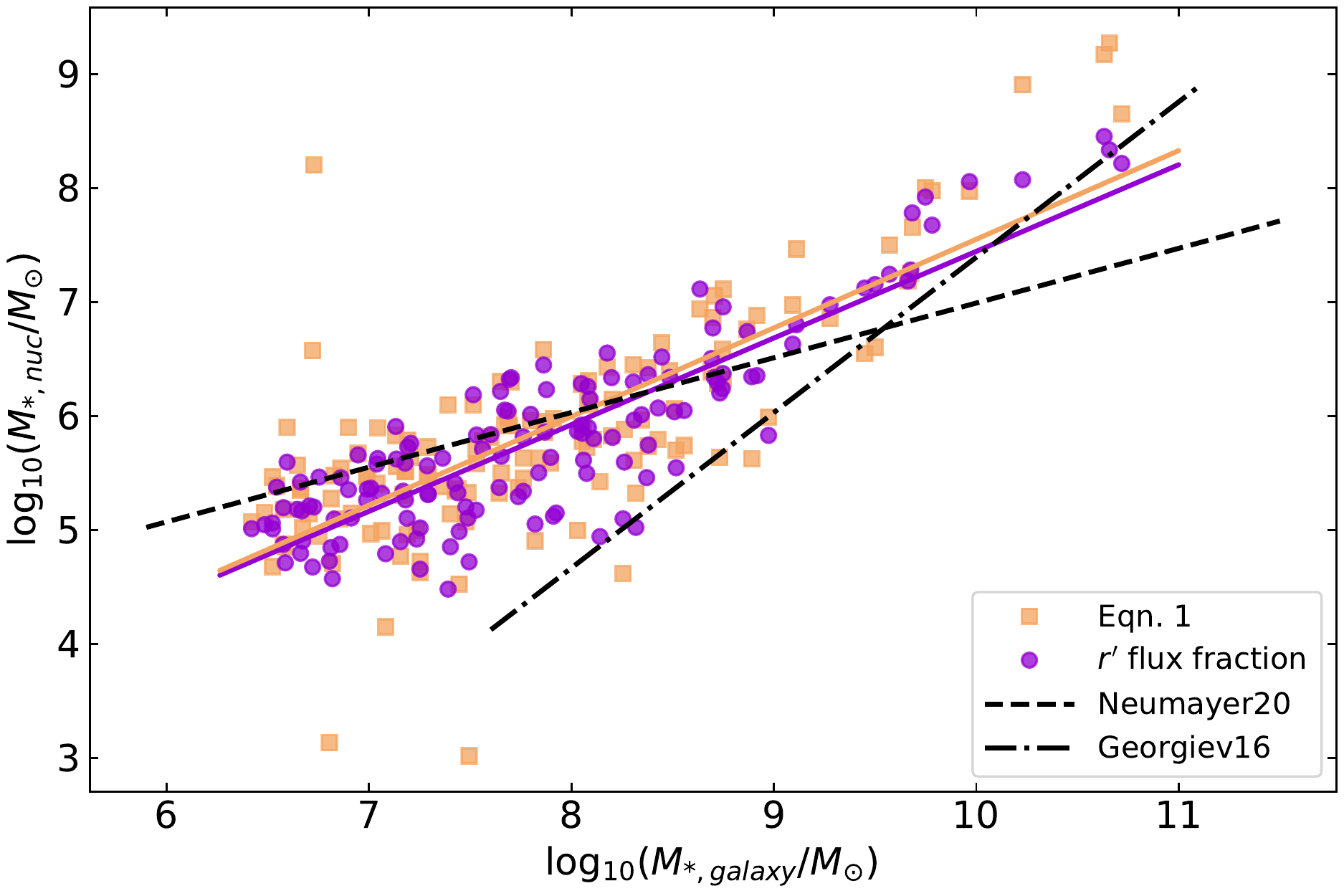}
\caption{Nucleus stellar mass as a function of the host galaxy stellar mass. We applied two estimates for the nucleus stellar mass: from $r'$ band nucleus flux fractions (violet) and Eq.~\ref{eqn:mstar} (light brown). The linear fits for both estimates are shown as solid lines. For comparisons we also include the relations from \citet{neumayer2020} (their Eq.~1) and \citet{georgiev2016} (their early-type sample) as dashed and dash-dotted black lines. The nucleus stellar mass based on Eq.~\ref{eqn:mstar} for FDS12\_0367 and FDS15\_0232 have been omitted due to too faint $g'$ and $i'$ band nucleus magnitudes.}
\label{fig:nuc_mstar_multi}
\end{figure}

To estimate the mass of the nuclei we test two methods: through the nucleus to host galaxy flux fraction in the $r'$ band only (and multiplying by the host galaxy stellar mass), and by applying the $g'$, $r'$, and $i'$ band magnitudes of each nuclei from multi-component decompositions to Eq.~\ref{eqn:mstar}. In Fig.~\ref{fig:nuc_mstar_multi} we show both stellar mass estimates as a function of the host galaxy stellar mass, which overall appear to be similar to each other. We find that the nucleus mass estimates from Eq.~\ref{eqn:mstar} have a larger scatter towards the low-mass end than the mass inferred from the $r'$ band flux fraction, although the linear fits appear to be similar. From the linear fits we obtain the relations 
\begin{equation}
\begin{split}
    \log_{10}(M_{\rm *,nuc} / M_{\odot}) = \,&(0.78\pm 0.06) \; \times \; \log_{10}(M_{\rm *,galaxy} / M_{\odot}) \\
    &- (0.23\pm 0.44)
\end{split}
\end{equation}
for the Eq.~\ref{eqn:mstar} nucleus stellar mass, and 
\begin{equation}\label{eqn:mnuc_1linfit_ff}
\begin{split}
    \log_{10}(M_{\rm *,nuc} / M_{\odot}) = \,&(0.76\pm 0.04) \; \times \; \log_{10}(M_{\rm *,galaxy} / M_{\odot}) \\
    &- (0.16\pm 0.29)
\end{split}
\end{equation}
for stellar mass based on $r'$ band flux fractions. Comparing Eq.~\ref{eqn:mnuc_1linfit_ff} to that of \citet{neumayer2020} (their Eq.~1), we see that our relation has a steeper gradient ($M_{\rm *,nuc} \propto M_{\rm *,galaxy}^{0.76}$ against $M_{\rm *,nuc} \propto M_{\rm *,galaxy}^{0.48}$), despite covering a comparable stellar mass range ($10^6 M_{\odot} \lesssim M_{\rm *,galaxy} \lesssim 10^{11} M_{\odot}$). The largest difference occurs at the highest mass end, where we have much more massive nuclei than their relation predicts. Conversely, the relation from \citet{georgiev2016} for early-type host galaxies ($10^8 M_{\odot} \lesssim M_{\rm *,galaxy} \lesssim 10^{11} M_{\odot}$) has a steeper gradient (i.e. $M_{\rm *,nuc} \propto M_{\rm *,galaxy}^{1.36}$) than our relation, which fits with our massive galaxies but clearly deviates for low-mass galaxies. As evident in Fig.~\ref{fig:nuc_mstar_multi}, the $\log$ nucleus stellar mass relation is not linear with $\log$ galaxy stellar mass, which can lead to varying gradients depending on the sample. Using two linear fits with a split at $M_{\rm *,galaxy} = 10^{8.5} M_{\odot})$ for nucleus stellar masses based on the $r'$ band flux fraction, we find that 
\begin{equation}
\begin{split}
    \log_{10}(M_{\rm *,nuc} / M_{\odot}) = \,&(0.52\pm 0.07) \; \times \; \log_{10}(M_{\rm *,galaxy} / M_{\odot}) \\
    &+ (1.58\pm 0.50) \label{eqn:nucmstar_lowmass}
\end{split}
\end{equation}
for the lower mass galaxies, and 
\begin{equation}
\begin{split}
    \log_{10}(M_{\rm *,nuc} / M_{\odot}) = \,&(1.08\pm 0.09) \; \times \; \log_{10}(M_{\rm *,galaxy} / M_{\odot}) \\
    &- (3.09\pm 0.82) \label{eqn:nucmstar_highmass}
\end{split}
\end{equation}
for the higher mass galaxies in our sample. 

\begin{figure}
\centering
\includegraphics[width=\hsize]{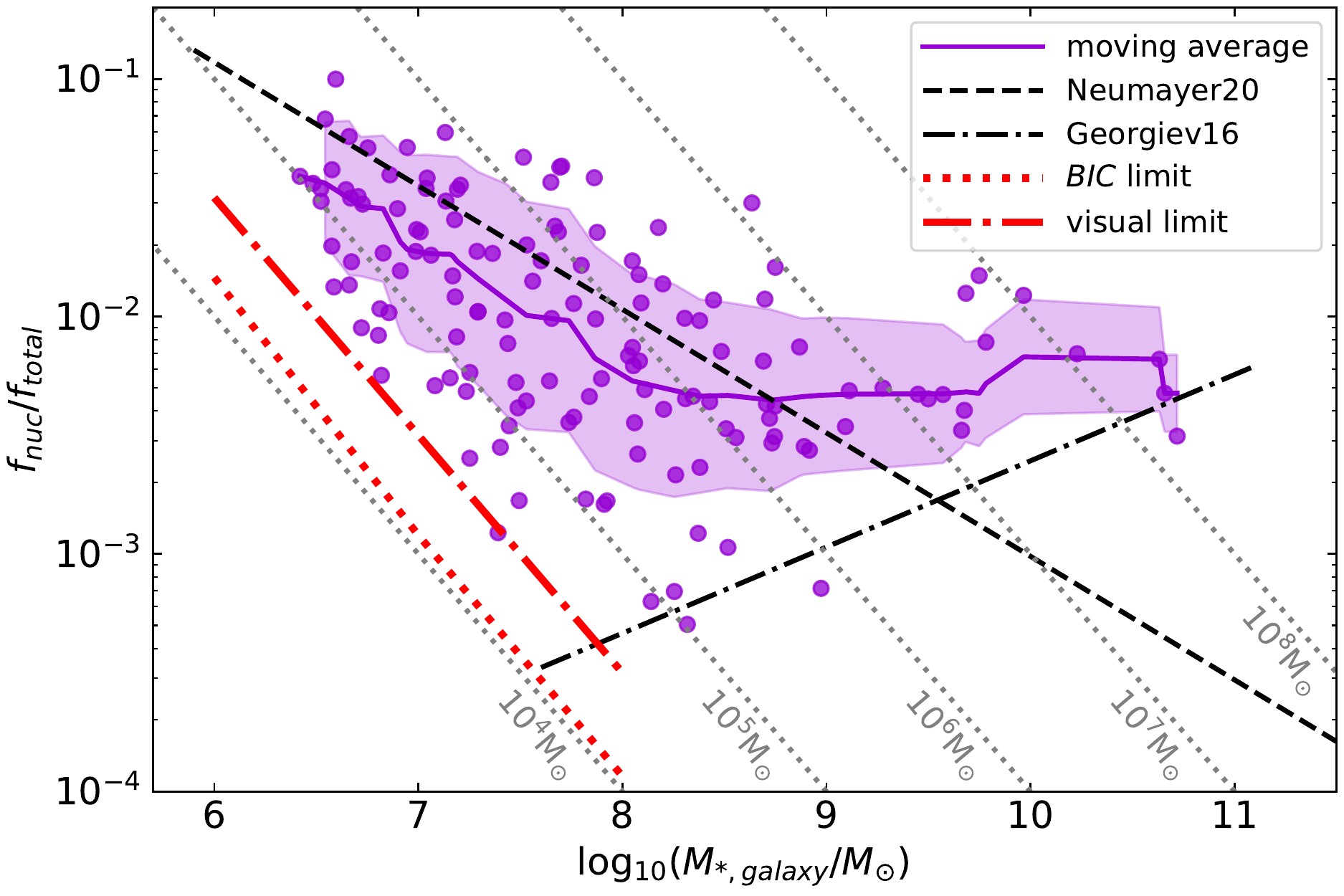}
\caption{Nucleus flux fraction as a function of host galaxy stellar mass for all nucleated early-type galaxies. The moving average (violet) was calculated using a bin width of 2\,dex, and moved in steps of 0.2\,dex (based on $\log_{10}(f_{\textup{nuc}}/f_{\textup{total}})$ instead). The shaded region denotes the $1\sigma$ uncertainty within the moving average bins. The dashed and dot-dashed black lines denote the relation from \citet{neumayer2020} (their Eq.~1) and \citet{georgiev2016} (their early-type sample), respectively. The dotted and dot-dashed red lines denote the estimated detection limits based on $BIC$ and visual inspection, respectively. The dotted grey lines denote constant nucleus stellar masses.}
\label{fig:nuc_ff_multi}
\end{figure}

The deviation from linearity in nucleus stellar mass can be seen more clearly when we consider the nucleus flux fraction as a function of host stellar mass (see Fig.~\ref{fig:nuc_ff_multi}). Here we estimate the nucleus flux fraction based on the multi-component decomposition models in the $r'$ band. From the moving average, we find that the nuclei with host masses of $M_* \lesssim 10^9 M_{\odot}$ clearly follow a trend where the nucleus flux fraction decreases with increasing host mass. However, the trend changes for galaxies with $M_* \gtrsim 10^{8.5} M_{\odot}$, reaching a minimum at $f_{\rm nuc}/f_{\rm total} \sim 0.005$ and roughly plateauing for increasing host stellar mass. This feature for high-mass galaxies is reminiscent of what was observed in \citet{sanchezjanssen2019_nsc}, who observed an 'uptick' in the ratio of nucleus and host stellar mass (i.e. the nucleus mass fraction) for $M_* \gtrsim 10^{9.5} M_{\odot}$. Interestingly, \citet{neumayer2020} found a 'bump' instead of an uptick, in that the nucleus mass fraction increases around $10^{9.5} M_{\odot}$, but decreases again with increasing galaxy stellar mass (see their Fig.~12). They attributed this difference to the inclusion of galaxies from \citet{lauer2005} in their sample. Despite the difference at the high mass end, from Fig.~\ref{fig:nuc_ff_multi} we find that the moving average at lower galaxy masses ($M_* \lesssim 10^8 M_{\odot}$) coincides with the \citet{neumayer2020} relation except being lower by a factor of 0.2\,dex, as both have an approximate $f_{\textup{nuc}}/f_{\textup{total}} \propto M_{\rm *,galaxy}^{0.5}$ dependence.

\subsection{Nucleated versus non-nucleated} \label{sect:nuc_vs_nonuc}

\begin{figure*}
\centering
\resizebox{\hsize}{!}{\includegraphics[width=\hsize]{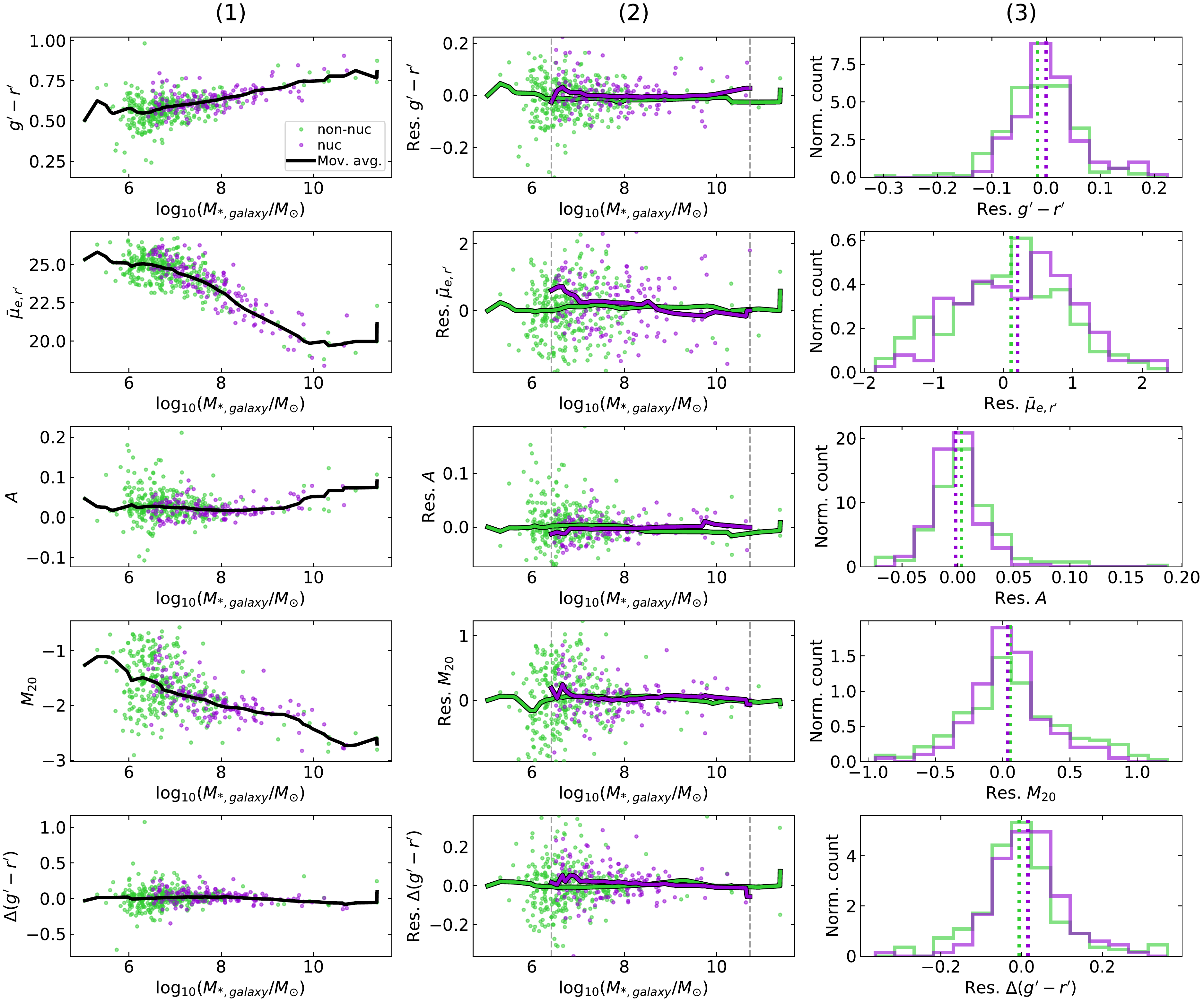}}
\caption{Structural quantities of the nucleated (violet) and non-nucleated (green) early-type galaxies in the Fornax main cluster and Fornax A group. Column~1: The structural quantities as functions of the galaxy stellar mass. The black lines denote the moving averages (median). Column~2: The residual quantities, calculated as the measured quantities minus moving average from column~1. The solid lines denote the moving averages of the residual quantities. The dashed grey vertical lines denote the range where both samples overlap in stellar mass. Column~3: Histograms of the residual quantities between nucleated and non-nucleated galaxies within the overlapping stellar mass range. The coloured dotted lines denote the median value of each sub-sample. For brevity here we show quantities of interest. The full plot with all considered quantities can be found in Appendix~\ref{app:host_pars_nosig}.}
\label{fig:host_pars_etg}
\end{figure*}

\begin{table*}[!ht]
    \centering
    \caption{p-values from hypothesis testing for residual galaxy properties}
    \input{fds_nuc_etg_stats.txt}
    \tablefoot{Test the null hypothesis that the residual galaxy properties~(1) between nucleated and non-nucleated galaxies in our sample are drawn from the same distribution. The alternative hypothesis is that the null hypothesis is false. The Kolmogorov-Smirnov~(2), Anderson-Darling~(3), Lepage~(4), and Cucconi~(5) statistics were used for each galaxy quantity. The p-values were calculated for the galaxies within overlapping stellar masses ($10^{6.4} M_{\odot} < M_* < 10^{10.7} M_{\odot}$; column~3 from Fig.\ref{fig:host_pars_etg}). p-values below the significance level $\alpha=0.05$ are shown in bold. We also describe the difference in residual properties for nucleated galaxies as compared to non-nucleated galaxies (6) for those which have significant test statistic(s). }
    \tablefoottext{a}{The significance of $C$ is likely due to the fact that the non-parametric indices are calculated including the nucleus.}
    \label{tab:fds_stats_etg}
\end{table*}

Given that nucleation is dependent on the host galaxy property (i.e. their stellar mass; see Fig~\ref{fig:nuc_frac_maingroup}), we investigate whether the structural properties of galaxies also differ with nucleation. We use a combination of quantities derived from the Sérsic component of Sérsic+PSF decompositions ($g'-r'$, $R_e$, $n$, and $\bar{\mu}_{e,r'}$) as well as non-parametric morphological indices ($C$, $A$, $S$, $G$, $M_{20}$, and $\Delta(g'-r')$) to measure the global properties of the host galaxies. We choose these quantities since they allow for a more homogeneous treatment of our galaxies, as opposed to directly comparing the internal structures such as bulges or bars, which are not present in all galaxies. The quantities were defined in detail in \citet{su2021}, so here we only provide a brief summary. The $g'-r'$ colour was calculated based on the total magnitudes from the decomposition models. The $R_e$ and $n$, and $\bar{\mu}_{e,r'}$ were based on the $r'$ band decompositions. The concentration ($C$), asymmetry ($A$), and clumpiness ($S$) indices quantify the central concentration, rotational asymmetry, and the amount of small substructures in a galaxy. The Gini coefficient ($G$) denotes the (in)equality in the distribution of flux across a galaxy, whereas the second order moment of the brightest 20\% of flux ($M_{20}$) indicates the variance in the brightest parts of a galaxy. Finally, $\Delta(g'-r')$ is the difference in colour between the outer (1$R_e$ to 2$R_e$) minus inner (0 to 0.5$R_e$) region of a galaxy, so a positive $\Delta(g'-r')$ implies a bluer inner region than its outskirts. The non-parametric quantities are calculated including the nucleus, such that we would expect $C$ to be higher for nucleated galaxies. 

Previous studies have noted the differences in the stellar masses and structures between nucleated and non-nucleated early-type galaxies \citep[e.g.][]{eigenthaler2018, venhola2019, sanchezjanssen2019_nsc, poulain2021}. However, given the clear dependence between structural quantities and the galaxy stellar mass (see the first column of Fig.~\ref{fig:host_pars_etg}), and the difference in the stellar mass distributions between nucleated and non-nucleated galaxies (see Fig.~\ref{fig:hist_nucleation}), it is not clear how much the differences in structural quantities can be attributed to the stellar mass and how much to nucleation. Hence, following \citet{su2021}, we remove the mass dependence for each quantity by subtracting the moving averages from the measured values, which we refer to as the residual quantities. This allows us to compare the difference between the two sub-samples across the range of stellar mass (see column~2 of Fig.~\ref{fig:host_pars_etg}). 

To test whether the differences in the distributions of residual quantities between the two sub-samples (if any) are significant, we calculate the (two sample) test statistics of the Kolmogorov-Smirnov (KS) test, the Anderson-Darling (AD) test, the Lepage (LP) test, and the Cucconi (CU) test, under the null hypothesis that the two sub-samples are drawn from the same distribution\footnote{We use the \texttt{ks\_2samp} and \texttt{anderson\_ksamp} functions from the \texttt{SciPy} library \citep{scipy2020} to calculate the KS and AD test statistics, respectively, and their corresponding p-values. For the LP and CU statistics we construct user defined functions in Python following \citet{marozzi2009}.}. We choose these four test statistics as they are: non-parametric, meaning they do not make any explicit assumption that the data follows a certain distribution; and sensitive to different characteristics of a distribution, so they provide independent ways of assessing the differences between the sub-samples. Comparing to the KS test, the AD test statistic is more sensitive to the tails of the cumulative distributions, rather than the centres. The LP and CU statistics test the location and scale (analogous to the mean and standard deviation, respectively, for a Gaussian distribution) of the two samples and compare their differences. As a result, all four test statistics are sensitive to a difference in the averages, dispersion (and indeed both) between the two samples. For the null hypotheses we utilise a nominal significance level of $\alpha = 0.05$ such that they can be rejected for p-values less than $\alpha$. In Table~\ref{tab:fds_stats_etg} we tabulate the p-values based on the residual quantities for nucleated and non-nucleated early-type galaxies with overlapping stellar masses (i.e. $10^{6.4} M_{\odot} < M_* < 10^{10.7} M_{\odot}$; column~3 of Fig.~\ref{fig:host_pars_etg}). 

Regarding the early-type galaxies, we find that the nucleated hosts tend to be redder in $g'-r'$ colour, less asymmetric, and exhibit redder outskirts relative to their own inner regions compared to their non-nucleated counterparts at a given stellar mass. In the case of $M_{20}$ the significant difference between nucleated and non-nucleated galaxies is likely due to the higher scatter rather than any difference in the mean. The redder $g'-r'$ colour suggests that nucleated galaxies generally have more evolved stellar populations. Additionally, the larger colour difference between the inner and outer region (i.e. higher $\Delta(g'-r')$) for nucleated galaxies can be interpreted as stronger signs of ram pressure stripping (RPS), which affect a galaxy's outskirts first \citep[see e.g.][]{vollmer2009}. Finally, the nucleated galaxies are on average less asymmetric, which suggests that non-nucleated galaxies are potentially more susceptible to disruptions and interactions. It is possible that the difference in the host properties between nucleated and non-nucleated galaxies could be tied to the higher nucleation fraction towards the centre of the cluster/group. We explore the dependence of nucleation on the environment of the host galaxy in Sect.~\ref{sect:env_dependence}. 

From column 2 of Fig.~\ref{fig:host_pars_etg} there appears to be a change in the moving averages below and above $M_* \sim 10^{8.5} M_{\odot}$ for $\bar{\mu}_{e,r'}$, $A$, and $\Delta(g'-r')$. Restricting the stellar mass range to $M_* < 10^{8.5} M_{\odot}$, we find significant differences for the KS test ($p=0.047$) and the AD test ($p=0.044$) for residual $\bar{\mu}_{e,r'}$. Moreover, we find that the moving averages differ the most at the low mass end. This difference could be due to a bias in the detection of nuclei, where nucleated galaxies with low surface brightness would have a higher nucleus contrast than their high surface brightness counterparts. For $A$ and $\Delta(g'-r')$, we found that all test statistics remain significant: $p_{\textup{KS}} = 0.009$, $p_{\textup{AD}}=0.006$, $p_{\textup{LP}}=0.006$, $p_{\textup{CU}}=0.005$ for residual $A$, and $p_{\textup{KS}} = 0.021$, $p_{\textup{AD}}=0.009$, $p_{\textup{LP}}=0.029$, $p_{\textup{CU}}=0.028$ for residual $\Delta(g'-r')$. 

Recently, studies found that nucleated cluster dwarfs tend to have higher axial ratios ($q$) than their non-nucleated counterparts \citep[][]{lisker2007, eigenthaler2018, venhola2019, sanchezjanssen2019_shape, poulain2021}. Furthermore, by modelling the dwarfs as triaxial ellipsoids, \citet{sanchezjanssen2019_shape} inferred that the nucleated dwarfs from the cores of the Virgo and Fornax clusters have higher intrinsic (vertical) thickness than the non-nucleated galaxies. Here we address this difference between nucleated and non-nucleated galaxies with our sample, considering also the surface brightness of our dwarfs with $10^{6.2} M_{\odot} \lesssim M_{\rm *,galaxy} \lesssim 10^{7.8} M_{\odot}$\footnote{Given the slight differences between the $\bar{\mu}_{e,r'}$ moving averages between nucleated and non-nucleated galaxies (see Fig.~\ref{fig:host_pars_etg_nosig}), we also tested different intervals in stellar mass where the differences were smaller (e.g. $10^{6.7}\,M_{\odot} - 10^{7.3}\,M_{\odot}$), which did not alter our results and interpretation.} (equivalent to the limits set in \citealt{sanchezjanssen2019_shape}). 

\begin{figure}
\centering
\includegraphics[width=\hsize]{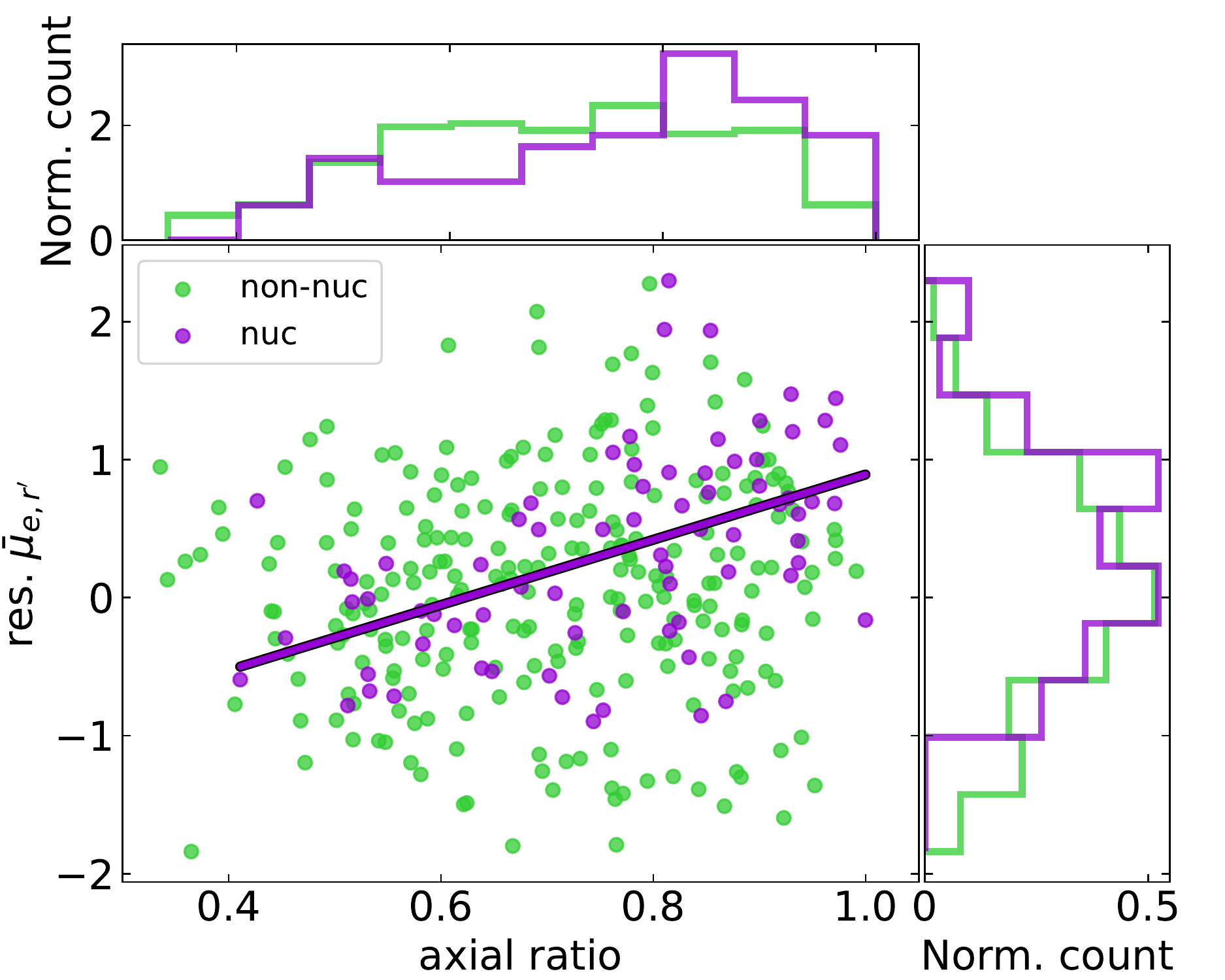}
\caption{Residual $\bar{\mu}_{e,r'}$ as a function of the axial ratio of the Sérsic component from Sérsic+PSF models. The points denote nucleated (violet) and non-nucleated (green) galaxies within the stellar mass range of $10^{6.2} < M_* < 10^{7.8}$. The purple line shows the linear fit to the nucleated galaxies. The upper and left subplots show the distributions of the axial ratio and residual $\bar{\mu}_{e,r'}$, respectively. }
\label{fig:muebar_vs_q}
\end{figure}

In Fig.~\ref{fig:muebar_vs_q} we show the residual $\bar{\mu}_{e,r'}$ as a function of the axial ratio for galaxies within the stellar mass range, as well as their distributions as histograms in the subplots. We confirm that our nucleated galaxies tend to have higher axial ratio, with the peak of the distribution around 0.85. Comparatively, we do not find a clear peak for the non-nucleated galaxies, with the majority spanning from 0.5 to 0.9, and the maximum occurring around 0.75. We do not find two peaks for the nucleated sample as \citet{sanchezjanssen2019_shape} reported, although this may be due to the differences in our sample (i.e. their sample consists of galaxies from both the cores of the Virgo ($<300$\,kpc) and Fornax ($<350$\,kpc) clusters). In comparison, our axial ratio distributions are similar to the Virgo dwarfs of \citet{lisker2007} (see their Fig.~3), which showed a peak at an axial ratio of 0.9 for the nucleated galaxies, and a flatter plateau of axial ratios between approximately 0.5 to 0.8. 

Intriguingly, using Spearman's rank correlation coefficient ($r_s$), we find a significant positive correlation between $\bar{\mu}_{e,r'}$ and axial ratio for the nucleated galaxies ($r_s=0.53$, p-value $\ll0.001$)\footnote{Using a linear fit to the nucleated galaxies, we find a slope of $2.36\pm 0.48$ and an intercept of $-1.47\pm 0.37$.} but not for the non-nucleated galaxies ($r_s=0.09$, p-value $=0.16$). Substituting $q=0.5$ into the linear relation and extrapolating to $q=1$, we would expect a nucleated galaxy to experience a change in (residual) $\bar{\mu}_{e,r'}$ of $1.18 \pm 0.24$\,mag\,arcsec$^{-2}$, or a factor of approximately 2.4--3.6 brighter. This is of the same order as the factor of $\sim 2$ that we would expect from oblate spheroids (when viewed from different orientations). Conversely, the lack of correlation for the non-nucleated galaxies suggests that they are likely better described as triaxial ellipsoids \citep[such as in][]{lisker2007,sanchezjanssen2019_shape}. Splitting the sub-sample of nucleated and non-nucleated at $M_{\rm *,galaxy} = 10^{7} M_{\odot}$ \citep[equivalent to $M_{g'} = -12.5$, as used in][]{sanchezjanssen2019_shape}, we find that the least massive non-nucleated galaxies are responsible for the lack of a significant correlation. Indeed, we find significant correlations for both nucleated and non-nucleated galaxies for $10^{7} M_{\odot} < M_{\rm *,galaxy} < 10^{8.5} M_{\odot}$. The trend between surface brightness and axial ratio is investigated in detail in \citet{venhola2021}. 


\section{Automatic model selection}\label{sect:auto_mod_select}
The process of conducting multi-component decompositions under human supervision can be rather time consuming, particularly in identifying morphological structures in the galaxies and considering the type of model which best fit them. Hence, while multi-component models exist for our galaxy sample, there is merit in developing methods which can provide insight into the structures of galaxies in an unsupervised manner. Here we focus on the nucleation of early-type galaxies, specifically whether a galaxy hosts a central nucleus. To do so, we use the single Sérsic and Sérsic+PSF decomposition models (conducted in the $r'$ band) to test the model selection criteria. These two models have been chosen due to the general versatility of the Sérsic function in fitting galaxy light profiles, whilst the PSF models fit the nuclei of our galaxies well without many free parameters, due to being unresolved in our images. Most importantly, both models are simple and can produce reasonable fits to a variety of galaxies with rudimentary initial values, which is desirable for an unsupervised methodology. In this section we test both formulations of the BIC used in Sect.~\ref{sect:nuc_detect} as the model selection criteria. 

\begin{figure}
\centering
\includegraphics[width=\hsize]{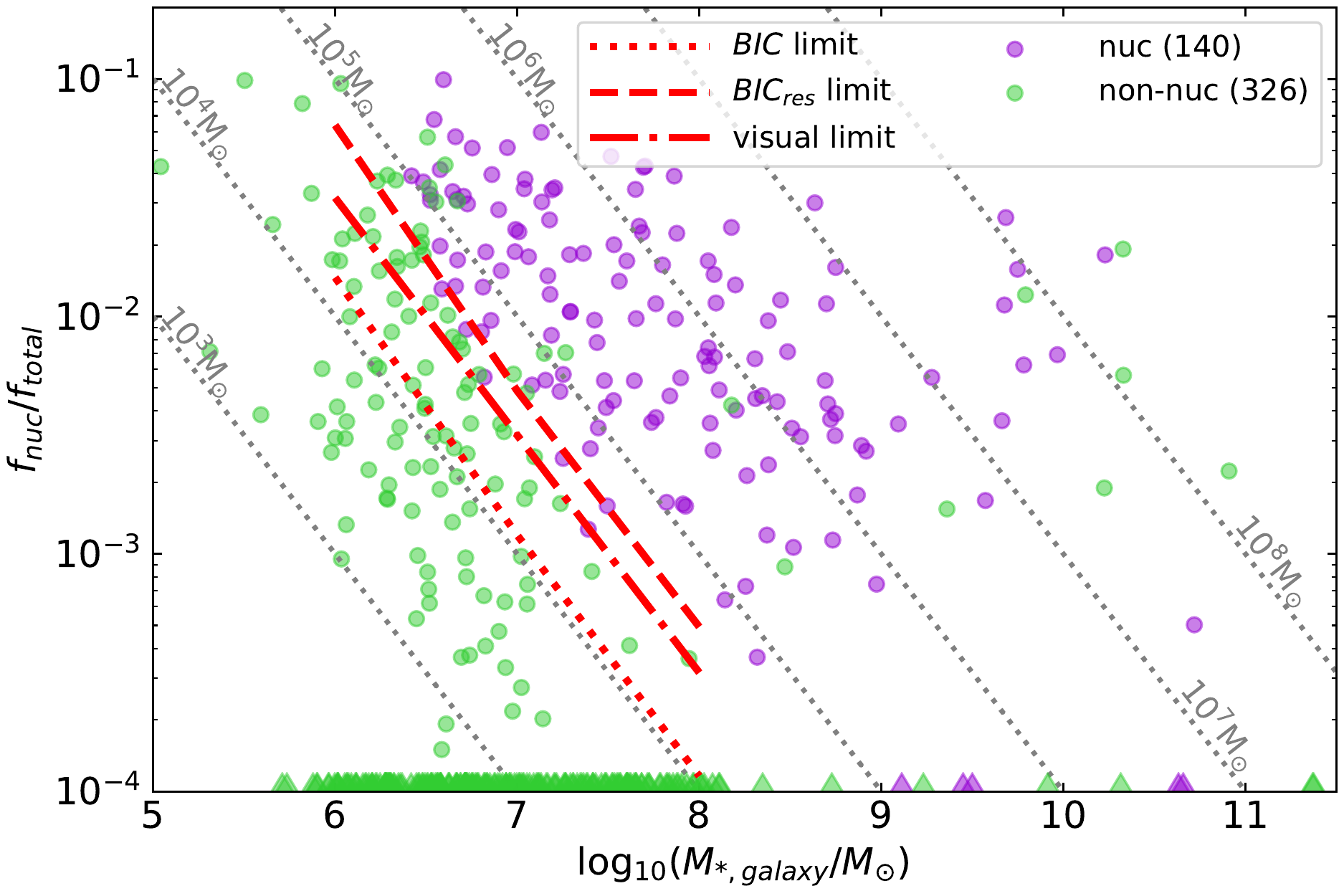}
\caption{Nucleus flux fraction as a function of host stellar mass, based on Sérsic+PSF models. Nucleated galaxies are shown in violet and non-nucleated galaxies are shown in green. Those with a fitted nucleus total magnitude of 35\,mag are shown as triangles along the x-axis. The dotted, dot-dashed, and dashed red lines are the estimated $BIC$, visual, and $BIC_{\textup{res}}$ detection limits, respectively, based on our synthetic nucleation tests (see Sect.~\ref{sect:det_limit}). }
\label{fig:deltabic_limits}
\end{figure}

In Fig.~\ref{fig:deltabic_limits} we show the nucleus flux fraction (in this case the total PSF flux divided by total Sérsic and PSF flux from Sérsic+PSF models) as a function of the galaxy stellar mass. Galaxies with $f_{\textup{nuc}}/f_{\textup{total}} < 10^{-4}$ are shown as triangles along the x-axis as their PSF components have reached the 35\,mag limit without excess central flux above the Sérsic component. A few galaxies which we identified as nucleated appear in this region due to their high central concentration from morphological structures (e.g. bulges, bars, barlens). Given that these galaxies are massive ($M_*>10^9M_{\odot}$), the nuclei represent a much smaller fraction of the galaxy light and are easily dominated by the Sérsic component. Nevertheless, we find that the general shape of the nucleated galaxies in Fig.~\ref{fig:deltabic_limits} to be very similar to that of Fig.~\ref{fig:nucff_contrast_multicomp}, where the former is based on Sérsic+PSF models and the latter is based on multi-component models. This similarity demonstrates the general robustness of our nucleus flux estimates, as on the whole they do not appear to heavily depend on the applied decomposition model (with only a few high-mass exceptions). 

\begin{figure}
\centering
\includegraphics[width=\hsize]{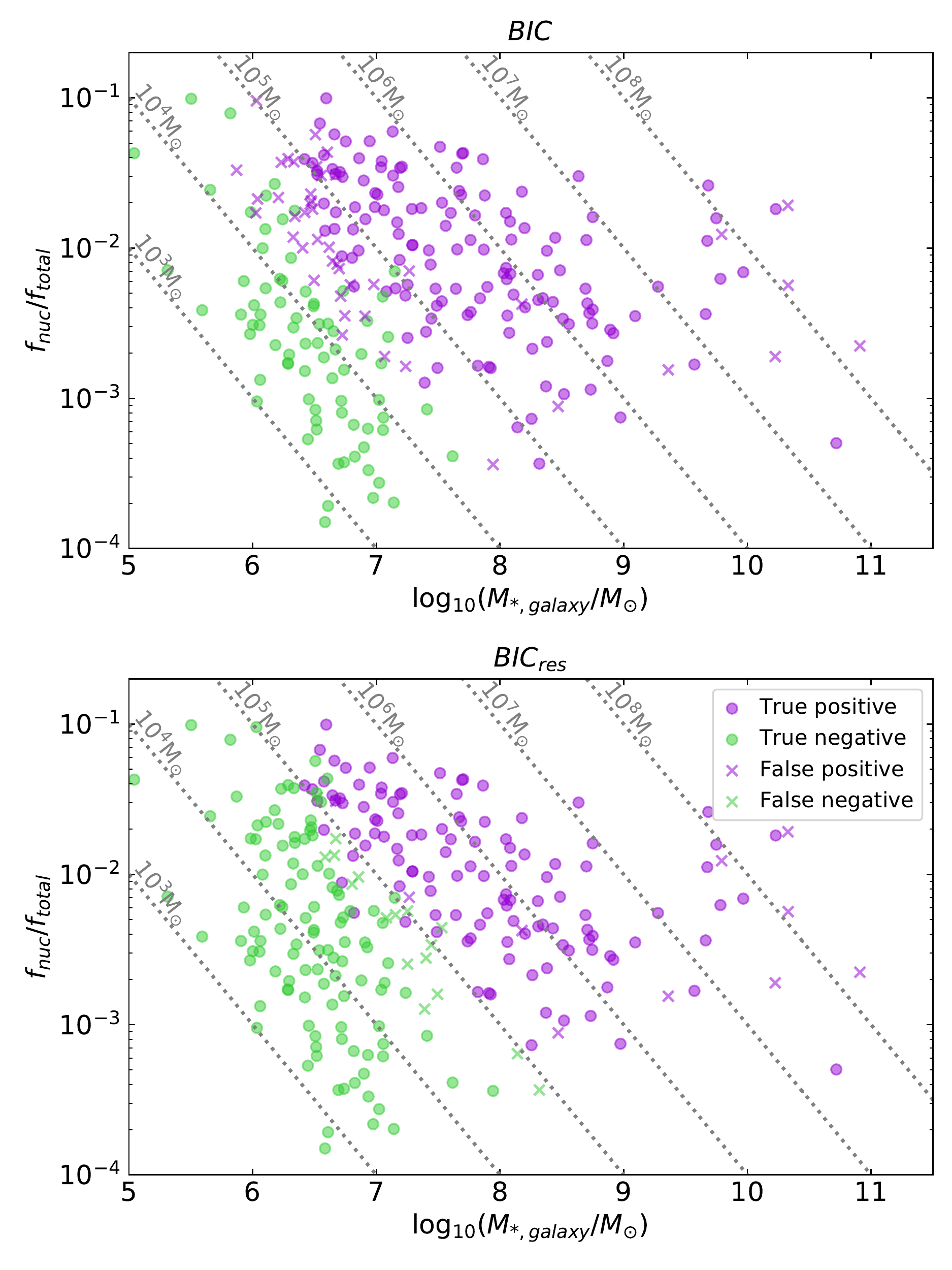}
\caption{Accuracy of $BIC$ (upper) and $BIC_{\textup{res}}$ (lower) in classifying nucleated and non-nucleated early-type galaxies. True (i.e. correct) and false (i.e. incorrect) classifications are denoted by circles and crosses, respectively. Galaxies where GALFIT fails to find a nucleus (i.e. the triangles from Fig.~\ref{fig:deltabic_limits}) have been omitted. The BIC classification accuracies are shown in Table~\ref{tab:bic_accuracy}. }
\label{fig:bic_accuracy}
\end{figure}

To test the performance of classifying nucleated and non-nucleated galaxies, we apply both $BIC$ and $BIC_{\textup{res}}$ on our sample of galaxies\footnote{Here we evaluate the accuracy using our nucleation labels. A small caveat is that these are not strictly ground truths, which should be used instead for the true accuracy. Nevertheless, our labels provide a comparison to the accuracy from visual inspections, which has been the de facto method for detecting nuclei.}. Here the BIC classifies a galaxy as nucleated if $\Delta$BIC$=$BIC$_{\textup{Sérsic}}-$BIC$_{\textup{Sérsic+PSF}}$$>$0 (i.e. a Sérsic+PSF model is preferred over a single Sérsic model). The $BIC$ and $BIC_{\textup{res}}$ classifications are then compared to our labels of nucleated and non-nucleated. To evaluate the accuracy of the classifications we assign each galaxy with one of four labels: true positive, true negative, false positive, and false negative. True and false denote whether the galaxies were correctly or incorrectly classified based on the labels, where as positive and negative denote nucleated and non-nucleated, respectively. From Fig.~\ref{fig:bic_accuracy} and Table~\ref{tab:bic_accuracy} we find that both BIC generally have a low number of incorrect classifications (i.e. false positive and false negatives). For all galaxies in our sample we find an overall accuracy of 89$\%$ and 93$\%$ for $BIC$ and $BIC_{\textup{res}}$, respectively. However, given that many of the higher mass galaxies tend to have multiple components (e.g. bulge, bar), we also exclude galaxies with $M_* > 10^9 M_{\odot}$ and found that the accuracies improve slightly for both $BIC$ and $BIC_{\textup{res}}$ (93\% and 97\%, respectively). 

\begin{table}[!ht]
    \centering
    \caption{Classification accuracies for $BIC$ and $BIC_{\textup{res}}$}
    \input{fds_bic_accuracies_etg.txt}
    \tablefoot{We show the number of early-type galaxies which were classed as true positives (TP), true negatives (TN), false positives (FP) and false negatives (FN), based on the BIC classifications relative to our nucleation classification from multi-component decompositions (see text for more detail). We calculate the accuracy as (TP$+$TN)$/$(TP$+$TN$+$FP$+$FN). }
    \label{tab:bic_accuracy}
\end{table}

From Table~\ref{tab:bic_accuracy}, we find the $BIC$ is able to correctly identify more nucleated galaxies (true positives), but also misidentify more non-nucleated galaxies as nucleated (false positives). This, along with the lack of misclassified nucleated galaxies (false negatives), provides evidence towards the sensitivity of the $BIC$ to perturbations at the centre of the galaxies\footnote{Here we specify the centre of the galaxies since the difference in our models is the inclusion of a central PSF component. In principle, the $BIC$ should be sensitive to perturbations anywhere, depending on the choice of decomposition models under consideration.}. This is also supported by the lower nucleus detection limit for the $BIC$ than for the $BIC_{\textup{res}}$ (Fig.~\ref{fig:bic_accuracy}). In the case of $BIC_{\textup{res}}$, there are much fewer false positives at the expense of more false negatives. Nonetheless, the total number of misclassified galaxies are lower for $BIC_{\textup{res}}$, hence its higher overall accuracy. 

In terms of the false positives (i.e. $\Delta BIC > 0$ but non-nucleated), for $M_{\rm *,galaxy} < 10^9 M_{\odot}$ we find a significant overlap (37/39) with the false positives identified in Sect.~\ref{sect:final_sample}, which were based on multi-component models. From inspection of the residual images, we confirm that these false positives can also be classified into the two types of false positives for the same reasons. Whilst it is unlikely, we do not entirely disregard the possibility that some of the low-mass false positives could actually host a nucleus. Future works could shed light on this, as currently there are no images with the required spatial resolution available.

\section{Discussions}\label{sect:discussion}
We find that the nucleation of Fornax galaxies is dependent on a number of factors, such as the galaxy stellar mass and the environment that they reside in. In this section we discuss the nature of the nuclei themselves, their formation mechanisms, and the role that the environment plays in nucleation. 

\subsection{Growth of NSCs}\label{sect:nsc_growth}
\begin{figure}
\centering
\includegraphics[width=\hsize]{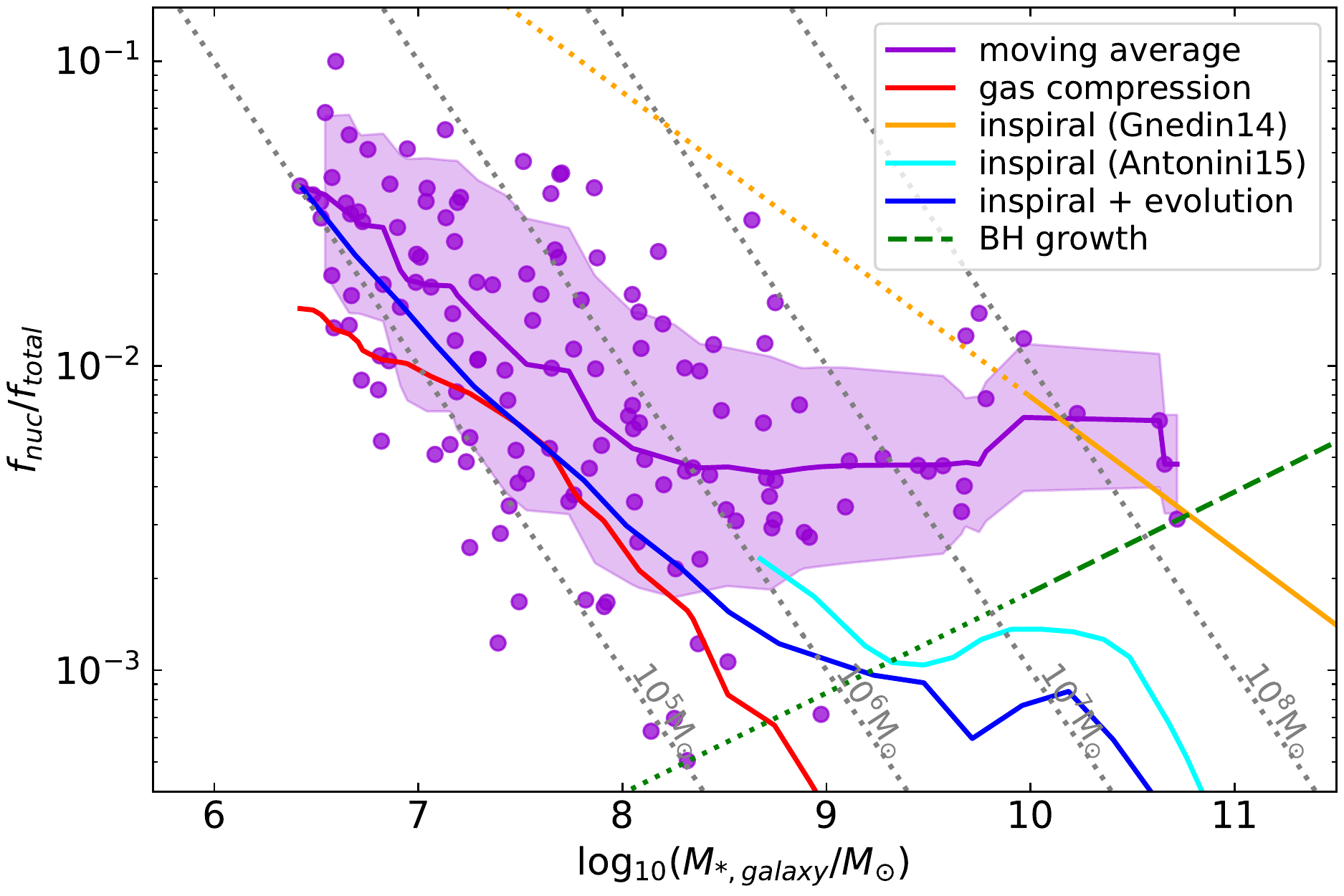}
\caption{Comparison of different NSC formation mechanisms with our nucleated early-type galaxies. The violet points, solid line, and shaded region denote our nucleated galaxies (same as in Fig.~\ref{fig:nuc_ff_multi}). We consider the maximum NSC mass via tidal compression of gas \citep[i.e. an upper limit; red,][]{emsellem2008}, GC in-spiral (orange and cyan, \citealt{gnedin2014} and \citealt{antonini2015}, respectively), and a combination of GC in-spiral and galaxy evolution models \citep[blue,][]{antonini2015}. The black hole scaling relation \citep[green,][]{greene2020} is included to illustrate the evolution of black holes. The dotted portion of the coloured lines denote an extrapolation of the original relation. The dotted grey lines denote constant nucleus stellar masses.}
\label{fig:nsc_formation_mechanisms}
\end{figure}

A prevailing mechanism of NSC growth is the infall of GCs to the minimum of the potential of a galaxy due to dynamical friction \citep{tremaine1975}. Recently, \citet{gnedin2014} modelled the evolution of GCs in galaxies and found that not only can the NSCs form rather quickly ($\lesssim 1$\,Gyr), but also that their mass correlates with the host galaxy mass: $M_{\rm *,NSC} \propto {M_{\rm *,galaxy}}^{0.5}$ (from their Eq.~13). The exponent of 0.5 is consistent with our moving average from Fig.~\ref{fig:nuc_ff_multi} for galaxies with $M_{\rm *,galaxy} \lesssim 10^{8.5} M_{\odot}$, as well as with the exponent of 0.48 from \citet{neumayer2020}. However, the models of \citet{gnedin2014} were based on much more massive galaxies ($M_{\rm *,galaxy} > 10^{10} M_{\odot}$), and the extrapolation of their scaling relation leads to much too massive NSCs at lower galaxy stellar masses. On the other hand, the model of \citet{antonini2015} finds that GC in-spiral alone is insufficient to reproduce the observed NSC stellar masses at high galaxy stellar masses. However, assuming a linear extrapolation, the model qualitatively appears to be in line with our moving average of $f_{\textup{nuc}}/f_{\textup{total}}$ with galaxy stellar mass (see Fig.~\ref{fig:nsc_formation_mechanisms}). It is therefore possible that the low-mass nucleated galaxies in our sample could have grown a significant portion of their nuclei via the in-spiral of GCs. 

An additional mechanism is the compression of gas in the central region of galaxies due to their local tidal field. In particular, \citet{emsellem2008} found that the radial tidal forces are naturally compressive in the central regions of galaxies with relatively shallow central density (i.e. Sérsic index $n \lesssim 3.5$). Furthermore, they found the relation $\log_{10}(M_+ / M_{\rm *,galaxy}) \sim -1.9 \times n - 0.4$, where $M_+$ is the limiting mass of the NSC, above which the tidal forces become disruptive. Applying this relation to our nucleated galaxies, we find that only $\sim 22\%$ of nuclei have masses below their $M_+$. Moreover, we find that those with nuclei masses below their $M_+$ have $M_{\rm *,galaxy} \lesssim 10^9 M_{\odot}$. This suggests that only a fraction of the nuclei could have grown purely based on tidal compression. A potential factor for this is the necessity of gas, for which many of our galaxies lack in the present day. Nevertheless, we do not rule out the possibility that these galaxies were gas-rich in the past and formed NSCs at an earlier epoch. Based on spectroscopic studies of NSCs from \citet{johnston2020} and \citet{fahrion2021}, there is evidence to suggest that the star formation histories of the NSCs not only vary from their host galaxies, but some even exhibit multiple episodes of star formation. This suggests that in-situ star formation also plays an important role in the growth of the NSCs, particularly for massive ($M_* > 10^9 M_{\odot}$) galaxies \citep[see also][]{neumayer2020}.

In Fig.~\ref{fig:nsc_formation_mechanisms} we overlay the various mechanisms discussed on top of our nucleated sample. We also include the model from \citet{antonini2015}, which combined both mechanisms of GC infall and in-situ star formation as well as the effects of central black holes on NSC growth. In particular, they found that the central black holes can inhibit NSC growth for massive ($M_* > 10^{10} M_{\odot}$) galaxies. Recently, the N-body simulations of \citet{askar2021} show that the merging of star clusters can potentially form black holes which grow via gas accretion to become supermassive black holes. Based on early-type galaxies, \citet{greene2020} found a trend of $\log_{10}(M_{\rm BH}) = (1.33\pm 0.12) \times \log_{10}(M_{\rm *,galaxy}/(3\times10^{10} M_{\odot})) + (7.89\pm 0.09)$ (from their supplement table~5). The gradient of this $\log$--$\log$ relation is steeper than the gradient we find in Eq.~\ref{eqn:nucmstar_highmass} for our massive galaxies, which suggests that the central black holes grow faster than NSCs. This difference in gradient could be interpreted as a sign of the disruptive effects of central black holes on NSC growth.

\subsection{Dependence on the environment}\label{sect:env_dependence}
\subsubsection{Nucleation fraction}\label{sect:env_dependence_nucfrac}
From Sect.~\ref{sect:nuc_frac} we find that the nucleation fraction depends on the galaxy environment, with most nucleated galaxies found at the centre of the Fornax main cluster and Fornax A group. Furthermore, we find the overall nucleation fraction of the Fornax main cluster to be higher than in the Fornax A group. This difference in nucleation fraction between environments is in line with the findings of \citet{sanchezjanssen2019_nsc} and \citet{carlsten2021_dwarf}, who found similar nucleation fractions in the Virgo and Fornax clusters, whereas the Coma cluster and the Local Volume had higher and lower nucleation fractions than the Fornax cluster, respectively. Similarly, \citet{zanatta2021} found a trend between the halo mass of the environment and the galaxy luminosity at a given nucleation fraction (their Fig.~7). In contrast, \citet{baldassare2014} found similar nucleation fractions between Virgo cluster galaxies and galaxies in the field. As \citet{neumayer2020} discusses, this apparent discrepancy could be due to the difference in the stellar mass of the samples (\citet{baldassare2014} sampled high-mass galaxies whereas \citet{sanchezjanssen2019_nsc} mainly sampled low-mass galaxies). According to \citet{poulain2021}, their field dwarfs have lower nucleation fractions than dwarfs from the core of the Virgo cluster\footnote{We note that the Virgo cluster may not be the most typical cluster \citep[see][]{janz2021}.}. Based on our sample of dwarf and massive galaxies, we find that the nucleation fraction for galaxies beyond the virial radius of the Fornax main cluster (and to some degree, the Fornax A group) is comparable to the nucleation fraction of galaxies in the field (see Sect.~\ref{sect:nuc_frac}). 

Recently, \citet{leaman2021} studied the link between the nucleation fraction of galaxies and the environment that they reside in. In particular, they focused on the GC in-spiral mechanism and defined a limiting GC mass above which they can survive mass loss during in-spiral, and hence contribute to the NSCs. Based on their model, they found that the observed shape of nucleation fraction as a function of galaxy mass (e.g. Fig.~\ref{fig:nuc_frac_maingroup}), and the location of the peak, can be reproduced, depending on the assumed galaxy size--mass scaling relation and the GC mass function. Furthermore, they argue that the observed difference in nucleation fractions in different halo mass environments \citep[e.g. in][]{sanchezjanssen2019_nsc, zanatta2021} are likely due to the preferential disruption of the most diffuse (low surface brightness) galaxies by the cluster potential, which also tend to be low-mass and non-nucleated. In this scenario, over time, those which survive the disruption are generally the more massive, higher surface brightness galaxies, which have relatively higher nucleation fractions. This can lead to two features in the observed nucleation fractions: the correlation between cluster mass and nucleation fraction at a given galaxy mass \citep[e.g.][]{zanatta2021}, as the strength of tidal disruption is correlated with the cluster mass; and the high nucleation fraction at the cluster centre, since the stronger tidal effects at the cluster centre would be more efficient in disrupting the galaxies. 

An important caveat to this scenario is the 'initial' galaxy size--mass relation (i.e. the population of galaxies in the cluster before they are affected by the cluster potential), specifically for galaxies at the low-mass end (e.g. $M_* < 10^7 M_{\odot}$). In \citet{leaman2021}, they adopted a size--mass relation which has an inflexion at $\sim 10^9 M_{\odot}$, such that, on average, galaxies below and above this mass tend to have larger sizes. Although we recognise that our observed size--mass relation is of the present day, and hence can differ from the size--mass relation of the past (in other words, the 'initial' galaxy size--mass relation), we do not find any signs which point to such a feature in our moving average (see column~1 of Fig.~\ref{fig:host_pars_etg_nosig}). For such a change in the size--mass relation, this would imply that the tidal effects must be very efficient to leave no trace of these very diffuse galaxies. Recently, \citet{marleau2021} studied ultra-diffuse galaxies (UDGs) in low density environments (as a part of MATLAS). They defined UDGs as those with $R_e \ge 1.5$\,kpc and a central surface brightness of $\mu_{0,g} \ge 24$\,mag\,arcsec$^2$. From inspection, their sample generally appears to be inline with our size--mass scaling relation, and they found roughly comparable nucleation fractions to our FDS sample ($20/59=0.34$ and $148/557=0.27$, respectively). Given that the sample from \citet{marleau2021} reside in low density environments, we do not expect the environment to be able to efficiently disrupt UDGs. As such, some of the faint, low-mass UDGs (which would appear as an upturn in the size--mass relation) should survive to the present day and be observable, so their absence from the sample of \citet{marleau2021} is intriguing. On the other hand, an argument could be made that these UDGs may be fainter than the detection limits from current surveys. 

It is possible that in the early Universe, GCs are more likely to form in dense environments. This naturally makes galaxies residing in the core of clusters ideal candidates to host more GCs, which over time can transform into NSCs via dynamical friction. Indeed, \citet{lisker2013} showed that early-type galaxies typically have resided in clusters for several Gyr, and \citet{sanchezjanssen2012} found that the properties of early-type galaxies in Virgo are not consistent with recent environmental transformations. Overall, in order to address the effects of tidal disruption in a more quantitative manner, simulations which model the cluster potential and account for the galaxy mass distribution, including the lowest mass dwarfs, are required. 

\subsubsection{Galaxy structures}
\begin{figure}
\centering
\includegraphics[width=\hsize]{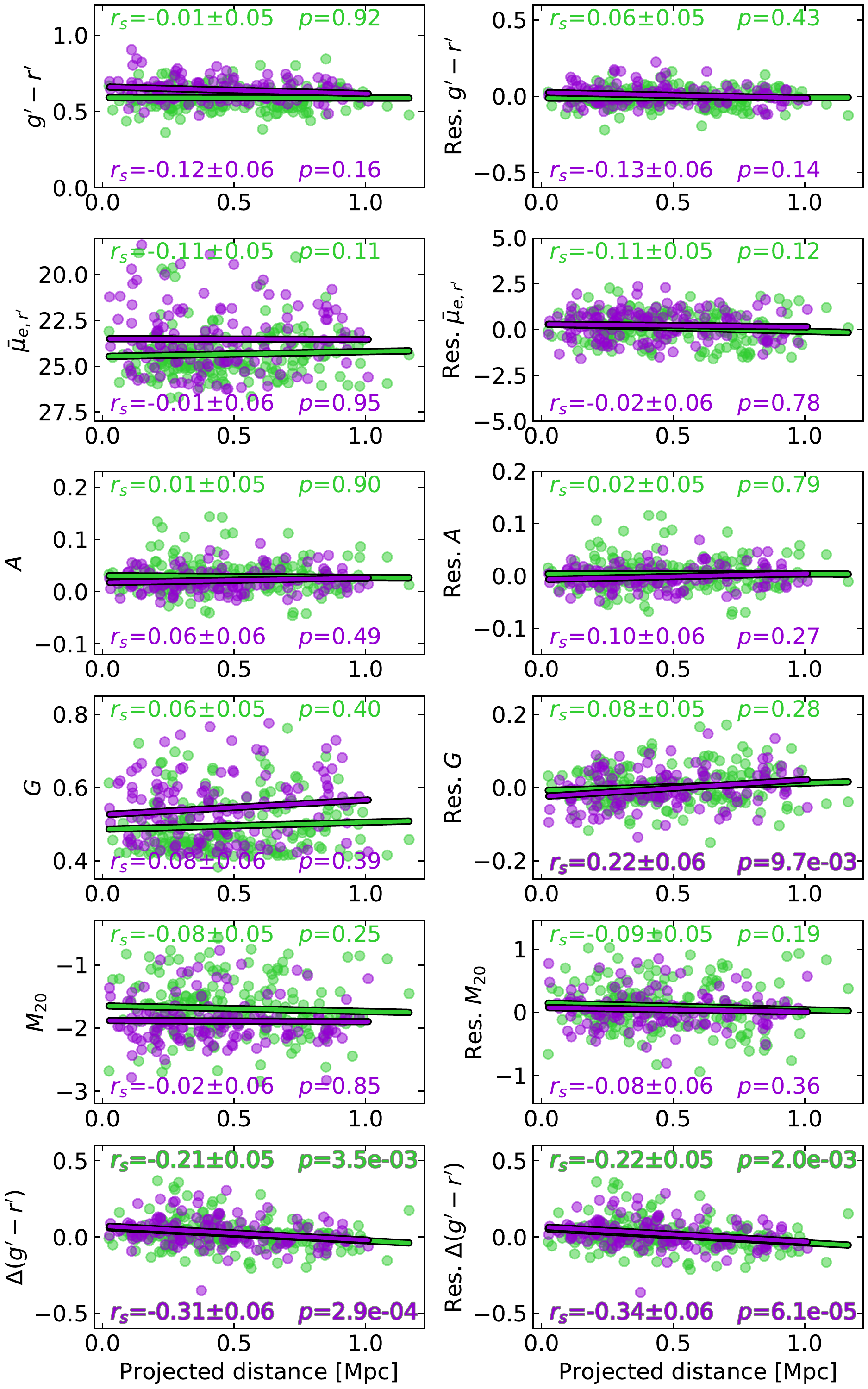}
\caption{Structural quantities of the nucleated (violet) and non-nucleated (green) early-type galaxies in the Fornax main cluster only, as a function of the projected distance. Only galaxies with overlapping stellar mass are shown (i.e. column 4 from Fig.~\ref{fig:host_pars_etg}). The measured (left) and residual (i.e. mass-trend corrected; right) quantities are shown. The solid lines are based on linear fits to the data and are only meant as visual guides to the Spearman's rank correlation coefficient, $r_s$. For each quantity the $r_s$ and associated p-values are annotated in the corresponding subplots. }
\label{fig:host_dists_sigonly}
\end{figure}

Given the apparent dependence of nucleation on projected distance, and the difference in galaxy properties between nucleated and non-nucleated galaxies (see Fig.~\ref{fig:host_pars_etg}), we also test whether environmental mechanisms (e.g. RPS, tidal interactions) could play a role in the observed differences in the structural quantities. Since the majority of our nucleated galaxies are from the Fornax main cluster, Fig.~\ref{fig:host_dists_sigonly} shows the galaxy properties of early-type Fornax main cluster galaxies (with overlapping stellar mass between nucleated and non-nucleated sub-samples) as a function of projected distance. Logically, one could argue that if the galaxies reside in the same environment, the environmental effects should apply to both nucleated and non-nucleated galaxies alike, and we should not expect any differences between the two sub-samples. However, it is not implausible that nucleated galaxies may have experienced more (or less) environmental effects in their past which lead to nucleation, or vice versa for non-nucleated galaxies. This can occur as the efficiencies of the environmental mechanisms can vary depending on the orbital parameters of the galaxies falling into the cluster \citep[e.g.][]{smith2015, bialas2015}. Alternatively, the conditions in high density regions in the past could have been more favourable towards NSC formation (or form GCs which then undergo in-spiral), which can lead to the observed higher nucleation fraction towards the cluster centre, and hence experience environmental effects for a longer period. In such cases, we might expect a difference in the projected distance trends between nucleated and non-nucleated galaxies. 

To test whether there are any significant differences in the projected distance trends between nucleated and non-nucleated galaxies, we calculate the Spearman's rank correlation coefficient $r_s$ and the associated p-value\footnote{Here, the null hypothesis is that there is no correlation with projected distance.} for each galaxy quantity. From Fig.~\ref{fig:host_dists_sigonly} we find that the quantities ($g'-r'$, $R_e$, $\bar{\mu}_{e,r'}$, $A$) which had significantly different distributions between nucleated and non-nucleated galaxies do not have significant projected distance trends. If we interpret the projected distance as a rough indicator of the time spent within the cluster for environmental mechanisms to act (as \citealt{su2021} did), this implies that the environmental mechanisms are unlikely to be the main driver in the observed differences in these galaxy quantities. 

Of all the structural quantities tested, we only find significant trends for the residual Gini coefficient ($G$) and the outer-to-inner colour difference ($\Delta (g'-r')$, see Fig.~\ref{fig:host_dists_sigonly}) as a function of projected cluster-centric distance. In the case of $\Delta (g'-r')$ both nucleated and non-nucleated galaxies show significant trends. The negative values of $r_s$ suggest that galaxies towards the centre of the cluster have bluer inner regions than those in the cluster outskirts. \citet{su2021} interpreted this as a result of RPS acting outside-in. Given the clear cluster-centric trends for both nucleated and non-nucleated galaxies, this suggests that the RPS affects both types of galaxies equally. However, for $G$ we find a significant trend for nucleated galaxies, but not for non-nucleated galaxies. The positive $r_s$ suggests that nucleated galaxies closer to the cluster centre have a more homogeneous distribution of light (i.e. the galaxy light is more evenly distributed amongst the pixels). Overall, the general absence of difference in the cluster-centric trends (or lack thereof) between the nucleated and non-nucleated galaxies suggests that the environmental effects affect both types of galaxies to a similar extent on average. Furthermore, the lack of cluster-centric trends with the structural quantities suggests that environmental mechanisms are unlikely to be directly responsible for the observed differences between the nucleated and non-nucleated galaxies (see Fig.~\ref{fig:host_pars_etg}). Nevertheless, the environment clearly plays an important role in the nucleation of galaxies in Fornax.

\subsection{Comparison to UCDs}\label{sect:ucds}
Ultra-compact dwarfs \citep[UCDs;][]{hilker1999, phillipps2001} are compact objects with typical half-light sizes of $< 100\,$pc and stellar masses of $\gtrsim 10^{6} M_{\odot}$ \citep[e.g.][]{drinkwater2003, evstigneeva2007, mieske2008}. They appear to be ubiquitous, having been observed in both clusters and in lower density environments \citep[see][and references therein]{liu2020}. UCDs appear quite similar to the brightest, most massive GCs in the GC luminosity functions of galaxies, as well as resemble isolated NSCs without a visible host galaxy. As such, the formation channels for UCDs are thought to follow two mechanisms: the stripping of NSCs from nucleated galaxies via mergers or tidal disruption \citep{bekki2003, goerdt2008, pfeffer2013, norris2015, janz2016_ucd}, or the coalescence of massive stellar clusters which form the most massive GCs \citep{fellhauer2002}. 

Here we would like to compare the observed properties of UCDs in Fornax with the nuclei from our sample. For our UCD sample we compiled the likely UCD candidates based on photometry in \citet{saifollahi2021} and the catalogue of spectroscopically confirmed Fornax compact objects (including GCs and UCDs) from \citet{wittmann2016}. The UCD candidates in \citet{saifollahi2021} were selected based on a limit of $m_{g'} < 21$\,mag ($M_{g'} = -10.5$\,mag), as well as a combination of colours spanning the visible wavelengths to near infrared. To ensure a homogeneous sample we applied the same $m_{g'}$ magnitude cut for the \citet{wittmann2016} sample to select the UCDs\footnote{In \citet{wittmann2016} they used a magnitude cut of $m_{\textup{Ve}}<21.5$\,mag, where $m_{\textup{Ve}}$ denotes the 'equivalent $V$' band magnitude (see their Sect.~2.4).}. This resulted in $44+65=109$ UCDs from the Fornax main cluster. We estimated the stellar mass of UCDs based on their colours and Eq.~\ref{eqn:mstar} and used the nucleus stellar masses based on $r'$ band flux fractions. 

\begin{figure}
\centering
\includegraphics[width=\hsize]{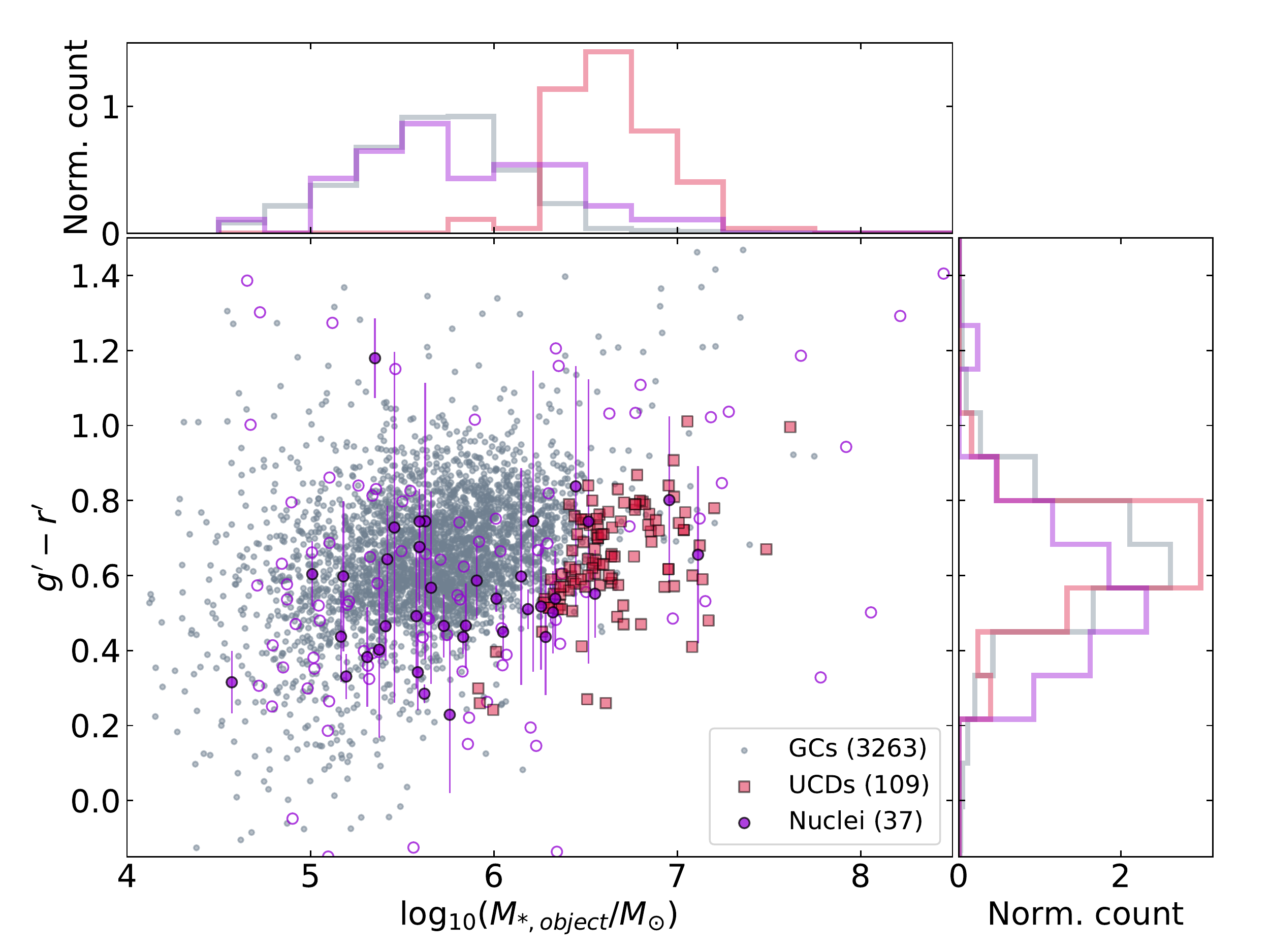}
\caption{$g'-r'$ colour as a function of stellar mass for GCs (grey), UCDs (crimson) and nuclei (violet). As in Fig.~\ref{fig:nuc_colour}, the nuclei with nucleus contrast $>1$ are shown as filled circles, whereas nuclei with nucleus contrast $<1$ are shown as open circles. The upper and right panels show the stellar mass and $g'-r'$ distributions, respectively. The nuclei uncertainties are the same as in Fig.~\ref{fig:nuc_colour}. The UCD uncertainties are generally within the size of the points, hence they are omitted. }
\label{fig:nuc_ucd_cmd}
\end{figure}

In Fig.~\ref{fig:nuc_ucd_cmd} we compare the $g'-r'$ colour of the nuclei and UCDs as a function of their stellar masses. As in Sect.~\ref{sect:nuc_colours}, here we select nuclei with nucleus contrast $>1$ for the colour comparisons. For reference we also include the master GC catalogue of \citet{cantiello2020}, which includes photometrically and spectroscopically confirmed GCs (see their Table~5). At first glance, the plot seems to suggest that UCDs tend to be more massive than nuclei. However, this is due to the magnitude limit from the UCD compilation sample, so UCDs with lower stellar masses are not well represented. In terms of the $g'-r'$ colours, we find that UCDs have slightly redder mean colours compared to nuclei (0.64 and 0.56 for UCDs and nuclei, respectively) and similar standard deviations in colour for UCDs and nuclei (0.14 and 0.18 for UCDs and nuclei, respectively). Considering the nuclei and UCDs within an overlapping range in stellar mass ($10^{6}\,M_{\odot} \lesssim M_* \lesssim 10^{7.5}\,M_{\odot}$), the colour distributions are rather similar (mean and standard deviation in $g'-r'$ are 0.65 and 0.13 for UCDs, and 0.60 and 0.13 for nuclei, respectively). This comparability in colours potentially allude to similar stellar populations between UCDs and nuclei. 

\begin{figure}
\centering
\includegraphics[width=\hsize]{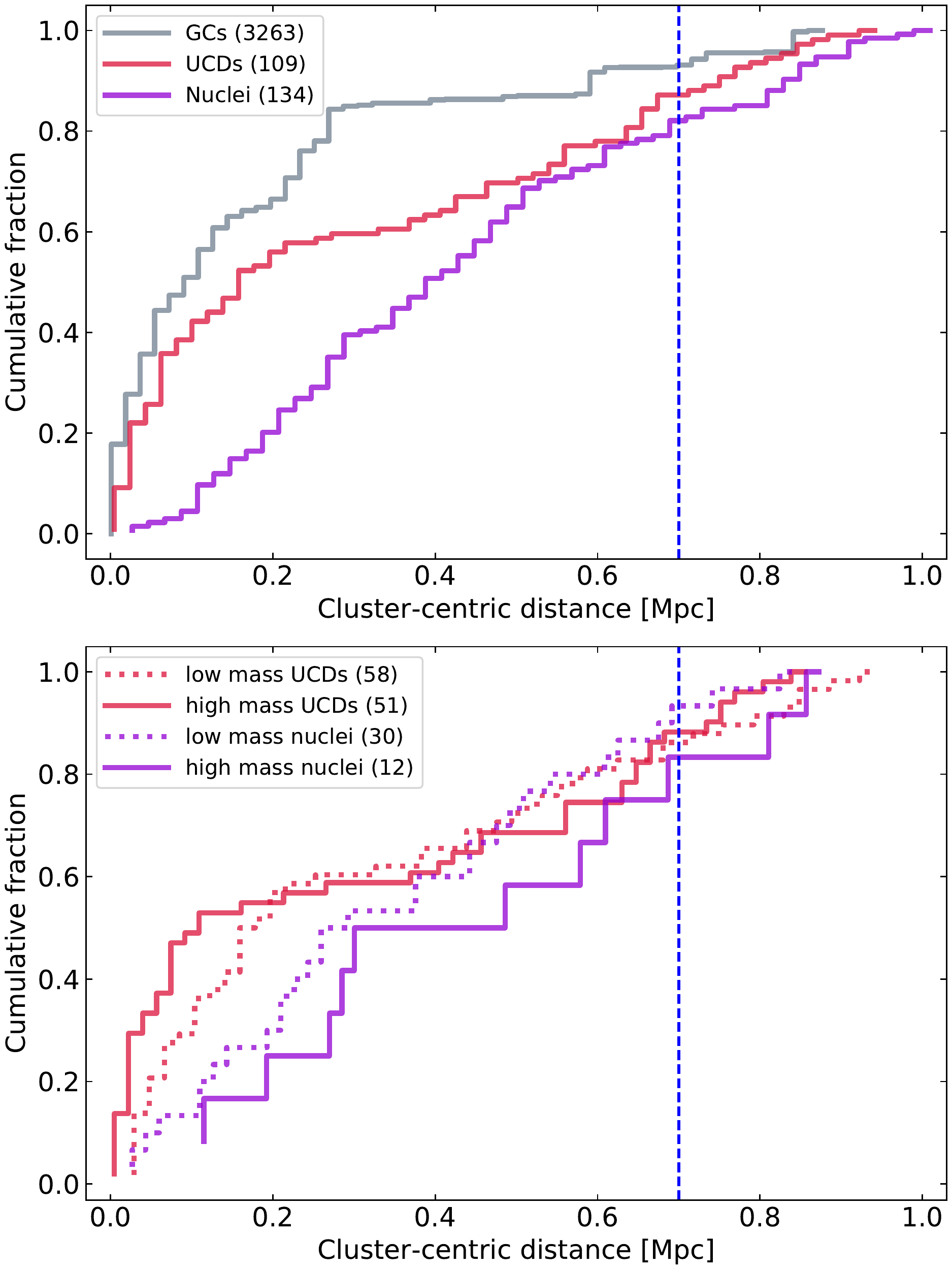}
\caption{Distribution of projected cluster-centric distance for GCs (grey), UCDs (crimson), and nuclei (violet). Here, only nuclei from early-type galaxies in the Fornax main cluster are shown. The upper panel shows the cumulative distribution of all the compact objects, respectively. The lower panel shows the cumulative distributions of the low (dotted) and high (dashed) mass sub-samples of UCDs split at $M_* = 10^{6.6} M_{\odot}$ and the nuclei within the accompanying UCD stellar mass ranges. The dotted blue line denotes the virial radius of the Fornax main cluster. }
\label{fig:nuc_ucd_dist}
\end{figure}

Additionally, we compare the projected distance distributions of the nuclei and UCDs. In Fig.~\ref{fig:nuc_ucd_dist} we show the nuclei (regardless of nucleus contrast), UCDs, and GCs in the Fornax main cluster. Overall, a significant fraction of UCDs are located at the central ($<0.2$\,Mpc) region of the Fornax main cluster. Towards the cluster outskirts and beyond ($>0.5$\,Mpc) we do not find any particular differences between nuclei and UCDs. Applying the four test statistics (KS, AD, LP, CU), we find that the distributions of projected distances are significantly different for the two samples (p-value $<0.001$). To test the significance further, we split our UCD sample by $M_* = 10^{6.6} M_{\odot}$ into a high and low-mass sub-samples and recalculated the hypothesis tests with respect to the nuclei. We find that both the low and high-mass UCDs are significantly different from our nuclei. When we also restrict the nuclei sample to within the same stellar mass range to match the UCDs sub-samples, we find that the high-mass UCDs and nuclei remain significantly different, but not for the low-mass UCDs and nuclei ($p>0.05$). This appears to be due to the prevalence of high-mass UCDs at the cluster centre, whereas the low-mass UCDs tend to be located beyond and are more evenly distributed with distance. Whilst the higher number of UCDs at the cluster centre compared to nuclei is consistent with the scenario that UCDs are the remaining nuclei from disrupted dwarf galaxies, it remains possible that some of the UCDs are, in fact, massive GCs, which are also concentrated at the centre of the cluster.  

We also test the scenario of UCDs as stripped nuclei with the cosmological simulations of \citet{goerdt2008}. By allowing M33-like haloes to evolve in a Virgo-like cluster for 5\,Gyr, they found a set of orbital parameters (pericentre of apocentre) where the galaxies (with and without a gas disk) would be completely disrupted, leaving only the nuclei. Of course, not all galaxies become disrupted, with those that survive appearing as nucleated galaxies in the cluster. Hence, the scenario of UCDs as stripped nuclei can be tested by comparing the predicted number of UCDs and nucleated galaxies from the simulation to those observed in the cluster. We follow \citet{goerdt2008} by determining the UCD fraction, $f_{\textup{UCD}} = N_{\textup{UCD}} / (N_{\textup{UCD}} + N_{\textup{nuc}})$, as a function of projected cluster-centric distance, and comparing to the observations. As a caveat, we recognise that not all UCDs form from disrupted nucleated galaxies, with the possibility that low-mass UCDs are, in fact, massive GCs. Nevertheless, we find the comparison between the observations and simulations to be illuminating for examining stripped nuclei as an UCD formation channel. 

\begin{figure}
\centering
\includegraphics[width=\hsize]{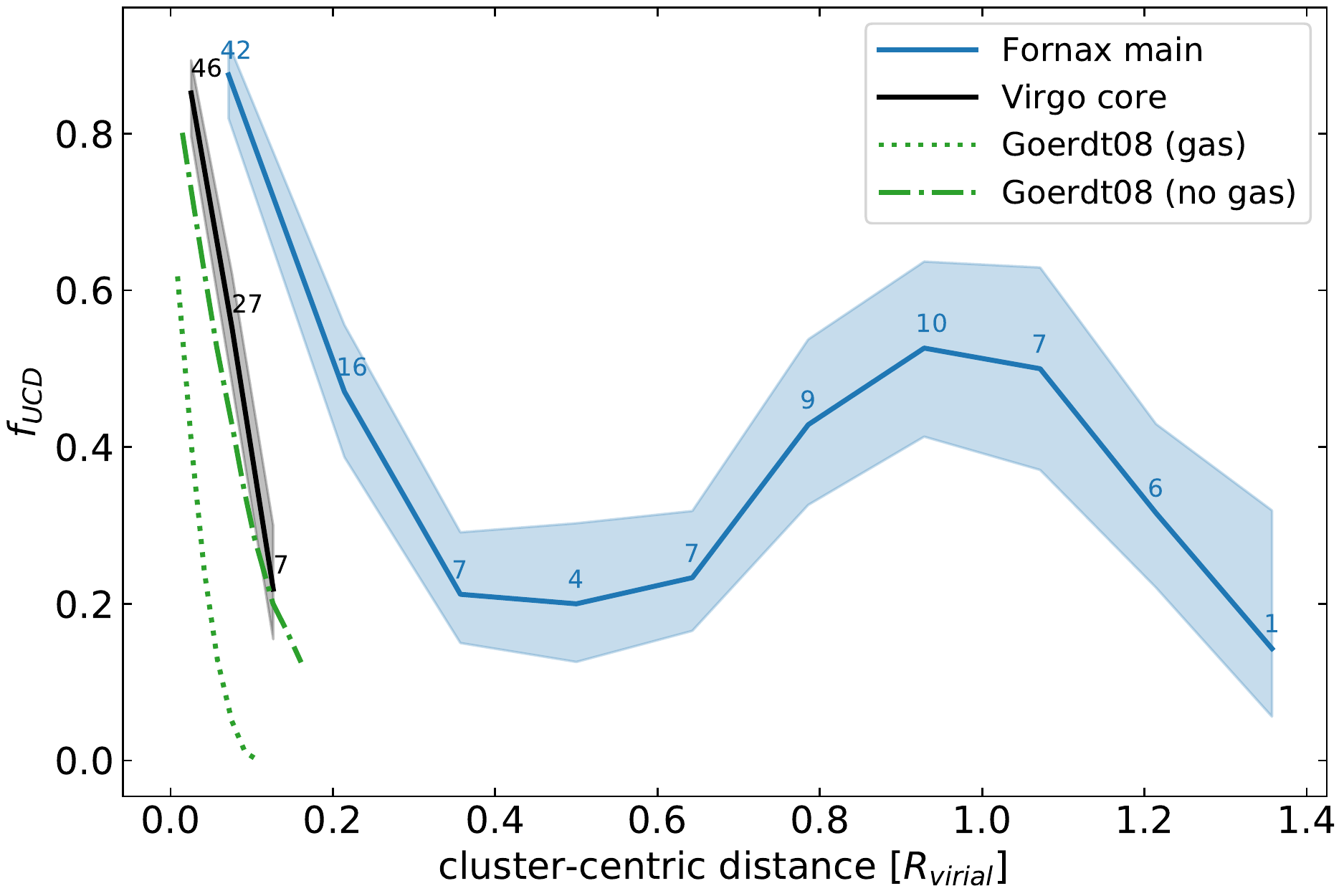}
\caption{UCD fraction ($f_{\textup{UCD}} = N_{\textup{UCD}} / (N_{\textup{UCD}} + N_{\textup{nuc}})$) as a function of projected cluster-centric distance for the Fornax (blue) and Virgo (black) clusters. We use a virial radius of 1.55\,Mpc for the Virgo cluster \citep{ferrarese2012}. The shaded regions show the $1\sigma$ binomial uncertainty. The dotted and dash-dot green lines show the predicted UCD fractions depending on whether the UCD progenitor galaxies contain gas or not, respectively. }
\label{fig:ucd_frac_dist}
\end{figure}

In Fig.~\ref{fig:ucd_frac_dist} we show the UCD fractions from the \citet{goerdt2008} simulations of galaxies with and without a gas disk. We compare their predictions with our Fornax main cluster sample, as well as using the nucleated galaxies \citep{ferrarese2020} and UCDs \citep{liu2020} from the core of the Virgo cluster from the NGVS. To ensure a reasonable comparison, we only select nucleated galaxies and UCDs within the overlapping region of both datasets. This restricts the Virgo sample to the cluster core ($\lesssim 200$\,kpc). We select the Virgo nucleated galaxies as those which were classed as certain or possible Virgo members and are nucleated or host an offset nucleus \citep['Class'=0 or 1 and 'Nuc ID'=1 or 2,][see their Tables~4 and 5]{ferrarese2020}. As for the Virgo UCDs, we select those classified as 'probable UCD' \citep['Class'=1,][see their Table~4]{liu2020} and we apply the same magnitude limit from \citet{saifollahi2021}. We find that the UCD fraction peaks at the centre of the Virgo cluster and drops steeply with increasing distance. Similarly, the UCD fraction peaks at the centre of the Fornax cluster, although the decrease with distance appears shallower. The predicted UCD fractions of the galaxies without gas better match the observed UCD fractions in the Virgo core. However, neither models match well with the observed UCD fractions from the Fornax main, which warrants further investigation. Interestingly, in Fig.~\ref{fig:ucd_frac_dist} we find a second, smaller peak around the virial radius of the Fornax main cluster, which appears to be consistent with the overdensities of UCDs identified in \citet{saifollahi2021} (see their Fig.~23). 

\section{Conclusions}\label{sect:conclusion}
In this work we studied the properties of nucleated galaxies from the Fornax main cluster and the Fornax A group, as well as the role that the environment plays. From the \citet{su2021} compilation of dwarfs \citep[from FDSDC,][]{venhola2018} and massive galaxies \citep[from][]{iodice2019, raj2019, raj2020}, we selected 557 galaxies which were labelled as nucleated and non-nucleated based on visual inspection of the galaxy and residual images from multi-component decompositions. To ensure we did not miss any faint nuclei, we compiled an additional candidate list of galaxies labelled as non-nucleated from \citet{su2021} which showed a better fit with a Sérsic+PSF model rather than a single Sérsic when using the model selection criterion $BIC$. This allowed us to find 20 additional galaxies which indeed hosted faint nuclei. 

From our nucleation labels, we tested the accuracies of the two BIC formulations, $BIC$ and $BIC_{\textup{res}}$, in determining whether a single Sérsic model is preferred over a Sérsic+PSF model, and hence infer whether a galaxy is nucleated or not (Sect.~\ref{sect:auto_mod_select}). We investigated the nucleation fraction as a function of the galaxy stellar mass and the projected distance (i.e. distance from the centre of the environment) between the Fornax main cluster and Fornax A group (Sect.~\ref{sect:nuc_frac}). We also utilised the multi-component decompositions to determine the total magnitude of the nuclei in the $g'$, $r'$, and $i'$ bands to determine their colours (Sect.~\ref{sect:nuc_colours}) and stellar masses (Sect.~\ref{sect:nuc_stellar_mass}). We also compared the structural properties of the nucleated and non-nucleated early-type galaxies through their Sérsic-derived quantities and non-parametric morphological indices \citep[from][]{su2021}, and calculated the statistical significance in their differences (Sect.~\ref{sect:nuc_vs_nonuc}). 

From the analyses in this work and in the literature, the big picture of NSCs (specifically, but not exclusively to Fornax) is their dependence on their host galaxies, particularly the internal and external effects they experience. The internal mechanism for NSC formation and growth is dependent on the mass of the galaxy, with different mechanisms dominating at different mass regimes (e.g. GC in-spiral and in-situ star formation, see Fig.~\ref{fig:nsc_formation_mechanisms} and Sect.~\ref{sect:nsc_growth}). As a result, we see that the nucleation fraction (see Fig.~\ref{fig:nuc_frac_maingroup}) and the NSC stellar mass (see Figs.~\ref{fig:nuc_mstar_multi} and \ref{fig:nuc_ff_multi}) depend heavily on the host galaxy mass. On the other hand, external effects due to the cluster environment must also play a role, given that NSCs appear preferentially in high density environments (see Fig.~\ref{fig:nuc_frac_distance}). It is not yet clear how environmental effects could lead to the high nucleation fraction at the cluster centre. Given that nucleated galaxies tend to be more massive than non-nucleated galaxies (see Fig.\ref{fig:hist_nucleation} and Sect.~\ref{sect:env_dependence_nucfrac}), it is plausible that non-nucleated galaxies are more susceptible to tidal disruptions, the effects of which are strongest at the cluster centre. Naturally, any nucleated galaxy which reach the cluster centre are mostly stripped of gas, and, if disrupted, can be transformed into a UCD (see Fig.~\ref{fig:ucd_frac_dist}). 

Our main results can be summarised as the following:
\begin{itemize}
    \item The nucleation fraction in the Fornax A group is generally lower at all galaxy stellar masses than in the Fornax main cluster, particularly regarding galaxies with $M_{\rm *,galaxy} \sim 10^7 M_{\odot}$ (see Figs.~\ref{fig:nuc_frac_maingroup} and \ref{fig:nuc_frac_distance}). 
    \item We find that for nucleated galaxies with $10^{7} M_{\odot} \lesssim M_{\rm *,galaxy} \lesssim 10^{9} M_{\odot}$ their nuclei tend to be marginally bluer than the host galaxies themselves (see Fig.~\ref{fig:nuc_colour}). More massive galaxies have redder nuclei compared to their host galaxy.  
    \item The relation between the nucleus and the host galaxy stellar masses follow different trends depending on the stellar mass of the galaxy. Galaxies with $M_{\rm *,galaxy} < 10^{8.5} M_{\odot}$ follow $M_{*,\textup{nuc}} \propto {M_{*,\textup{galaxy}}}^{0.5}$, whereas higher mass galaxies follow $M_{*,\textup{nuc}} \propto M_{*,\textup{galaxy}}$ (see Fig.~\ref{fig:nuc_ff_multi}). The relation for lower mass galaxies is consistent with the relation found in \citet{neumayer2020}, but deviates at higher galaxies stellar masses. 
    \item There are significant (p-value $<0.05$) differences in the distributions of residual (i.e. stellar mass-trend removed) colour ($g'-r'$), asymmetry index ($A$), brightest second order moment of light ($M_{20}$), and outer--to--inner colour difference ($\Delta(g'-r')$) between nucleated and non-nucleated early-type galaxies (see Fig.~\ref{fig:host_pars_etg} and Table~\ref{tab:fds_stats_etg}). 
    \item For $M_{\rm *,galaxy} \lesssim 10^{8} M_{\odot}$, we find a significant trend between residual $\bar{\mu}_{e,r'}$ and the axial ratio for the nucleated galaxies, but not for the non-nucleated galaxies (see Fig.~\ref{fig:muebar_vs_q}). This implies that the fainter residual $\bar{\mu}_{e,r'}$ for nucleated galaxies are likely due to inclination effects which reflect the fact that the nucleated galaxies are well described as oblate spheroids, whereas non-nucleated galaxies are better described as triaxial ellipsoids. 
    \item We do not find significant correlations between the residual quantities and projected distance except for $\Delta(g'-r')$ and $G$ (see Fig.~\ref{fig:host_dists_sigonly}). The lack of difference in the correlations between nucleated and non-nucleated galaxies suggests that their observed differences in residual quantities are not directly due to environmental effects. Nevertheless, we do not dismiss the role of the environment, particularly on nucleation of galaxies. 
    \item We find similar colours for Fornax UCDs and nuclei, in particular when considering similar stellar mass range (see Fig.~\ref{fig:nuc_ucd_cmd}). Furthermore, comparing the distributions of nuclei and UCDs as functions of their projected cluster-centric distances, we find the UCDs are more concentrated towards the cluster centre than the nuclei (see Fig.~\ref{fig:nuc_ucd_dist}). Additionally, we find that the UCD fraction ($f_{\textup{UCD}} = N_{\textup{UCD}} / (N_{\textup{UCD}} + N_{\textup{nuc}})$) as a function of projected cluster-centric distance peaks at the centre of the Fornax main cluster, which is consistent with the observed UCD fraction within the core of the Virgo cluster, as well as with the predicted UCD fractions from cosmological simulations of \citet{goerdt2008} (see Fig.~\ref{fig:ucd_frac_dist}). This is consistent with the idea that UCDs may be nuclei stripped from dwarfs. 
    \item We demonstrate that the BIC can be used as a method of model selection to differentiate between nucleated and non-nucleated early-type galaxies with high accuracy, particularly when only considering galaxies with $M_* < 10^9 M_{\odot}$ (up to 97\%, see Table.~\ref{tab:bic_accuracy}). In principle the scaling relations of NSCs (e.g. with NSC stellar mass and size) can be used in conjunction with the BIC to detect outliers and increase the overall detection accuracy and should be investigated in the future. 
\end{itemize}

\begin{acknowledgements}
      We thank the anonymous referee for the thoughtful and constructive comments which helped improve this work. We acknowledge financial support from the European Union’s Horizon 2020 research and innovation program under the Marie Skłodowska-Curie grant agreement No.~721463 to the SUNDIAL ITN network. HS and AV are also supported by the Academy of Finland grant No.~297738. 
\end{acknowledgements}

\bibliographystyle{aa}

\bibliography{references}

\appendix

\section{Newly identified nuclei}\label{app:new_nucleated}
\begin{figure*}
\centering
\resizebox{0.75\hsize}{!}{\includegraphics[width=\hsize]{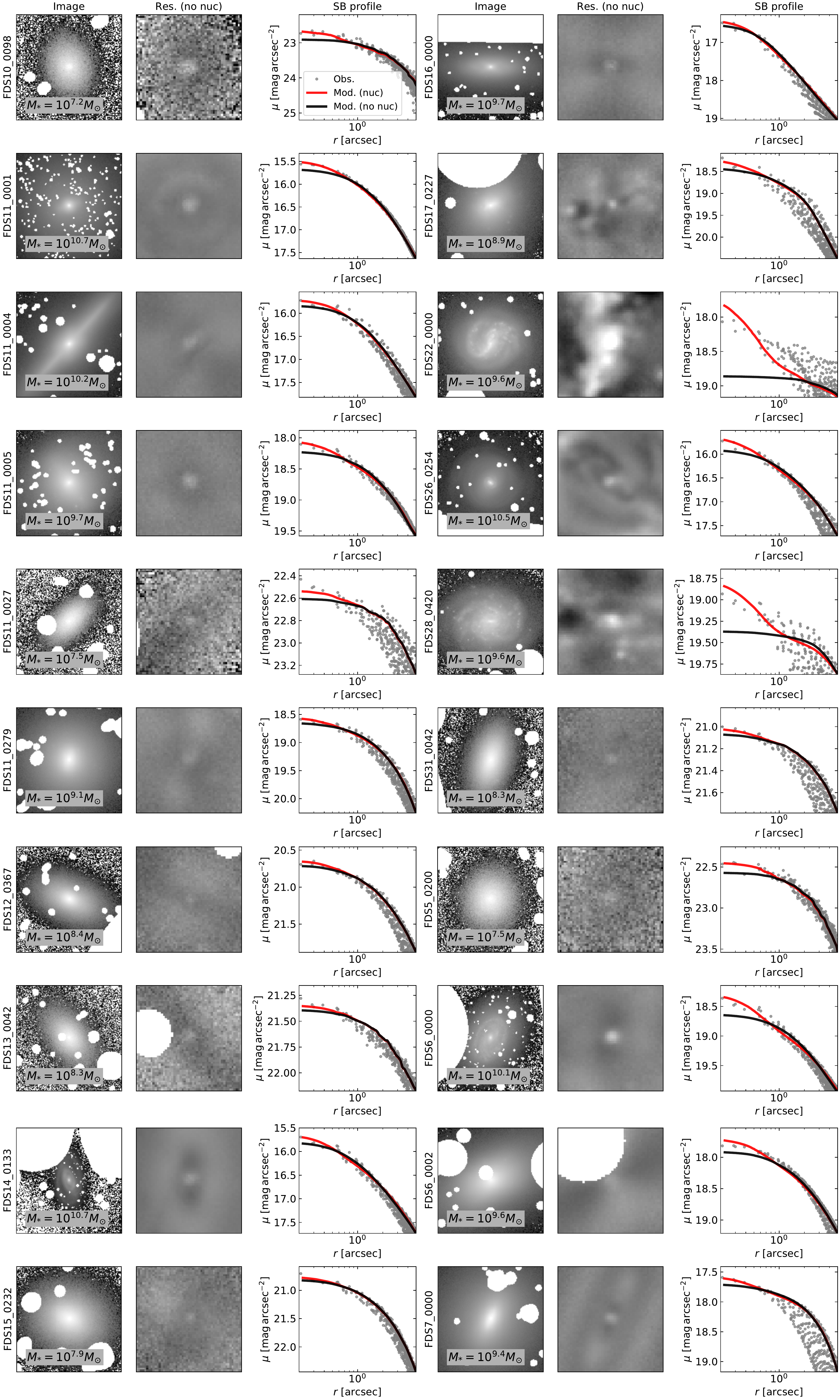}}
\caption{Twenty galaxies which are labelled as nucleated in this work (see Sect.~\ref{sect:final_sample}). Column 1 (4) shows the galaxy images with widths of $5R_e$, with the galaxy stellar masses annotated. Column 2 (5) shows the residual images (widths of 10\,arcsec) based on the multi-component models without a nucleus component. Column 3 (6) shows the galaxy surface brightness from individual pixels (grey) and the azimuthally averaged surface brightness profiles from multi-component models with (red) and without (black) a nucleus component. }
\label{fig:new_nucleated}
\end{figure*}

In Sect.~\ref{sect:final_sample} we identify 20 galaxies in our sample which were not labelled as nucleated in \citet{su2021} but are labelled nucleated in this work (based on a combination of $BIC$ and visual reassessment, as described in Sect.~\ref{sect:final_sample}). In Fig.~\ref{fig:new_nucleated} we show the galaxy images as well as the residual images based on multi-component decomposition models with and without a nucleus component. 

\section{Completeness}\label{sect:completeness}
Broadly speaking, the limit in detecting nuclei boils down to the surface brightness contrast of the nucleus to its host galaxy: the higher the nucleus contrast, the easier it is to detect. It is worth noting that the nucleus contrast is dependent on the seeing of the data: as the nuclei are unresolved in the FDS, the true nucleus contrast will be much higher than what we obtain in this work. In data with better seeing, the light of the nuclei will be spread over less pixels whilst the underlying galaxy light will not differ drastically, leading to higher nucleus contrasts. As we will discuss in Sect.~\ref{sect:det_limit}, the dependence of nucleus contrast on the spatial resolution impacts the lowest nucleus mass which can be detected. 

Another point to consider is that the nucleus contrast is affected by certain factors which dominate at different masses. In massive galaxies, the nucleus tends to reside alongside bright stellar structures (e.g. bulges, bars). As a result, the surface brightness profiles typically rise sharply in the inner regions of the galaxies. The brighter the structures, the lower the nucleus contrasts, making them more difficult to detect. Indeed, of the 20 galaxies which we identify as nucleated in this work, the 13 more massive galaxies ($M_* > 10^9 M_{\odot}$) exhibit bright, complex structures\footnote{Additionally, we compare our nucleation label with the Advanced Camera for Surveys Fornax Cluster Survey (ACSFCS) galaxies of \citet{turner2012}. We find that the nucleation labels agree for the majority (29/34) of the matched galaxies and discuss them in more detail in Appendix~\ref{app:acsfcs_nuc_comparison}.}. On the low mass end of the spectrum, dwarfs are generally not as centrally concentrated (Sérsic $n \lesssim 1$) and do not exhibit as much stellar structures as massive galaxies. However, both the dwarfs and their nuclei tend to be very faint, so the signal-to-noise ratio ($S/N$) becomes the limiting factor in detecting nuclei. Given the importance of the contrast in detecting nuclei, here we explore the contrast of our sample of nuclei. 

To estimate the apparent nucleus contrast, we define it as the flux ratio of the nucleus to the underlying central galaxy flux within a circular aperture with a FWHM radius, 
\begin{equation}\label{eqn:nuc_contrast}
    \textup{nuc contrast} \approx \frac{f_{\rm nuc}}{f_{\rm galaxy,aper}},
\end{equation}
where $f_{\rm nuc}$ is the nucleus total flux, and $f_{\rm galaxy,aper}$ is the galaxy flux within the FWHM aperture (excluding the nucleus component). In order to compare the apparent nucleus contrast between the nucleated and non-nucleated galaxies in our sample, we use their multi-component models (which include the added nucleus component for the non-nucleated galaxies) to calculate the apparent nucleus contrasts.   

\begin{figure}
\centering
\includegraphics[width=\hsize]{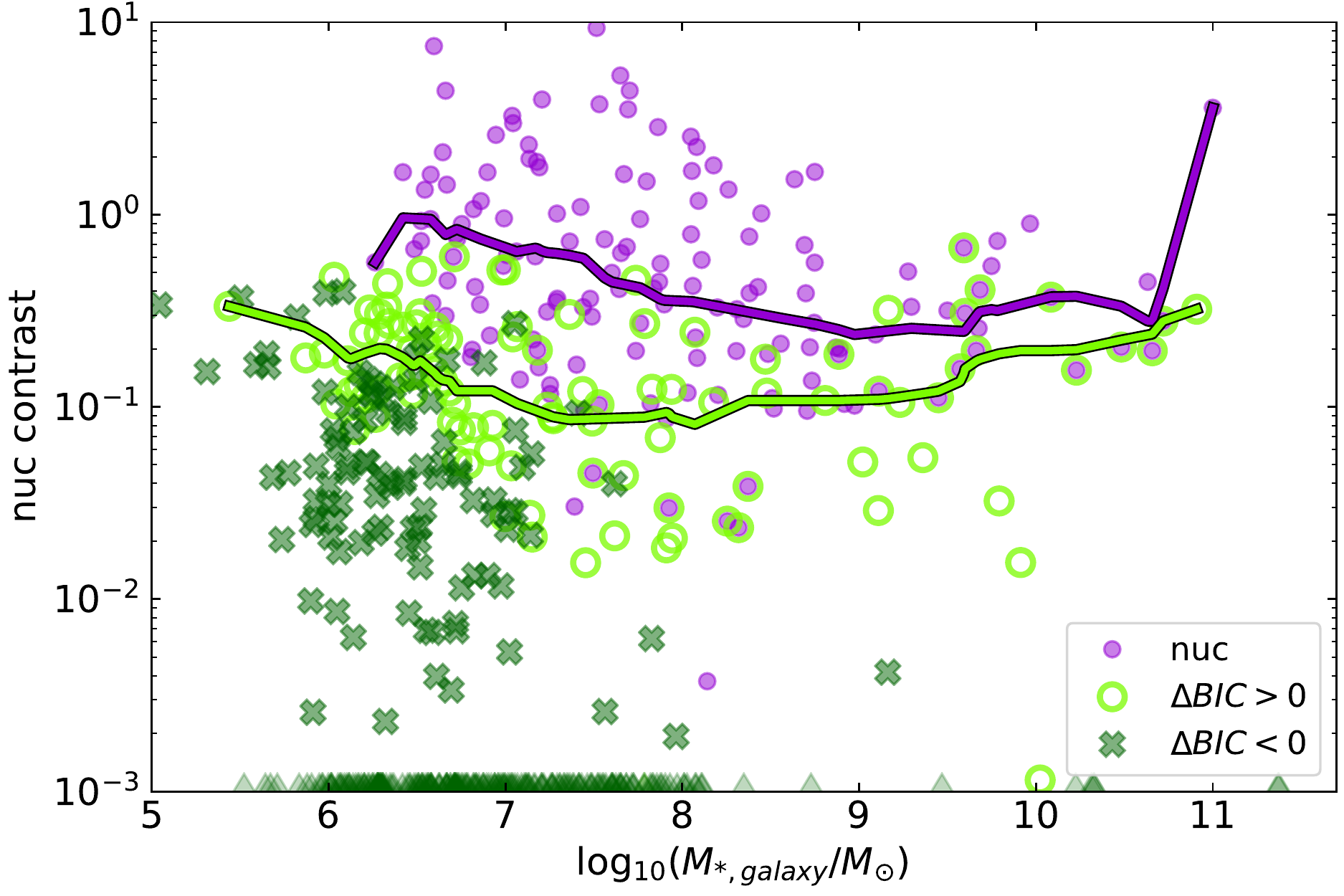}
\caption{Nucleus contrast as a function of galaxy stellar mass. The violet dots show the nucleated galaxies, whereas the green circles and crosses denote the non-nucleated galaxies with $\Delta BIC>0$ and $\Delta BIC<0$, respectively. The newly identified nucleated galaxies possess both the $\Delta BIC>0$ and nucleated symbols. For comparisons the solid lines show the moving averages for the nucleated and $\Delta BIC>0$ sub-samples. Those with nucleus contrast below $10^{-3}$ are shown as triangles along the x-axis. }
\label{fig:nuc_contrast}
\end{figure}

\begin{figure}
\centering
\includegraphics[width=\hsize]{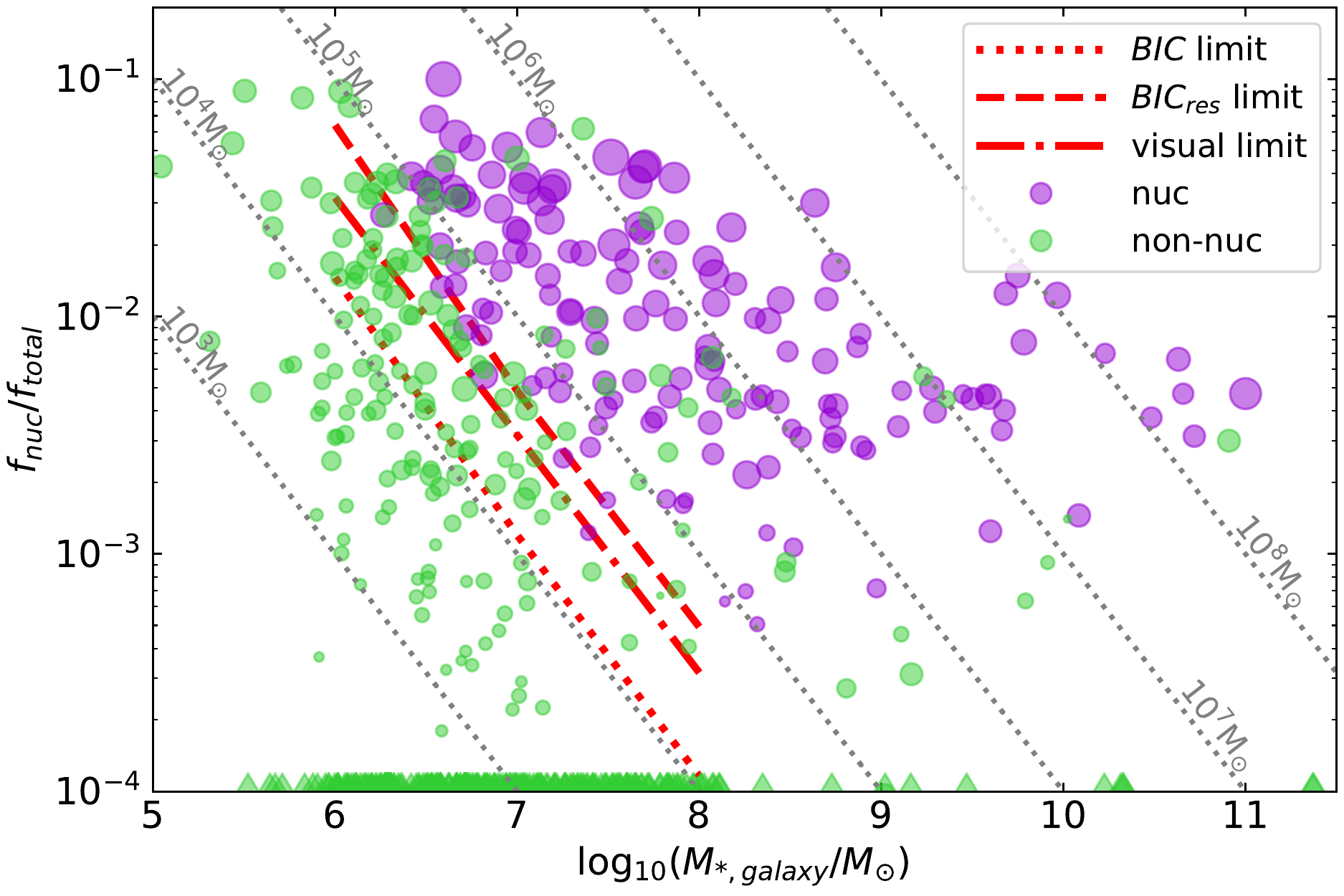}
\caption{Similar to Fig.~\ref{fig:deltabic_limits}, but based on multi-component models. The marker size denotes the nucleus contrast (i.e. larger markers denote higher contrasts). }
\label{fig:nucff_contrast_multicomp}
\end{figure}

In Fig.~\ref{fig:nuc_contrast} we find that, as expected, the galaxies which we labelled as nucleated have higher nucleus contrast than the non-nucleated galaxies. Moreover, we find that the average nucleus contrast is higher for low-mass nucleated galaxies than their higher mass counterparts, although the spread is larger at lower masses and the sample size is more modest at higher masses. Furthermore, of the non-nucleated galaxies, those with $\Delta BIC>0$ generally have higher nucleus contrast than those with $\Delta BIC<0$. The higher average values for $\Delta BIC>0$ compared to $\Delta BIC<0$ implies that the BIC most likely (indirectly) take the nucleus contrast into account. Indeed, for a galaxy which has a very low apparent nucleus contrast, the additional nucleus component would not affect the residual (and hence $\chi ^2$) as much as the penalisation for including the additional component. For $\Delta BIC<0$, the rapid decrease in the moving average is due to a lack of a significant nucleus component in the majority of their multi-comp models. Despite the differences in the moving averages, there is a clear overlap between the sub-samples at low galaxy stellar masses ($< 10^7 M_{\odot}$). There is not a clear constant limit which can delineate the nucleated from the non-nucleated sub-samples. Instead, the apparent limit appears to increase with decreasing galaxy stellar mass, which is likely a result of the $S/N$ on the apparent nucleus contrast, before increasing at the highest mass end. In Fig.~\ref{fig:nucff_contrast_multicomp} we show the nucleus flux fraction as a function of galaxy stellar mass and overlay the detection limits due to the $S/N$ (see Sect.~\ref{sect:det_limit}). 

\subsection{Detection limit}\label{sect:det_limit}

\begin{figure}
\centering
\includegraphics[width=\hsize]{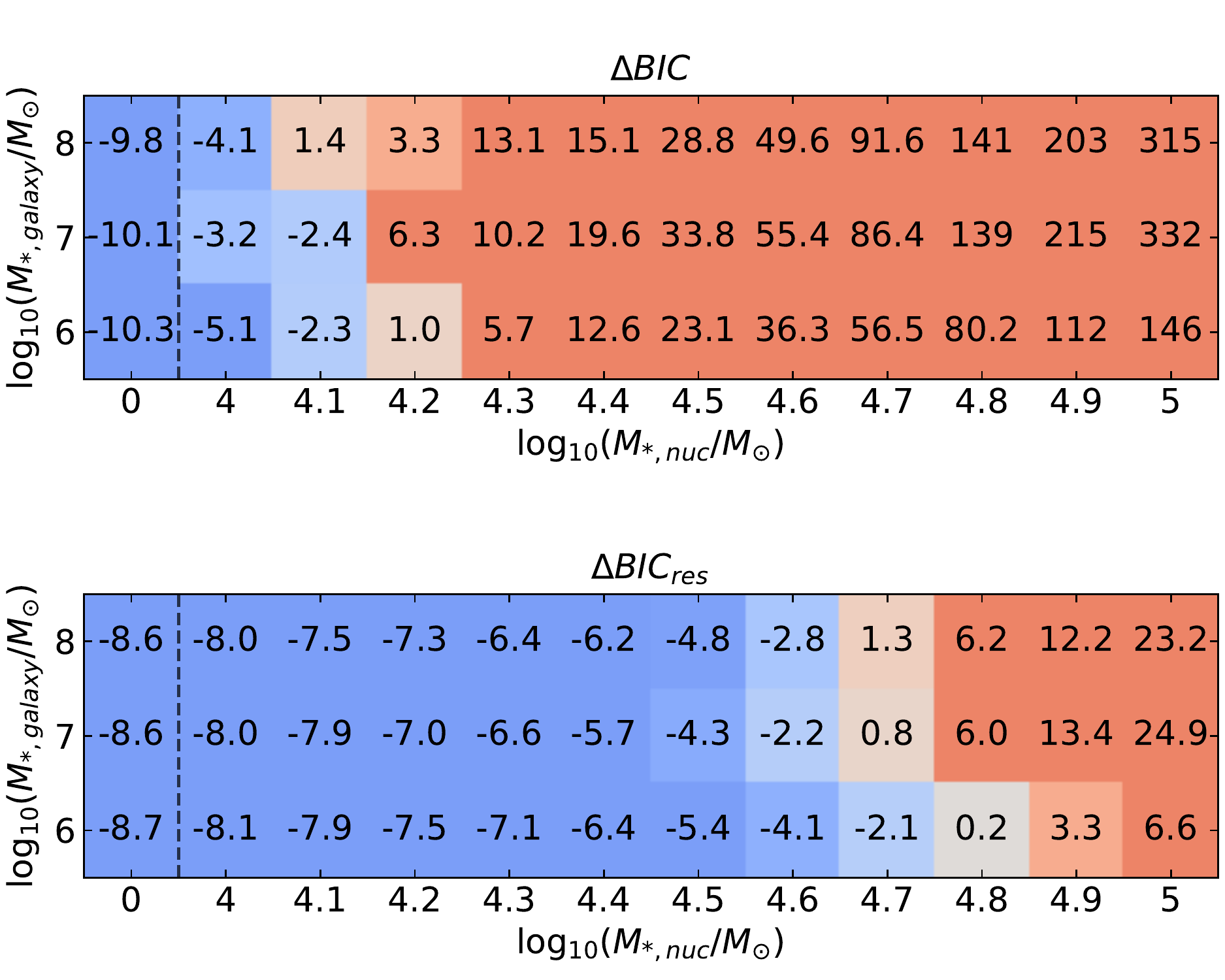}
\caption{$\Delta$BIC values for $BIC$ (upper) and $BIC_{\textup{res}}$ (lower) of synthetic nucleated galaxies with varying nucleus and galaxy stellar masses. The $\Delta$BIC values are annotated in each square. $\Delta$BIC$>0$ (red) implies that a nucleus of the given stellar mass would be detected, whereas $\Delta$BIC$<0$ (blue) would not be detected. The dashed vertical lines denote the jump between nucleated and non-nucleated synthetic galaxies. }
\label{fig:bic_completeness}
\end{figure}

\begin{figure}
\centering
\includegraphics[width=\hsize]{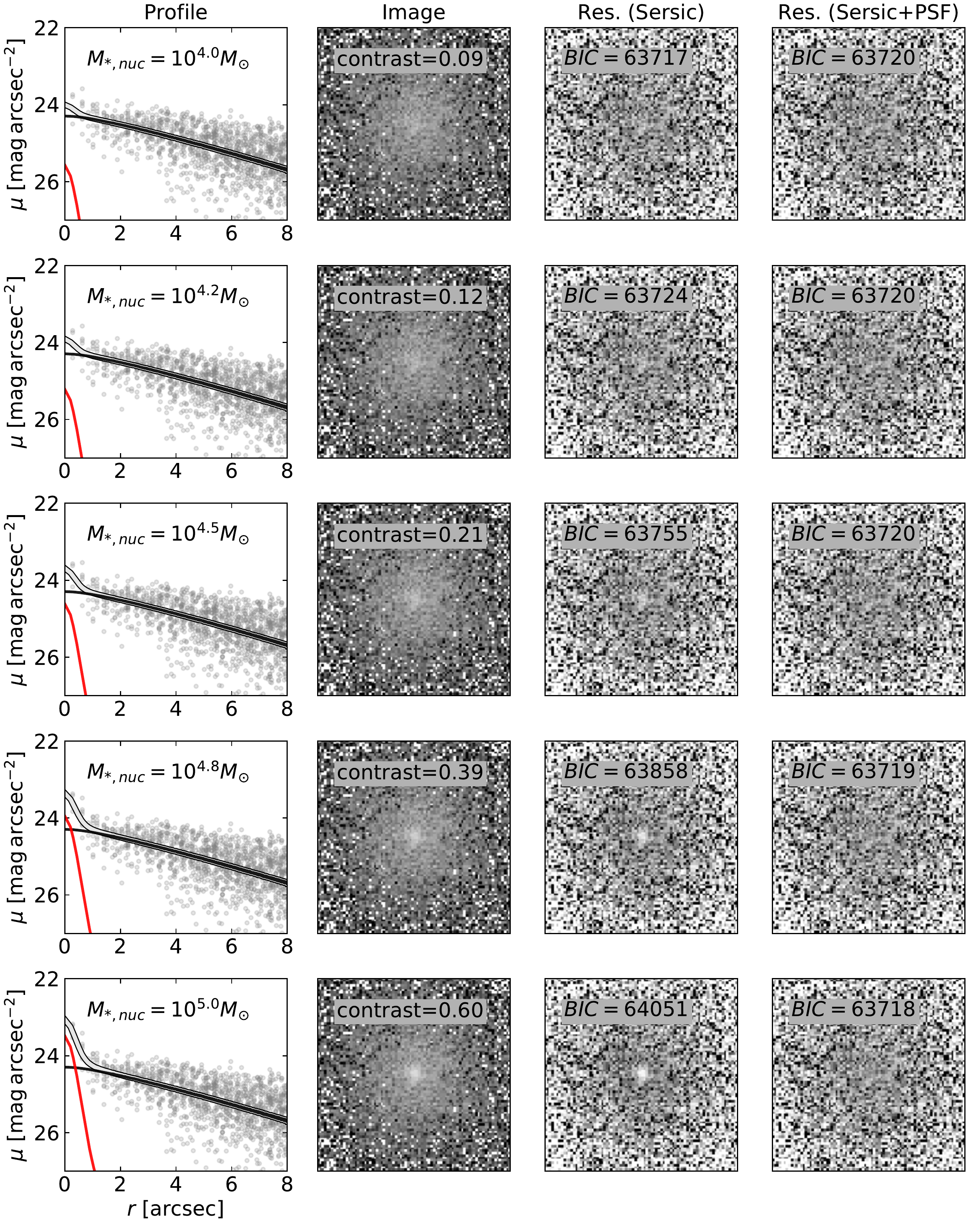}
\caption{Examples of synthetic nucleated galaxies with $M_{\rm *,galaxy} = 10^7 M_{\odot}$ and various $M_{\rm *,nuc}$ (see the middle rows of Fig.~\ref{fig:bic_completeness}). The first column shows the surface brightness profiles, where the grey points denote the surface brightness from individual pixels of the images. The galaxy (black), nucleus (red), and combined (white) profiles come from the Sérsic+PSF models. The second column shows the inner $R_e$ of the galaxy with $27 > \mu > 22$\,mag\,arcsec$^{-2}$, with the nucleus contrast (Eq.~\ref{eqn:nuc_contrast}) annotated. The third and fourth columns show the residual images from subtracting the single Sérsic and Sérsic+PSF models, respectively. The $BIC$ values for each model is annotated in the images. The detection limit for the $BIC$, visual inspection, and $BIC_{\textup{res}}$ is estimated as $10^{4.2}\,M_{\odot}$, $10^{4.5}\,M_{\odot}$, and $10^{4.8}\,M_{\odot}$, respectively.}
\label{fig:example_completeness}
\end{figure}

To quantify the effects of low $S/N$ in detecting nuclei, we first explore the $BIC$ and $BIC_{\textup{res}}$ by testing their effectiveness on detecting nucleus components in a controlled manner. To do so, we first create synthetic galaxy images with a range of galaxy stellar masses ($10^6 M_{\odot} < M_{*,\textup{galaxy}} < 10^8 M_{\odot}$) and nucleus stellar masses ($10^4 M_{\odot} < M_{*,\textup{nuc}} < 10^5 M_{\odot}$). We model each synthetic galaxy with a single Sérsic for the disk component and a PSF for the nucleus component. To ensure the synthetic galaxies appear realistic for its given galaxy stellar mass, we calculate and use the scaling relations of Sérsic parameters based on the Sérsic+PSF decomposition models (i.e. the moving averages in Fig.~\ref{fig:host_pars_etg_nosig}). The galaxy stellar mass is converted to total $r'$ band magnitude via moving averages, whereas the nucleus total magnitude is calculated using the approximation 
\begin{equation}\label{eqn:nucff_approx}
    \frac{M_{\rm *,nuc}}{M_{\rm *,galaxy}} \approx \frac{f_{\textup{nuc}}}{f_{\textup{galaxy}}},
\end{equation}
where $M_{\rm *,nuc}/M_{\rm *,galaxy}$ denotes the nucleus stellar mass fraction and $f_{\textup{nuc}}/f_{\textup{galaxy}}$ is the ratio of nucleus to galaxy total fluxes. Each nucleus component is placed at the centre of the galaxy, creating synthetic images of nucleated galaxies. All the synthetic galaxies have a face-on orientation and include realistic FDS noise based on their sigma images (constructed from the expected instrumental and photon noises). 

For each galaxy we conduct single Sérsic and Sérsic+PSF decompositions to calculate their $\Delta$BIC values for $BIC$ and $BIC_{\textup{res}}$. The point at which $\Delta$BIC$ < 0$ implies that the method does not find significant improvement in fitting the Sérsic+PSF model over the single Sérsic model at the given nucleus stellar mass. In other words, the method does not find a nucleus and a nucleus of this stellar mass would not be detected in our sample. In Fig.~\ref{fig:bic_completeness} we show the $\Delta$BIC values for a grid of galaxy and nucleus stellar masses. We find a weak trend in the nucleus and galaxy stellar mass: $\Delta$BIC$=0$ for $BIC$ and $BIC_{\textup{res}}$, such that the detection limit can be approximated as $M_{\rm *,nuc} \approx 10^{4.2} M_{\odot}$ for the $BIC$, and $M_{\rm *,nuc} \approx 10^{4.8} M_{\odot}$ for $BIC_{\textup{res}}$ for dwarfs. In Fig.~\ref{fig:example_completeness} we show the synthetic nucleated galaxies with $M_{\rm *,galaxy} = 10^7 M_{\odot}$ and different $M_{\rm *,nuc}$, including the limiting cases. In principle, the $BIC$ method of nucleus detection alone should be complete down to $M_{\rm *,nuc} \gtrsim 10^{4.2} M_{\odot}$. However, this should be considered as a lower limit, due to the simplicity of the synthetic galaxies and the closeness of the galaxy structures to the functions used in the decompositions. In practice, the detection limit is likely closer to the limit from visual inspections. From visual inspection of the images and residuals of the synthetic galaxies, we estimate the limit to be $M_{\rm *,nuc} \approx 10^{4.5} M_{\odot}$. It is worth noting that we do not account for other factors such as extinction due to dust, which can increase the nucleus detection limit. Moreover, our estimated detection limits are likely constrained by the seeing of our data. For high resolution data, such as from the HST, it is likely that the higher nucleus contrast would be able to push the nucleus detection limit to lower masses than in this work. In the context of the GC in-spiral formation channel, the lowest mass nuclei in our sample are already comparable to low mass GCs. However, it is uncertain if lower mass GCs can survive the process of dynamical friction towards the galaxy centre before they evaporate \citep[due to tidal stripping of stars; see e.g.][]{fujii2006}. Whether lower mass nuclei can exist in low mass dwarfs requires further investigation with high resolution data.

\section{Comparison of nucleation with ACSFCS}\label{app:acsfcs_nuc_comparison}
\begin{figure}
\centering
\includegraphics[width=\hsize]{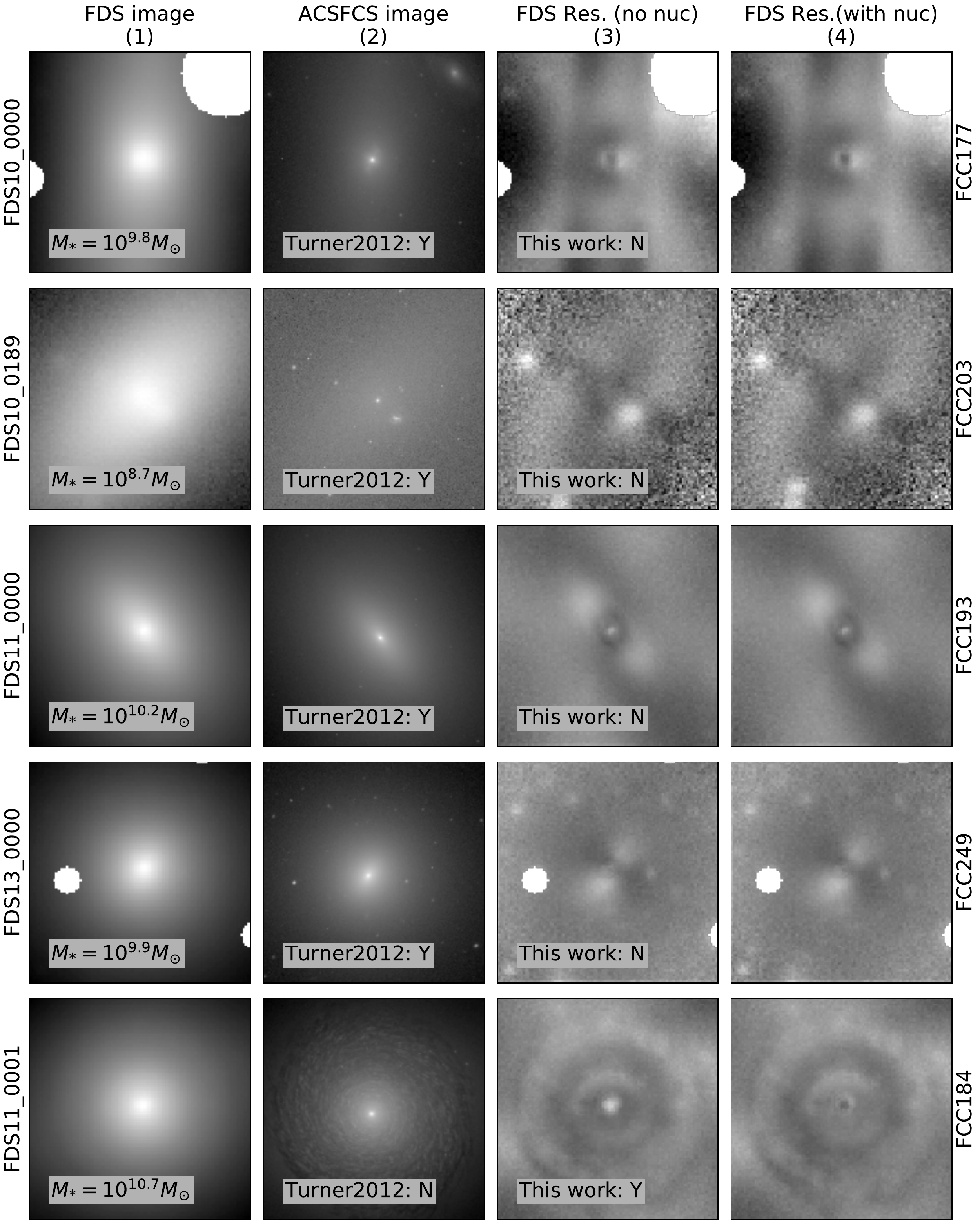}
\caption{Five galaxies which had opposing nucleation labels between \citet{turner2012} and this work. We show our galaxy images from the FDS (column~1) and the ACSFCS (column~2), as well as the residuals based on the multi-component models with (column~3) and without (column~4) a nucleus component. The nucleation labels from \citet{turner2012} and this work are annotated in column~2 and column~3, respectively (Y=nucleated, N=non-nucleated). The white regions denote masked regions. All the images have widths of 20\,arcsec. }
\label{fig:acsfcs_notmatched}
\end{figure}

Here we compare our nucleation label with that of the ACSFCS, specifically the work of \citet{turner2012}, who focused on the nucleation of 43 early-type galaxies with $B_T \leq 15.5$ from the FCC. We first cross-match the galaxies with our sample and find that 9 of the ACSFCS galaxies are not in our catalogue (FCC026, FCC043, FCC063, FCC095, FCC152, FCC204, FCC324, IC2006, NGC1340), either because they are not within the FDS coverage, or they were not deemed cluster members in the selection cuts of \citet{venhola2018}. Of the remaining 34 matched galaxies, we find that they are massive ($10^{8.7} M_{\odot} < M_* < 10^{11.4} M_{\odot}$) and a majority (26) exhibit stellar structures, such as bulges and bars, based on our multi-component models. In total, we find that the nucleation labels agree for 29/34 galaxies, and of those which are nucleated, their nuclei are relatively massive ($<M_{\rm *,nuc}> = 10^{7.1\pm 0.6}\,M_{\odot}$). In Table~\ref{tab:acsfcs_notmatched} we list the 5 galaxies which had differing nucleation labels and our remarks for each case. Additionally, we show their galaxy images as well as the residuals from the multi-component models with and without a nucleus component in Fig.~\ref{fig:acsfcs_notmatched}. Overall, the agreement between our nucleation label and those of \citet{turner2012} suggests that our nucleation detection is robust. 

\begin{table}
    \caption{Comparison of nucleation with \citet{turner2012}.}
    \centering
    \input{tab_acsfcs.txt}
    \label{tab:acsfcs_notmatched}
\end{table}

\section{Uncertainty of nucleus colours}\label{app:nuc_uncertainties}
To estimate the uncertainty in the nucleus magnitudes we considered two sources: the $S/N$, and the PSF model. The first source is most prominent for faint nuclei, where the low $S/N$ can affect the best-fit nucleus magnitude. We estimated this uncertainty by taking the formal GALFIT uncertainties. The second source of uncertainty comes from the PSF model used to fit the nuclei. To estimate this uncertainty, we created a new set of PSFs in the $g'$, $r'$, and $i'$ bands for each nucleated galaxy to use in decompositions. The new PSFs were created by first producing a catalogue of sources using Source Extractor \citep{bertin1996} based on the galaxy postage stamp images. The catalogue was then used as an input for PSF Extractor \citep[PSFEx,][]{bertin2011}, which selects suitable sources to model the PSF from. We configured PSFEx to select sources with 3\,pix$<\verb|SAMPLE_FWHMRANGE|<12$\,pix, $\verb|SAMPLE_VARIABILITY|=0.2$, $\verb|SAMPLE_MINSN|=10$, and $\verb|SAMPLE_MAXELLIP|=0.2$, and the resulting PSF is over-sampled by a factor of 2 (i.e. $\verb|PSF_SAMPLING|=0.5$). In principle the resultant PSFs from PSFEx can be used for decompositions, but the limited number of selected sources from some postage stamp images led to clear non-axisymmetric variations. To proceed, we fitted 2D Gaussians to the resultant PSFs from PSFEx to create smooth, axisymmetric PSFs, with which the multi-component decompositions were re-ran. 

\begin{figure}
\centering
\includegraphics[width=\hsize]{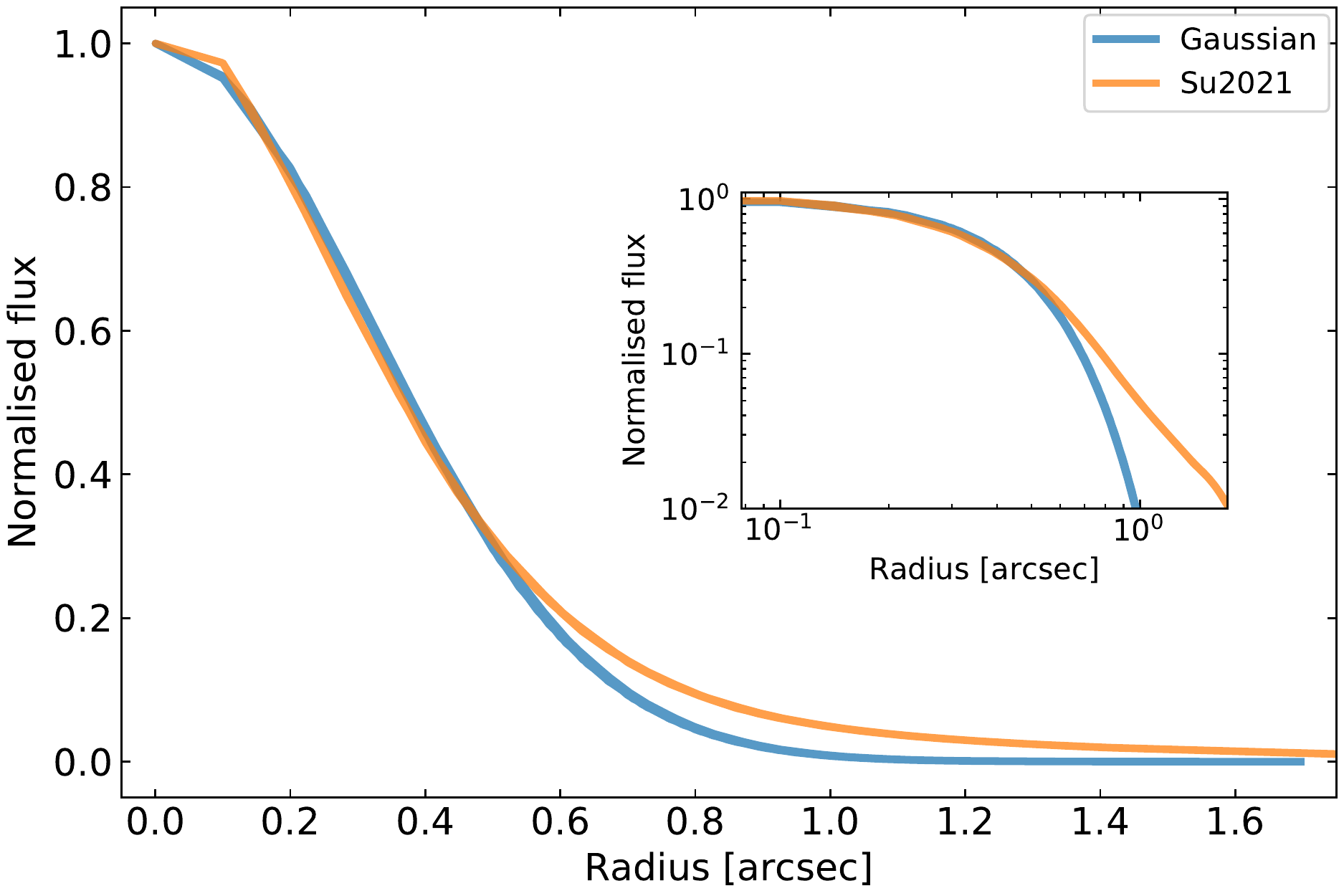}
\caption{Comparison of PSF profiles from the FDS25 field \citep[orange; from][]{su2021} and the Gaussian fits from the postage stamp image of FDS25\_0000 (blue) in the $r'$ band. The small inset plot shows the profiles in logarithmic axes. }
\label{fig:psf_comparison}
\end{figure}

\begin{figure}
\centering
\includegraphics[width=\hsize]{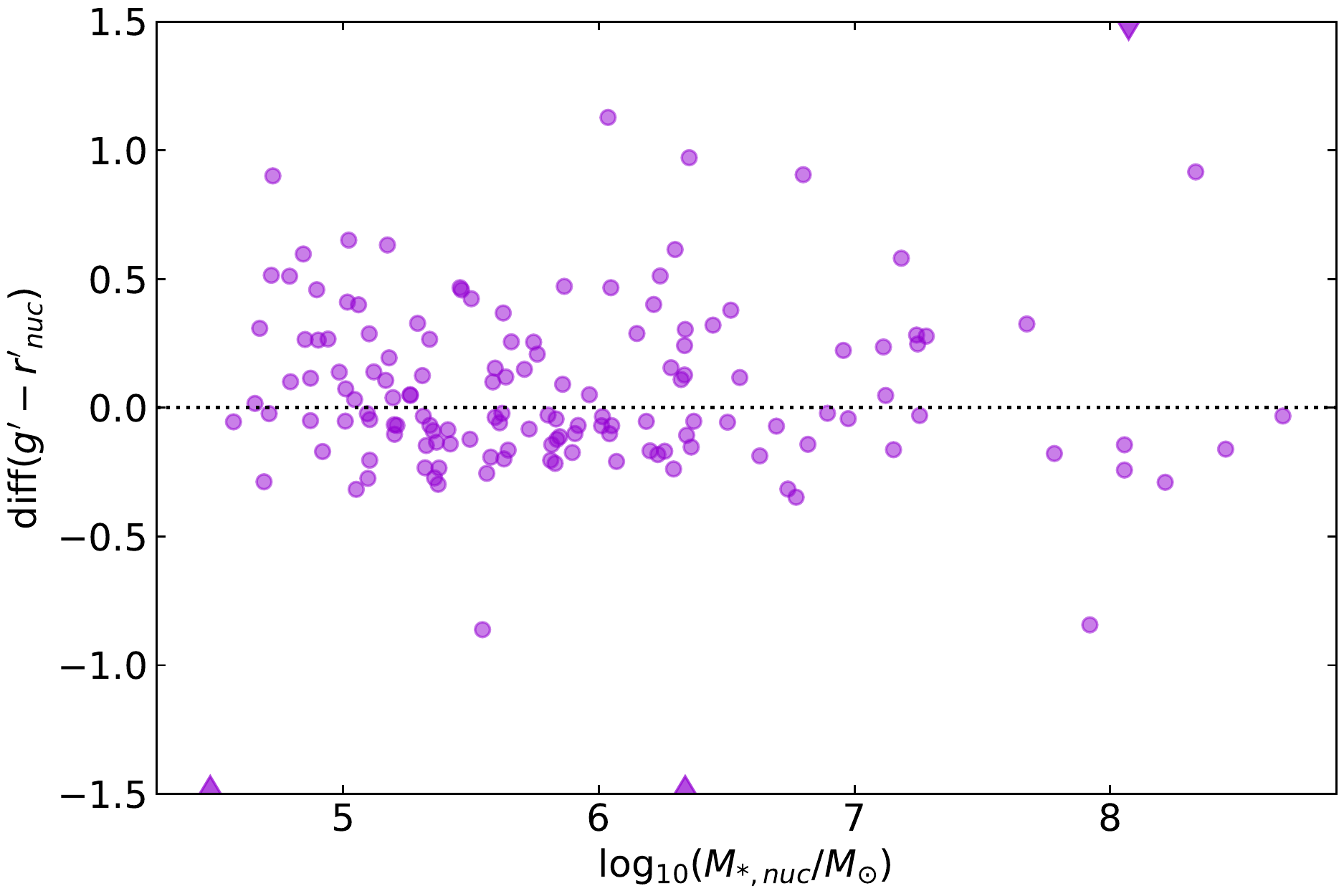}
\caption{Nucleus $g'-r'$ colour derived from the \citet{su2021} PSFs minus the colour derived from the Gaussian PSFs for each nucleated galaxy, as a function of nucleus stellar mass (same as in Fig.~\ref{fig:nuc_ucd_cmd}) for all nucleated galaxies in our sample. Nuclei with values beyond the axis limits are denoted as triangles along the x-axes.}
\label{fig:nuc_colour_uncertainty}
\end{figure}

In Fig.~\ref{fig:psf_comparison} we compare the PSFs created in \citet{su2021} and from the Gaussian fits. Both PSFs are axisymmetric and appear to be very similar within the HWHM, but differ at larger radii due to the extended tail of the \citet{su2021} PSF. As a result, the nucleus magnitudes derived from Gaussian PSFs are systematically fainter than the magnitudes from the extended PSFs. However, the systematic offsets cancel out when we calculate the colours (see Fig.~\ref{fig:nuc_colour_uncertainty}), so we estimate the uncertainty in nucleus colours as the difference in colours derived from the decompositions using the two different PSFs. 

\section{Early-type galaxy properties}\label{app:host_pars_nosig}
\begin{figure*}
\centering
\resizebox{\hsize}{!}{\includegraphics[width=\hsize]{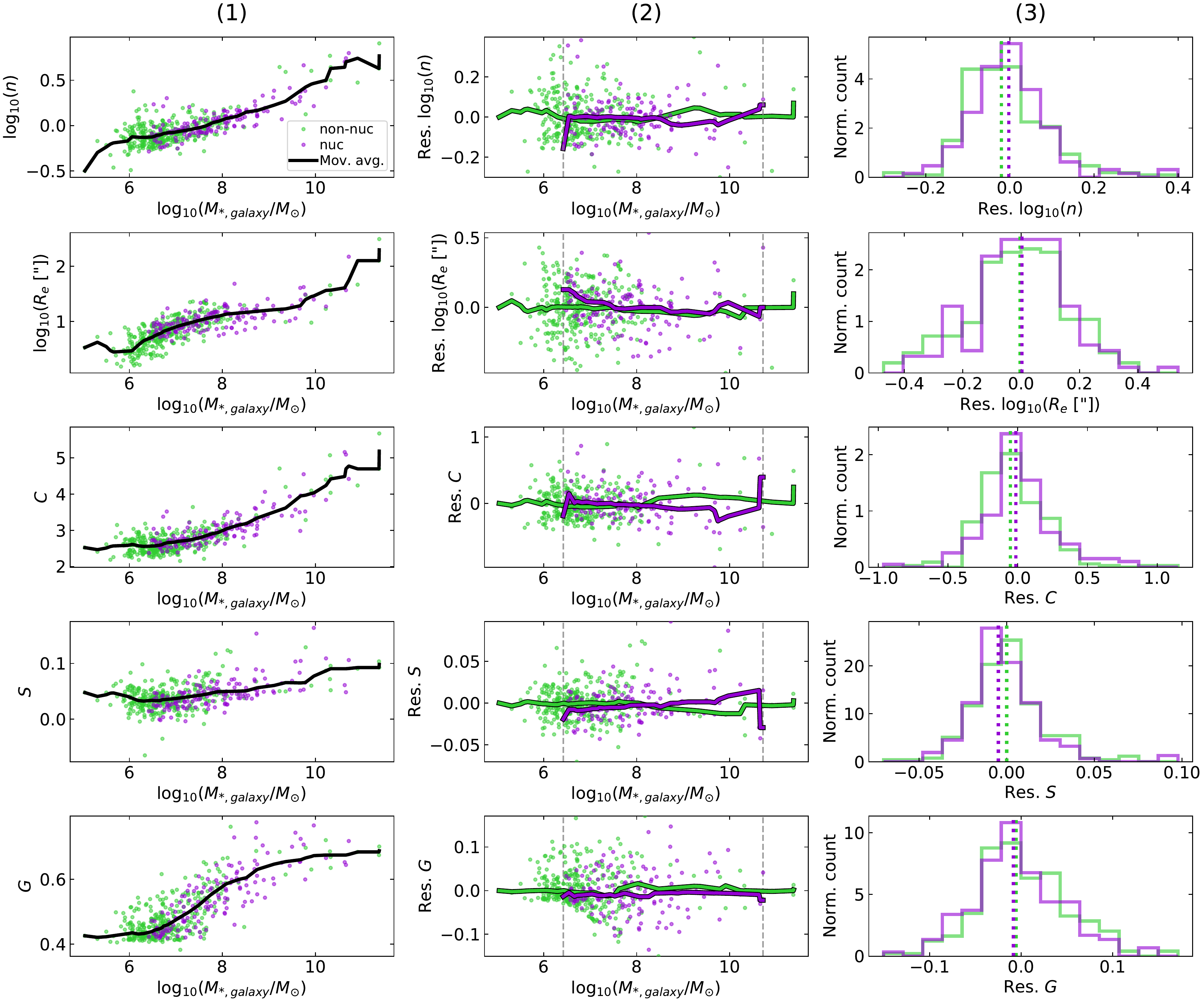}}
\caption{Same as Fig.~\ref{fig:host_pars_etg}, but for the host properties for which the p-values $>\alpha$. }
\label{fig:host_pars_etg_nosig}
\end{figure*}

In Fig.~\ref{fig:host_pars_etg_nosig} we show the host galaxy quantities of early-type nucleated and non-nucleated galaxies for which the p-values $>\alpha$, which implies that the null hypothesis (that the two sub-samples are drawn from the same distribution) cannot be rejected with confidence for these quantities. We also include the concentration index $C$ which did have p-values $<\alpha$, but this is most likely due to the fact that the nuclei were included in the calculations of $C$. 

\end{document}

%% file: table_galaxy_short.txt
\begin{tabular}{ccccccccc}
\hline\hline
FDS ID & FCC & RA\,[deg] & Dec\,[deg] & $\log_{10}(M_{*}/M_{\odot})$ & $M_{r'}$\,[mag] & $g'-r'$\,[mag] & $g'-i'$\,[mag] & Nucleation \\
(1) & (2) & (3) & (4) & (5) & (6) & (7) & (8) & (9) \\
\hline
FDS11\_0003  &    FCC213 & 54.6209 & -35.4504 &                        11.37 &          -23.19 &                  0.83 &                  1.17 &      False \\
FDS26\_0001  &    FCC021 & 50.6823 & -37.1931 &                        11.37 &          -23.55 &                  0.81 &                  1.02 &      False \\
FDS17\_0365        &    FCC121 & 53.4015 & -36.1408 &                        11.00 &          -22.52 &                  0.72 &                  1.03 &       True \\
FDS11\_0006        &    FCC167 & 54.1150 & -34.9760 &                        10.91 &          -21.99 &                  0.79 &                  1.16 &      False \\
FDS14\_0133        &    FCC088 & 52.7835 & -33.6284 &                        10.72 &          -21.66 &                  0.78 &                  1.11 &       True \\
FDS11\_0001        &    FCC184 & 54.2376 & -35.5066 &                        10.66 &          -21.30 &                  0.88 &                  1.22 &       True \\
FDS25\_0000        &    FCC029 & 50.9848 & -36.4644 &                        10.63 &          -21.37 &                  0.75 &                  1.12 &       True \\
FDS26\_0254        &    FCC022 & 50.6889 & -37.1051 &                        10.48 &          -21.04 &                  0.79 &                  1.12 &       True \\
FDS16\_0001        &    FCC147 & 53.8191 & -35.2263 &                        10.33 &          -20.85 &                  0.73 &                  1.03 &      False \\
FDS12\_0003        &    FCC179 & 54.1925 & -35.9993 &                        10.33 &          -20.67 &                  0.74 &                  1.09 &      False \\
\ldots & \ldots & \ldots & \ldots & \ldots & \ldots & \ldots & \ldots & \ldots \\
\hline
\end{tabular}

%% file: table_nucleus_short.txt
\begin{tabular}{ccccccc}
\hline\hline
FDS ID & FCC & $\log_{10}(M_{\rm *,galaxy}/M_{\odot})$ & $\log_{10}(M_{\rm *,nuc}/M_{\odot})$ & $(g'-r')_{\rm nuc}$\,[mag] & $(g'-i')_{\rm nuc}$\,[mag] & $f_{\rm nuc}/f_{\rm total}$ \\
(1) & (2) & (3) & (4) & (5) & (6) & (7) \\
\hline
FDS14\_0133  &  FCC088 &                     10.72 &                      8.22 &               1.29 &               1.69 &                     0.003 \\
FDS11\_0001  &  FCC184 &                     10.66 &                      8.33 &               1.53 &               2.31 &                     0.005 \\
FDS25\_0000  &  FCC029 &                     10.63 &                      8.45 &               1.41 &               2.02 &                     0.007 \\
FDS11\_0004  &  FCC170 &                     10.23 &                      8.07 &               1.91 &               2.27 &                     0.007 \\
FDS15\_0002  &  FCC153 &                      9.97 &                      8.06 &               0.50 &               0.98 &                     0.012 \\
FDS7\_0737   &  FCC310 &                      9.78 &                      7.67 &               1.19 &               1.55 &                     0.008 \\
FDS20\_0000  &  FCC047 &                      9.75 &                      7.92 &               0.94 &               1.17 &                     0.015 \\
FDS16\_0000  &  FCC148 &                      9.68 &                      7.78 &               0.33 &               0.65 &                     0.013 \\
FDS12\_0002  &  FCC176 &                      9.68 &                      7.28 &               1.04 &               1.18 &                     0.004 \\
FDS11\_0005  &  FCC190 &                      9.66 &                      7.18 &               1.02 &               1.13 &                     0.003 \\

\ldots & \ldots & \ldots & \ldots & \ldots & \ldots & \ldots \\
\hline
\end{tabular}

%% file: fds_nuc_etg_stats.txt
\begin{tabular}{cccccc}
\hline\hline
Residual &              KS &              AD &              LP &               CU & Nuc w.r.t. non-nuc \\
(1) & (2) & (3) & (4) & (5) & (6) \\
\hline
$g'-r'$\,[mag]                &  \textbf{0.003} &  \textbf{0.003} &  \textbf{0.003} &  \textbf{0.007} & redder \\
$\log_{10}(n)$         &           0.334 &           0.250 &           0.458 &           0.352 & - \\
$\log_{10}(R_e$\,[arcsec]$)$ &           0.653 &           0.250 &           0.555 &           0.430 & - \\
$\bar{\mu}_{e,r'}$\,[mag\,arcsec$^{-2}$]     &  0.083 &           0.096 &           0.180 &           0.216 & - \\
$C$                    &  \textbf{0.015} &  \textbf{0.040} &           0.096 &           0.121 & more concentrated\tablefootmark{a} \\
$A$                    &  \textbf{0.012} &  \textbf{0.008} &  \textbf{0.002} &  \textbf{0.002} & less asymmetric \\
$S$                    &           0.141 &           0.208 &           0.386 &           0.342 & - \\
$G$                    &           0.129 &           0.250 &           0.390 &           0.483 & - \\
$M_{20}$               &           0.096 &  \textbf{0.038} &  \textbf{0.004} &  \textbf{0.001} & lower scatter \\
$\Delta(g'-r')$\,[mag]        &  \textbf{0.026} &  \textbf{0.010} &  \textbf{0.027} &  \textbf{0.018} & redder outskirts \\
\hline
\end{tabular}

%% file: fds_bic_accuracies_etg.txt
\begin{tabular}{lcccc}
\hline
\hline
\multicolumn{1}{c}{} & \multicolumn{2}{c}{$BIC$} & \multicolumn{2}{c}{$BIC_{\textup{res}}$}\\
\cmidrule(r){2-3}\cmidrule(l){4-5}
{} &               {All} &               {$<10^9M_{\odot}$} &               {All} &                {$<10^9M_{\odot}$} \\
\hline

TP & 135 & 124 & 118 & 107 \\
TN & 281 & 276 & 316 & 311 \\
FP & 45 & 39 & 10 & 4 \\
FN & 5 & 0 & 22 & 17 \\
Accuracy & 89\% & 93\% & 93\% & 97\% \\

\hline
\end{tabular}

%% file: tab_acsfcs.txt
\begin{tabular}{p{0.18\linewidth}p{0.12\linewidth}p{0.12\linewidth}p{0.1\linewidth}p{0.3\linewidth}}
\hline\hline
\multicolumn{1}{c}{} & \multicolumn{1}{c}{} & \multicolumn{2}{c}{Nucleation} & \multicolumn{1}{c}{}\\
\cmidrule(r){3-4}
FDS ID &     FCC & Turner et al. & This work & Remarks \\
\hline
FDS10\_0000 &  FCC177 &             Y &          N &  Our multi-component model fitted the central structure as a bulge rather than a nucleus. \\
FDS10\_0189 &  FCC203 &             Y &          N &  No clear sign of a nucleus at the centre in our residual images; possibly an offset nucleus. \\
FDS11\_0000 &  FCC193 &             Y &          N &  Sign of positive residuals in our model without a nucleus component, although the model including a nucleus component does not offer any improvement in the residuals ($\Delta BIC<0$). \\
FDS13\_0000 &  FCC249 &             Y &          N &  No clear sign of a nucleus at the centre in our residual images \\
FDS11\_0001 &  FCC184 &             N &          Y &  Our residual images clearly shows a nucleus and is well fitted by including a nucleus component. \\
\hline
\end{tabular}